\documentclass[aps,prd,reprint,twocolumn,superscriptaddress,preprintnumbers,nofootinbib]{revtex4-2}

\usepackage{amsthm}
\usepackage{amsmath}
\usepackage{slashed}
\usepackage{amssymb}
\usepackage{graphicx,color,xcolor}
\usepackage{amsmath,amssymb,amsfonts}
\usepackage[caption=false]{subfig}
\usepackage[colorlinks=True, citecolor=blue, urlcolor=blue, linkcolor=blue]{hyperref}
\usepackage{bbm}
\usepackage{scalerel,stackengine}
\usepackage{wasysym}
\usepackage{qcircuit}
\usepackage{physics}
\usepackage{bbold}
\usepackage{orcidlink}
\usepackage{comment}

\newcommand{\nn}{\nonumber}

\newcommand{\be}{\begin{eqnarray}}
\newcommand{\ee}{\end{eqnarray}}

\newcommand{\ml}{\mathcal}
\newcommand{\bs}{\boldsymbol}

\usepackage{soul}

\begin{document}

\title{Classical and Quantum Computing of Shear Viscosity for $2+1D$ SU(2) Gauge Theory}

\author{Francesco Turro \orcidlink{0000-0002-1107-2873}}
\email{fturro@uw.edu}
\thanks{These two authors contributed equally.}
\affiliation{InQubator for Quantum Simulation, University of Washington, Seattle, WA 98195, USA}

\author{Anthony Ciavarella \orcidlink{0000-0003-3918-4110}}
\email{anciavarella@lbl.gov}
\affiliation{Lawrence Berkeley National Laboratory, Berkeley, California 94720, USA}

\author{Xiaojun Yao \orcidlink{0000-0002-8377-2203}}
\email{xjyao@uw.edu}
\thanks{These two authors contributed equally.}
\affiliation{InQubator for Quantum Simulation, University of Washington, Seattle, WA 98195, USA}

\date{\today}
\preprint{IQuS@UW-21-073}
\begin{abstract}
We perform a nonperturbative calculation of the shear viscosity for $(2+1)$-dimensional SU(2) gauge theory by using the lattice Hamiltonian formulation. The retarded Green's function of the stress-energy tensor is calculated from real time evolution via exact diagonalization of the lattice Hamiltonian with a local Hilbert space truncation, and the shear viscosity is obtained via the Kubo formula. When taking the continuum limit, we account for the renormalization group flow of the coupling but no additional operator renormalization. We find the ratio of the shear viscosity and the entropy density $\frac{\eta}{s}$ is consistent with a well-known holographic result $\frac{1}{4\pi}$ at several temperatures on a $4\times4$ honeycomb lattice with the local electric representation truncated at $j_{\rm max}=\frac{1}{2}$. We also find the ratio of the spectral function and frequency $\frac{\rho^{xy}(\omega)}{\omega}$ exhibits a peak structure when the frequency is small. 

Both the exact diagonalization method and simple matrix product state classical simulation method beyond $j_{\rm max}=\frac{1}{2}$ on bigger lattices require exponentially growing resources. So we develop a quantum computing method to calculate the retarded Green's function and analyze various systematics of the calculation including $j_{\rm max}$ truncation and finite size effects, Trotter errors and the thermal state preparation efficiency. Our thermal state preparation method still requires resources that grow exponentially with the lattice size, but with a very small prefactor at high temperature. We test our quantum circuit on both the Quantinuum emulator and the IBM simulator for a small lattice and obtain results consistent with the classical computing ones.
\end{abstract}

\maketitle

\section{Introduction}
The scientific goal of relativistic heavy ion collisions is to study the deconfined phase of nuclear matter at finite temperature and/or density, known as the quark-gluon plasma (QGP). The most striking property of the QGP created in current heavy ion collision experiments is its small shear viscosity, as shown by the good agreement between the experimental data on various particles' yields and azimuthal distributions and a description that is mainly based on relativistic hydrodynamic equations with small shear viscosity~\cite{Song:2010mg,Schenke:2010rr}. The smallness of the shear viscosity indicates the QGP created in current heavy ion collision experiments is strongly coupled. Interestingly, the current value of the ratio between the shear viscosity and the entropy density $\frac{\eta}{s}$ extracted from experimental data~\cite{Bernhard:2019bmu,Nijs:2020ors} is consistent with the AdS/CFT calculation result for a $\ml{N}=4$ supersymmetric Yang-Mills plasma in the strong coupling limit, which is $\frac{1}{4\pi}$~\cite{Policastro:2001yc}.

Theoretically, shear viscosity can be calculated from real time two-point correlation functions of stress-energy tensors via the Kubo formula. However, this computation is hard for both perturbative and nonperturbative approaches in QCD~\cite{Moore:2020pfu}. Perturbatively, certain diagrams have to be resummed due to the existence of ``pinching poles''~\cite{Jeon:1994if,Arnold:2000dr,Arnold:2003zc} and the convergence of the perturbative series is poor when the temperature of the QGP is below $1$~GeV~\cite{Ghiglieri:2018dib}, which is the temperature range of most interest in current collision experiments. Nonperturbatively, Euclidean lattice QCD methods have been applied to calculate the relevant two-point correlation functions in imaginary time~\cite{Meyer:2007ic,Mages:2015rea,Altenkort:2022yhb}. However, extraction of the shear viscosity from the imaginary time correlation function involves an ill-defined ``inverse problem,'' and thus is not under good theoretical control. Different frequency dependence of the real time correlation function can give the same Euclidean correlation function in imaginary time. These limitations of current theoretical studies urgently demand a new technique for transport coefficients calculations, since a fully theoretical determination of the shear viscosity provides an independent check of the experimental extraction and is thus of great value. These calculations can test the current hydrodynamical framework for heavy ion collisions and deepen our understanding of nuclear matter in conditions that are unachievable in experiments at the moment.

In this paper, we consider the Hamiltonian formulation of lattice gauge theory and investigate the calculation of shear viscosity from the retarded correlation function obtained from real time Hamiltonian evolution. Our study is motivated by recent developments in quantum computing for lattice gauge theories~\cite{banuls2020simulating,klco2022standard,Bauer_2023,beck2023quantum,bauer2023quantum,Lamm:2019bik}, which follow Feynman's idea~\cite{feynman} to use quantum computers and hopefully will be able to tame the exponential growth of the Hilbert space and perform more efficient quantum simulations than classical devices. A previous work studied the construction of lattice operators for stress-energy tensors and quantum algorithms for thermal state preparation in the same context~\cite{Cohen:2021imf}. Here, more specifically, we will carry out detailed calculations for the SU(2) pure gauge theory in $2+1$ dimensions ($2+1D$) on a small lattice via both classical and quantum methods and analyze various systematics. The paper is organized as follows: In Sec.~\ref{sec:eta}, we will briefly review the Kubo formula for the shear viscosity calculation in the context of the SU(2) pure gauge theory in $2+1D$. Then, in Sec.~\ref{sec:lattice}, we will explain the lattice Hamiltonian formulation for the calculation, followed by an introduction of a quantum algorithm in Sec.~\ref{sec:qc}. Various systematics of the calculation will be discussed in Sec.~\ref{sec:sys}. We will show both classical and quantum results in Sec.~\ref{sec:results} and draw conclusions in Sec.~\ref{sec:conclusions}.

\section{Shear Viscosity in $2+1$\texorpdfstring{$D$}{} SU(2) Gauge Theory}
\label{sec:eta}
The Lagrangian density of the continuum $2+1D$ SU(2) gauge theory can be written as
\be
\ml{L} =  - \frac{1}{4g^2}F_{\mu\nu}^aF^{a\mu\nu} \,,
\ee
where $g$ is the coupling constant, and $F^a_{\mu\nu} = \partial_\mu A^a_\nu - \partial_\nu A^a_\mu + f^{abc}A_\mu^b A_\nu^c$ is the non-Abelian field strength tensor with $A_\mu^a$ as the gauge field. In particular, $F_{0i}^a$ denotes the electric field along the $i$-th spatial direction and $F_{ij}^a$ is related to the non-Abelian magnetic field. The alphabet indices $a,b,c\in[1,2,3]$ label the SU(2) adjoint indices.

Stress-energy tensors of the theory are given as
\begin{align}
\label{eqn:Tmunu}
T^{\mu\nu} &= -\frac{1}{g^2} F^{a\mu\rho}F^{a\nu}_{~~~\rho} + \frac{1}{4g^2} \eta^{\mu\nu} F^{a\rho\sigma}F^a_{\rho\sigma} \,.
\end{align}
Standard linear response analysis and gradient expansion of the stress-energy tensor for relativistic hydrodynamics in the Minkowski spacetime with a small metric perturbation $h_{xy}(t)$ lead to~\cite{Baier:2007ix,Moore:2010bu}\footnote{The total metric of the spacetime is $g_{\mu\nu} = \eta_{\mu\nu} + h_{\mu\nu}(t)$, where $\eta_{\mu\nu}={\rm diag}(+1,-1,-1)$ is the Minkowski metric and the only nonzero elements of the perturbation are $h_{xy}$ and $h_{yx}$.}
\be
\label{eqn:eta}
\eta = \lim_{\omega\to0} \frac{\partial}{\partial\omega} G_r^{xy}(\omega) \,,
\ee
where $G_r^{xy}(\omega)$ can be expressed in terms of the retarded Green's function of $T^{xy}$\footnote{Our definition of the retarded Green's function has no $-i$, which leads to the prefactor $+1$ in Eq.~\eqref{eqn:eta} instead of $i$, compared with Ref.~\cite{Moore:2010bu}. Our convention has been used in recent perturbative calculations for quarkonium transport~\cite{Binder:2021otw}.}
\begin{align}
\label{eqn:Gr1}
G_r^{xy}(\omega) &= \int {\rm d}t\, e^{i\omega t} G_r^{xy}(t) \equiv \int {\rm d}t\, {\rm d}^2x \, e^{i\omega t} G_r^{xy}(t,{\boldsymbol x}) \nn\\
G_r^{xy}(t,{\boldsymbol x}) &\equiv \theta(t) {\rm Tr}\big( [T^{xy}(t,{\boldsymbol x}), T^{xy}(0,{\boldsymbol 0})] \rho_T \big) \,.
\end{align}
The density matrix in the definition of $G_r^{xy}(t,{\boldsymbol x})$ describes the thermal state at temperature $T=\beta^{-1}$
\be
\label{eqn:rho_T}
\rho_T = \frac{1}{Z}e^{-\frac{H}{T}} = \frac{e^{-\beta H}}{{\rm Tr}\, e^{-\beta H}} \,.
\ee
By utilizing translation invariance, the retarded correlation function can also be written as
\begin{align}
\label{eqn:Gr2}
G_r^{xy}(\omega) &=
\frac{1}{\mathcal{A}}\int {\rm d}t\, e^{i\omega t} \theta(t) {\rm Tr}\big([ \widetilde{T}^{xy}(t) , \widetilde{T}^{xy}(0) ] \rho_T\big) \nn\\
\widetilde{T}^{xy}(t) &= \int {\rm d}^2x \, T^{xy}(t,{\bs x}) \,,
\end{align}
where $\mathcal{A}$ denotes the area of the system.

Combining everything together gives
\begin{align}
\eta = - \int_0^{\infty} t\,{\rm d}t \, {\rm Im} G_r^{xy}(t)\,.
\end{align}
We have two ways to evaluate $G_r^{xy}(t)$ as shown in Eqs.~\eqref{eqn:Gr1} and~\eqref{eqn:Gr2}.
If we know all the eigenstates $|n\rangle$ of the system and their corresponding eigenenergies $E_n$, we can write
\begin{align}
\label{eqn:eta_to_use}
\eta & = \lim_{t_f\to\infty} \tilde{\eta}(t_f) \nn\\
\tilde{\eta}(t_f) &\equiv - \int_0^{t_f} t\,{\rm d}t \, {\rm Im} G_r^{xy}(t) \nn\\
& = -\frac{2}{Z}  \sum_n \sum_{m\neq n} \langle n|\widetilde{T}^{xy}|m \rangle \langle m|{T}^{xy}|n \rangle e^{-\beta E_n} f(t_f) \nn\\
& = -\frac{2}{Z\ml{A}}  \sum_n \sum_{m\neq n} |\langle n|\widetilde{T}^{xy}|m \rangle|^2 e^{-\beta E_n} f(t_f) \nn\\
f(t_f) &\equiv \frac{\sin((E_n-E_m)t_f)}{(E_n-E_m)^2} - \frac{t_f\cos((E_n-E_m)t_f)}{E_n-E_m} \,.
\end{align}
When the system has an infinite number of states, as continuum quantum field theories do, the symbol $\sum_n$ means $\int {\rm d}E_n\rho(E_n)$, where $\rho(E_n)$ is the eigenstate density at energy $E_n$.

Rigorously speaking, Eq.~\eqref{eqn:eta} is the tree-level matching condition between the hydrodynamic effective theory and the full theory. In this sense, $\eta$ can be thought of as a Wilson coefficient of the hydrodynamic effective theory. The matching condition between $\eta$ and the retarded Green's function becomes more complicated once nonlinear terms are taken into account in $T^{xy}$ of the hydrodynamic effective theory. These nonlinear terms contribute at one-loop level and generate the so-called long-time tails~\cite{Kovtun:2003vj}, which lead to a logarithmic divergence in two spatial dimensions~\cite{Kovtun:2012rj,Romatschke:2021imm}. As a result, $\eta$ becomes scale dependent; it is some function of $\eta(\omega)$ that equals $G_r^{xy}(\omega)$ when $\omega$ is small \{see, e.g., Eq.~(52) in Ref.~\cite{Romatschke:2021imm} as the one-loop matching condition\}. In this work, we will only consider the tree-level matching and use Eq.~\eqref{eqn:eta_to_use} to calculate the shear viscosity for simplicity. 
Classically, Eq.~\eqref{eqn:eta_to_use} can be evaluated on a lattice by solving all eigenenergies and eigenstates. We will also take the continuum limit along the renormalization group flow of the coupling.

\section{Lattice Hamiltonian Formulation}
\label{sec:lattice}
\subsection{General Setup}

\begin{figure}
\includegraphics[width=0.44\textwidth]{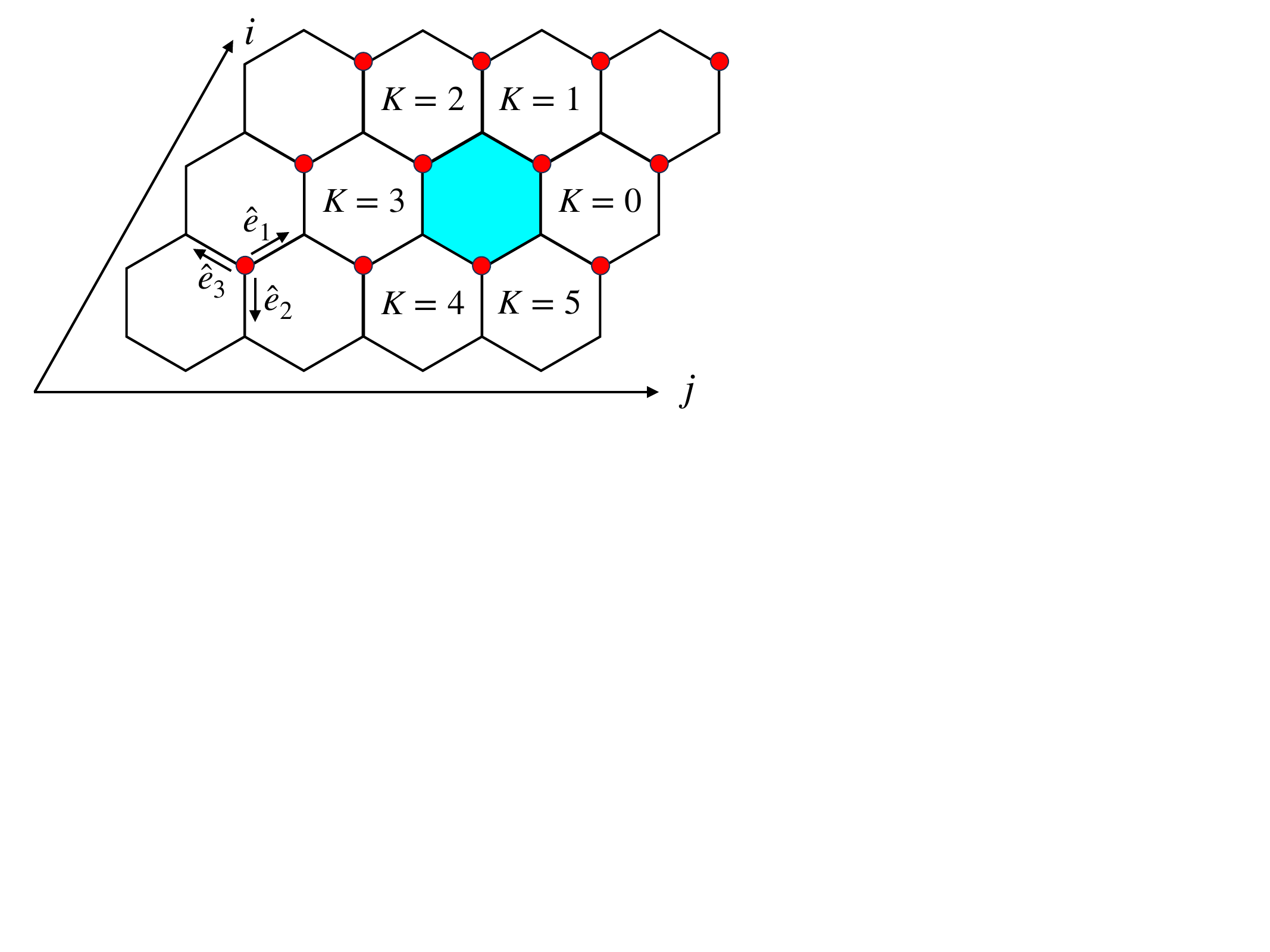}
\caption{Honeycomb lattice on which the central position of each plaquette is labeled by $(i,j)$ along the two axes shown, starting from $i=j=0$. For example, the plaquette marked in blue is located at $(i=1,j=2)$. The red dots represent the positions at which the stress-energy tensors are evaluated. The location of a red dot can be identified by the coordinates of the three plaquettes sharing the same red dot. The six plaquettes labeled by the $K$ values around the blue-colored plaquette at $(i=1,j=2)$ are used when the magnetic term at $(i=1,j=2)$ is written out explicitly, as in
Eq.~\eqref{eq:H_Ising}.}
\label{fig:honeycomb_lattice}
\end{figure}

The Kogut-Susskind Hamiltonian~\cite{PhysRevD.11.395} of the 2+1$D$ SU(2) gauge theory can be discretized on a honeycomb lattice as shown in Fig.~\ref{fig:honeycomb_lattice}
\begin{align}
\label{eq:HKS}
H &=   \frac{3\sqrt{3}g^2}{4}  \sum_{\rm links} E_i^a E_i^a
- \frac{4\sqrt{3}}{9 g^2a^2}  \sum_{\rm plaqs} \varhexagon  \nn\\
\varhexagon &\equiv {\rm Tr} \bigg( \prod_{({\bs x},\hat{i})\in {\rm plaq}} U({\bs x},\hat{i}) \bigg)\,,
\end{align}
where $a$ in the denominator is the side length of the honeycomb, and we have shifted the energy reference point. The honeycomb plaquette operator $\varhexagon$ is defined as the trace of the product of the six Wilson lines $U({\bs x},\hat{i})$ on the edges of one honeycomb. The two-vector ${\bs x} = (i,j)$ labels the position of a honeycomb on the lattice plane along the directions specified as in Fig.~\ref{fig:honeycomb_lattice}. The electric field 
\be
\label{eqn:def_E}
{\bs E}^a = (E_x^a,E_y^a)\equiv \frac{a}{g^2}(F^a_{0x}, F^a_{0y})\,,
\ee
is projected along three unit directions $E_i^a \equiv \hat{e}_i\cdot {\bs E}^a$, where the three unit vectors are defined as in Fig.~\ref{fig:honeycomb_lattice}. On each link, only one type of projected electric field lives, i.e., $i$ is $1$, $2$ or $3$. More details can be found in Ref.~\cite{Muller:2023nnk}. Physical states satisfy Gauss's law
\begin{align}
\sum_{i=1}^3 E_{i}^a |\psi_{\rm phy}\rangle = 0 \,,
\end{align}
at each vertex for every $a$.

We use the electric basis that labels each link by the quantum number $j$. In this basis, the electric energy is diagonal~\cite{Byrnes:2005qx,Zohar:2014qma,Liu:2021tef}
\begin{align}
\langle J | E^a_iE^a_i | j\rangle = j(j+1)\delta_{Jj} \, .
\end{align}
The matrix element of the plaquette term (magnetic energy) has been worked out to be~\cite{Zache:2023dko,Hayata:2023puo,Muller:2023nnk} (see Refs.~\cite{Klco:2019evd,ARahman:2021ktn,Hayata:2021kcp,ARahman:2022tkr,Yao:2023pht} for the square plaquette case):
\begin{align}
\label{eq:6j}
&\langle \{J\} | \varhexagon |\{j\} \rangle = \\
&\prod_{V=1}^{6} (-1)^{j_a+J_b+j_x} \sqrt{(2J_a+1)(2j_b+1)}
\bigg\{ \begin{array}{ccc}  j_x & j_a & j_b \\ \frac{1}{2} & J_b & J_a   \end{array}  \bigg\} \, ,\nn
\end{align}
where $\{j\}$ ($\{J\}$) labels the states on the six links of the honeycomb plaquette before (after) the action of the $\varhexagon$ operator. The product is over all the vertices $V$ of the honeycomb plaquette, attached to which are two internal links labeled by the subscripts $a$ and $b$ and an external link labeled by $x$.

For the 2+1$D$ SU(2) lattice gauge theory, gauge invariant states can be uniquely represented by the $j$ values if each vertex has at most three links joined. On a square lattice, one has to introduce extra labels besides the $j$ values in order to represent gauge invariant states, which leads to additional computational cost. This is an advantage of using the honeycomb lattice~\cite{Muller:2023nnk}.

From Eqs.~\eqref{eqn:Tmunu} and~\eqref{eqn:def_E}, we find
\be
T^{xy} = -\frac{g^2}{a^2} E_x^a E_y^a\,.
\ee
Using the electric field projection, we find $E_1^a - E_3^a = \sqrt{3} E_x^a$ and $E_2^a = -E_y^a$. Combining with Gauss's law $E_1^a+E_2^a+E_3^a = 0$, we can express $T^{xy}$ as
\be
\label{eqn:Txy}
T^{xy} = - \frac{g^2}{\sqrt{3}a^2} \big( (E_1^a)^2 - (E_3^a)^2 \big) \,.
\ee
In this expression, we need to specify the position where $T^{xy}$ is defined, since the two electric fields $E_1^a$ and $E_3^a$ are defined on different links. We use the convention that the vertex joining the two electric fields represents the position of $T^{xy}$. On a $3\times4$ lattice as shown in Fig.~\ref{fig:honeycomb_lattice}, we should specify 12 positions for different $T^{xy}$'s. We choose the 12 red points in Fig.~\ref{fig:honeycomb_lattice} as our convention, which can be easily generalized to bigger lattices. Summing over all red points gives
\begin{align}
\widetilde{T}^{xy} &= \frac{3\sqrt{3}}{2}a^2 T_{\rm sum}^{xy} \equiv \frac{3\sqrt{3}}{2}a^2 \sum_{\rm red\  dots} T^{xy} \nn\\
\mathcal{A} &= \frac{3\sqrt{3}}{2}a^2 N_{\rm plaq} \,,
\end{align}
where $N_{\rm plaq} $ is the total number of honeycomb plaquettes on the lattice and is equal to the number of red dots, as shown in Fig.~\ref{fig:honeycomb_lattice}. Using Eqs.~\eqref{eqn:Gr1} and~\eqref{eqn:Gr2} leads to
\begin{align}
\label{eq:Gtot_defin}
G^{xy}_{r}(t) =  \frac{3\sqrt{3} a^2}{2N_{\rm plaq}} \theta(t) {\rm Tr} \big( [  {T}^{xy}_{\rm sum}(t),  {T}^{xy}_{\rm sum}(0) ] \rho_T \big)\,.
\end{align}

\subsection{Truncation at \texorpdfstring{$j_{\rm max}=\frac{1}{2}$}{Lg}}
For quantum computation discussed later, we need to decompose the Hamiltonian and $T^{xy}$ in terms of tensor products of Pauli matrices, for which quantum circuits of implementation are known. This decomposition has been done for the case with the local Hilbert space truncated at $j_{\rm max}=\frac{1}{2}$.

Under this truncation, the Hamiltonian can be represented as a 2D Ising-like model~\cite{Muller:2023nnk}
\begin{align}
\label{eq:H_Ising}
aH & = H^{\rm el} + H^{\rm mag}  \\
H^{\rm el} & = h_+ \sum_{(i,j)} \Pi^+_{i,j} \nn\\
&\ - h_{++} \sum_{(i,j)} \Pi^+_{i,j} ( \Pi^+_{i+1,j} + \Pi^+_{i,j+1} + \Pi^+_{i+1,j-1} ) \nn\\
H^{\rm mag} & = h_x \sum_{(i,j)} \sigma_{i,j}^x \prod_{K=0}^5 \bigg[ \Big(\frac{1}{2}-\frac{i}{2\sqrt{2}}\Big) \sigma_K^z\sigma_{K+1}^z + \frac{1}{2}+\frac{i}{2\sqrt{2}}\bigg] \,,\nn
\end{align}
where $\Pi^\pm_{i,j} = (1 \pm \sigma^z_{i,j})/2$ are the projection operators onto the spin-up and spin-down states that represent the plaquette state at $(i,j)$ as labeled in Fig.~\ref{fig:honeycomb_lattice}. $H^{\rm el}$ represents the electric part of the Hamiltonian and $H^{\rm mag}$ stands for the magnetic part. We have multiplied the Hamiltonian by the lattice spacing $a$ such that every quantity is unitless
\begin{align}
h_+ & = \frac{27\sqrt{3}}{8}ag^2 \,, \quad 
h_{++} = \frac{9\sqrt{3}}{8}ag^2 \,,\quad 
h_x = \frac{4\sqrt{3}}{9ag^2} \,.
\end{align}
The index $K$ comes from a periodic ($K \!\mod 6$) chain $\{K=0: (i,j+1),\ K=1: (i+1,j),\ K=2: (i+1,j-1), \ K=3: (i, j-1), \ K=4: (i-1,j), \ K=5: (i-1, j+1) \}$, as shown in Fig.~\ref{fig:honeycomb_lattice}.

The $xy$ component of the stress-energy tensor for the honeycomb plaquette located at $(i,j)$ is calculated from Eq.~\eqref{eqn:Txy} where the two electric fields are those attaching the red vertex, which is at the upper right corner of the honeycomb, as shown in Fig.~\ref{fig:honeycomb_lattice}
\begin{align}
\label{eqn:T12_spin}
T^{xy}_{ij} &= - \frac{g^2}{\sqrt{3}a^2} \frac{3}{4} \bigg(  \frac{1-\sigma^z_{i,j+1}\sigma^z_{i+1,j}}{2} - \frac{1-\sigma^z_{i,j}\sigma^z_{i+1,j}}{2} \bigg)\nn\\
&= \frac{\sqrt{3} g^2}{8a^2} (\sigma^z_{i,j+1}\sigma^z_{i+1,j} - \sigma^z_{i,j}\sigma^z_{i+1,j}) \,.
\end{align}

\subsection{Closed Boundary Condition}
For a finite lattice, we use a closed boundary condition in which all the links outside the lattice boundary are in $j=0$ states. In other words, no electric fluxes go out of the lattice. For the case with $j_{\rm max}=\frac{1}{2}$ truncation, the imposed closed boundary condition is equivalent to setting all the spins outside the boundary to be pointing down.

We choose the closed boundary condition since it makes the quantum circuit construction more convenient for the $j_{\rm max}=\frac{1}{2}$ case. The periodic boundary condition results in an overall spin-flipping degeneracy in the spin representation of physical states~\cite{Muller:2023nnk}. Lifting up the degeneracy will distort the expressions of the Hamiltonian and $T^{xy}$ away from the spin representations in Eqs.~\eqref{eq:H_Ising} and~\eqref{eqn:T12_spin}. Then the corresponding quantum circuits are unknown and thus need explicit constructing, which can be computationally expensive.\\

This concludes our discussion of the calculation setup. Classical computing results of the shear viscosity that are obtained from Eqs.~\eqref{eqn:eta_to_use} and~\eqref{eqn:T12_spin} will be shown in Sec.~\ref{sec:results}. In the next section, we will introduce a quantum algorithm to evaluate the retarded correlation function.

\section{Quantum Computation of \texorpdfstring{$G_r^{xy}(t)$}{Lg} \label{sec:qc}}
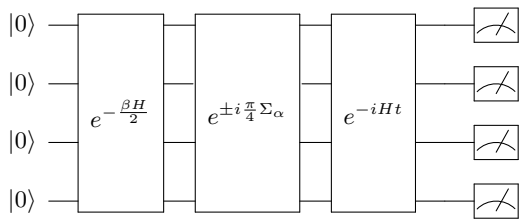
\begin{figure}[t!]
$\Qcircuit @C=1.2em @R=1em {
\lstick{\ket{0}}& \multigate{3}{e^{-\frac{\beta H}{2}}} &\multigate{3}{e^{\pm i \frac{\pi}{4} \Sigma_\alpha}} &\multigate{3}{e^{-i H t}}&\qw&\meter\\
\lstick{\ket{0}}& \ghost{e^{-\frac{\beta H}{2}}} &\ghost{e^{\pm i \frac{\pi}{4}} \Sigma_\alpha} &\ghost{e^{-i H t}}&\qw&\meter\\
\lstick{\ket{0}}& \ghost{e^{-\frac{\beta H}{2}}} &\ghost{e^{\pm i \frac{\pi}{4} \Sigma_\alpha}} &\ghost{e^{-i H t}}&\qw&\meter\\
\lstick{\ket{0}}& \ghost{e^{-\frac{\beta H}{2}}} &\ghost{e^{\pm i \frac{\pi}{4} \Sigma_\alpha}} &\ghost{e^{-i H t}}&\qw&\meter\\
}$
\caption{Schematic diagram of the implemented quantum circuit to calculate the retarded Green's function of $T^{xy}$ on a $2\times2$ lattice with $j_{\rm max}=\frac{1}{2}$. The $e^{-\frac{\beta H}{2}}$ part implements the algorithm of Ref.~\cite{franceschino_thermal2023} to prepare the thermal state, which is shown explicitly in Fig.~\ref{fig:qc_thermal}. The $e^{\pm i\frac{\pi}{4}\Sigma_\alpha}$ part is used to calculate the commutator, as will be explained in Sec.~\ref{sec:commutator}.}
\label{fig:final_qc}
\end{figure}

In this section, we present a quantum algorithm to calculate the retarded Green's function $G_r^{xy}(t)$. A schematic diagram of the quantum circuit for a four-qubit system (e.g., a $2\times2$ lattice with $j_{\rm max}=\frac{1}{2}$) is shown in Fig.~\ref{fig:final_qc}, which consists of four parts: thermal state preparation, unitary transformation driven by $T^{xy}$ for the evaluation of the commutator in $G_r^{xy}(t)$, real time evolution and measurements.

\subsection{Thermal State Preparation}
\label{sect:Tstate}
We first discuss the quantum circuit for thermal state preparation that uses the algorithm of Refs.~\cite{turro_QITP_2022,franceschino_thermal2023}. It is based on imaginary time propagation. The imaginary time propagation techniques are well-known methods for classically preparing ground or thermal states. However, for a quantum algorithm, we have to deal with the non-unitary nature of the imaginary time operator. In the algorithm we will use, this is overcome with a diluted operator using ancilla qubits~\cite{turro_QITP_2022}.

\begin{figure}[t!]
$\Qcircuit @C=1em @R=.7em {
\lstick{\ket{0}}& \gate{H}& \ctrl{1} &  \qw &\qw &  \multigate{8}{{\rm QITP}_{\rm th}} & \qw&  \qw& \\
\lstick{\ket{0}}& \qw & \targ &  \qw & \meter \\
\lstick{\ket{0}}& \gate{H}& \ctrl{1} &  \qw&\qw& \ghost{{\rm QITP}_{\rm th}} &  \qw&  \qw&\\
\lstick{\ket{0}}& \qw & \targ &  \qw & \meter\\
\lstick{\ket{0}}& \gate{H}& \ctrl{1} &  \qw&\qw& \ghost{{\rm QITP}_{\rm th}} &  \qw&  \qw&\\
\lstick{\ket{0}}& \qw & \targ &  \qw & \meter\\
\lstick{\ket{0}}& \gate{H}& \ctrl{1} &  \qw&\qw& \ghost{{\rm QITP}_{\rm th}} &  \qw&  \qw&\\
\lstick{\ket{0}}& \qw & \targ &  \qw & \meter\\
\lstick{\ket{0}}& \qw & \qw &  \qw &\qw& 
\ghost{{\rm QITP}_{\rm th}} &  \qw &\meter \gategroup{1}{2}{8}{5}{0.5em}{--}\\
}$
\centering
\caption{Quantum circuit for preparing the thermal state in quantum processors. The dashed box shows the gates for initializing the system qubits to be the maximally mixed state. The first, third, fifth and seventh qubits from the top represent the system while the second, fourth, sixth and eighth are used to initialize the maximally mixed state. The bottom qubit is the ancilla qubit used to implement the imaginary time evolution in a unitary way.}
\label{fig:qc_thermal}
\end{figure}
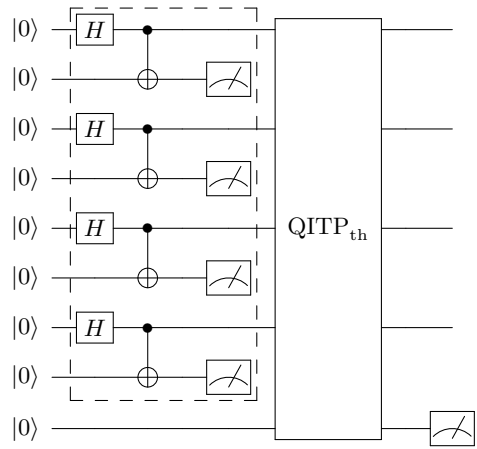

The algorithm is schematically shown in Fig.~\ref{fig:qc_thermal}. The physical qubits representing a state of the lattice gauge theory are the first, third, fifth and seventh qubits from the top of the figure. We first apply a Hadamard gate to each of them and then apply controlled-NOT (CNOT) gates with these four physical qubits as controls and four auxiliary qubits (second, fourth, sixth and eighth from the top) as targets. After the CNOT gates, the four auxiliary qubits are measured. The measurement outcomes are not needed for the remaining circuit. So effectively, the measurements serve as partial trace and the resulting physical state is a maximally mixed state
\begin{align}
\rho_s = \frac{1}{2^{n_s}} \mathbb{1}_{2^{n_s}\times2^{n_s}} \,,
\end{align}
where $n_s$ is the number of qubits in the physical system and equals to four in the example shown in the figure. Then one applies the quantum imaginary time propagation (${\rm QITP}_{\rm th}$) to the physical qubits (whose number is four in the figure) plus an additional ancilla qubit, which is the bottom qubit in the figure:
\begin{align}
    {\rm QITP}_{\rm th} = \begin{pmatrix}
    \sqrt{p}\,e^{-\tau (H-E_T)} & \sqrt{ 1- p\,e^{-2\tau (H-E_T)}}\\
    -\sqrt{ 1-p\, e^{-2\tau (H-E_T)}} &\sqrt{p}\,e^{-\tau (H-E_T)}\\
    \end{pmatrix}\,,\label{eq:thermal_QITP_op} 
\end{align}
where $p$ is a parameter to be tuned (in the limit $\tau \to 0$, $p$ is equal to the success probability), and $E_T$ is another parameter chosen to be smaller than or equal to the ground state energy. The operators $e^{-\tau H}$ and  $\sqrt{1-e^{-2\tau H}}$ act on the physical system qubits. If $p=1$ and $\tau=\frac{\beta}{2}$ are chosen and the measurement of the ancilla qubit returns $|0\rangle$, one can show the physical system is in the Gibbs state
\begin{align}
\rho_T = \frac{1}{2^{n_s} p_s}e^{-\beta(H-E_T)} = \frac{1}{Z}e^{-\beta H} \,,
\end{align}
where $p_s$ indicates the success probability of measuring the ancilla in $\ket{0}$ state.

The thermal state preparation algorithm used here is efficient for preparing high temperature thermal states. At low temperature, it becomes less efficient as the Hilbert space dimension increases.
In particular, the success probability, i.e., the probability of measuring the ancilla qubit in $\ket{0}$, decreases with the Hilbert space dimension at low temperature. The lower bound (that corresponds to $T=0$) is determined by the degeneracy of the ground state divided by the dimension of Hilbert space; in the worst scenario, it is given by $\frac{1}{2^{n_s}}$, with $n_s$ the number of system qubits. For future applications in QCD, we only need temperatures above the confinement-deconfinement crossover temperature, which is around $150$ MeV. So the reduced efficiency of the thermal state preparation algorithm at lower temperature may not be a severe problem. In Sec.~\ref{sect:ps}, we will perform a more detailed analysis of the thermal state preparation success probability as a function of the system size.

In practice, to explicitly construct the quantum circuit for the ${\rm QITP}_{\rm th}$ from the Hamiltonian in Eq.~\eqref{eq:H_Ising}, we apply Trotter decomposition at first order
\begin{align}
& {\rm QITP}_{\rm th}\left(H,\tau \right) \approx \\
& \prod_{i_\tau=1}^{N_\tau} \prod_{(i,j)} {\rm QITP}_{\rm th}\left(H^{\rm el}_{i j}, \Delta\tau \right) {\rm QITP}_{\rm th}\left(H^{\rm mag}_{i j}, \Delta\tau \right) \,, \nn
\end{align}
where $\Delta\tau=\frac{\tau}{N_\tau}$ is the imaginary time Trotter step, and $(i,j)$ denotes a plaquette position as in Eq.~\eqref{eq:H_Ising}. In the Trotter decomposition, we need to add one additional ancilla qubit for each non-commuting Hamiltonian term, unless the quantum hardware allows ancilla qubit reset in the middle of the circuit. More concretely, the ${\rm QITP}_{\rm th}$ can be written as
\begin{align}
{\rm QITP}_{\rm th}(H_{ij}^{\rm el/mag},\Delta\tau) = e^{i\sigma^y_{\rm anc} \otimes \arccos(e^{-\Delta\tau H_{ij}^{\rm el/mag}})} \,,
\end{align}
where $\sigma^y_{\rm anc}$ denotes the Pauli-$y$ matrix acting on the ancilla qubit. The $\arccos$ function is best computed in the diagonal basis of $H_{ij}^{\rm el/mag}$. The electric part is already diagonal in the computational basis. For the magnetic part at $(i,j)$, one needs to apply a Hadamard gate before and another one after the ${\rm QITP}_{\rm th}$ circuit to the qubit representing the state at $(i,j)$ in order to convert to a basis where $H_{ij}^{\rm mag}$ is diagonal. Once the $\arccos$ part is made diagonal, one can decompose it into sums of tensor products of identity and $\sigma^z$ matrices. The procedure to construct quantum circuits for tensor products of Pauli matrices is well known~\cite{Nielsen:2012yss}. Appendix~\ref{app:circuit} shows a quantum circuit for the magnetic Hamiltonian term on a $2\times2$ lattice with $j_{\rm max}=\frac{1}{2}$.

As we will show later in 
Sec.~\ref{sec:trotter_error}, we only need to use a single Trotter step in the imaginary time because its Trotter error is very small, i.e., $\Delta\tau = \frac{\beta}{2}$.

\subsection{Circuit for Commutator of \texorpdfstring{$T^{xy}$}{Lg}}
\label{sec:commutator}

Using Eq.~\eqref{eqn:T12_spin}, the commutator $[T^{xy}_{\rm sum}(t),T^{xy}_{ij}(0)] = [\sum_{k,l} T^{xy}_{kl}(t),T^{xy}_{ij}(0)]$ can be rewritten as the sum of two commutators:
\begin{align}
& [T^{xy}_{\rm sum}(t),T^{xy}_{ij}(0) ] \nn\\
& = \frac{\sqrt{3}g^2}{8a^2} [ T^{xy}_{\rm sum}(t),\sigma^z_{i,j+1}\sigma^z_{i+1,j} - \sigma^z_{i,j}\sigma^z_{i+1,j} ] \nn\\
&= \frac{\sqrt{3}g^2}{8a^2} \left([ T^{xy}_{\rm sum}(t),\Sigma_0] - [ T^{xy}_{\rm sum}(t),\Sigma_1] \right) \,,
\label{eqn:Txy_commutator}
\end{align}
where we have decomposed $T^{xy}_{ij}(0)$ operator in terms of two Pauli strings as in Eq.~\eqref{eqn:T12_spin}. To simplify the notations, we have introduced $\Sigma_\alpha$ as $\Sigma_{0}=\sigma^z_{i,j+1}\sigma^z_{i+1,j}$, and $\Sigma_{1}=\sigma^z_{i,j+1}\sigma^z_{i+1,j}$. 

The commutator between a Pauli string $A$ and a (generic) unitary operator $B$ can be easily evaluated in quantum processor using the following relation~\cite{Gradient_pi_4,Gradient_parameter_shift}
\begin{equation}
    [A,B] = i \left( e^{-i\frac{\pi}{4}A} B e^{i\frac{\pi}{4}A} - e^{i\frac{\pi}{4}A} B e^{-i\frac{\pi}{4}A}\right) \,. \label{eq:comm_eq}
\end{equation} 
Applying Eq.~\eqref{eq:comm_eq} to Eq.~\eqref{eqn:Txy_commutator} with $A$ and $B$ identified as $\Sigma_\alpha$ and $T^{xy}_{\rm sum}(t)$, respectively, we find
\begin{align}
\label{eqn:Txy_commutator_expand}
[ \Sigma_\alpha, T^{xy}_{\rm sum}(t) ] & = i  e^{-i\frac{\pi}{4}\Sigma_\alpha } e^{iHt} T^{xy}_{\rm sum} e^{-iHt} e^{i\frac{\pi}{4}\Sigma_\alpha}
 \nn\\
& - i e^{i\frac{\pi}{4}\Sigma_\alpha } e^{iHt} T^{xy}_{\rm sum} e^{-iHt} e^{-i\frac{\pi}{4}\Sigma_\alpha}  \,,
\end{align}
where we have used $\mathcal{O}(t) = e^{iHt}\mathcal{O}(0)e^{-iHt}$ and omitted the argument $t=0$ of $T^{xy}_{\rm sum}$.

\subsection{Real Time Evolution}
As just discussed, the quantum computation procedure requires real time evolution. Here, we briefly discuss its implementation by applying the Trotter decomposition at first order. We divide the total time length $t$ into $N_t$ steps with the step size $\Delta t = \frac{t}{N_t}$. The full real time propagator becomes a product of the short-time electric and magnetic Hamiltonian evolution operators as the following:
\begin{equation}
U_t \equiv e^{-iHt}
\approx \prod_{i_t=1}^{N_t} \prod_{(i,j)} e^{-i H^{\rm el}_{ij} \Delta t} e^{-i H^{\rm mag}_{ij} \Delta t} \,,
\end{equation}
where $H^{\rm mag}_{ij}$ ($H^{\rm el}_{ij}$) indicates the magnetic (electric) part of the Hamiltonian at the $(i,j)$ plaquette position. Each Hamiltonian piece just contains commuting strings of Pauli matrices, for which the construction of quantum circuits is well known~\cite{Nielsen:2012yss}. For example, the circuit for the magnetic part is discussed in Ref.~\cite{Muller:2023nnk}. A discussion of the Trotter errors in the real time evolution for the calculation of $G_r^{xy}(t)$ can be found in Sec.~\ref{sec:trotter_error}.
After implementing the real time evolution gates, we measure the quantum state in the computational basis in which the operator $T_{\rm sum}^{xy}(0)=\sum_{k,l} T^{xy}_{kl}(0)$ is diagonal.

\subsection{Post-Measurement Processing}

The last step to obtain the retarded correlation function is to perform a post-measurement analysis. We note that the stress-energy tensor operator $T_{\rm sum}^{xy}$ is a sum of Pauli-$z$ tensor products; so it is diagonal in the computational basis. Therefore, the thermal expectation value of the commutator $[T_{\rm sum}^{xy}(t),\Sigma_\alpha]$ is given by
\begin{equation}
\!{\rm Tr}( [T_{\rm sum}^{xy}(t), \Sigma_\alpha ] \rho_T) = i \sum_b \langle b |T_{\rm sum}^{xy}(0) | b \rangle [P^{-}_{\alpha}(b) - P^{+}_{\alpha}(b)] \,, \label{eq:comm_prob}
\end{equation}
where $|b\rangle$ denotes the computational basis states, which, e.g., on a $2\times2$ lattice with $j_{\rm max}=\frac{1}{2}$, are $|0000\rangle,|0001\rangle,\dots |1111\rangle$ in the spin representation.
The symbol 
$P^\pm_{\alpha}(b)$ indicates the measured probability of the $\ket{b}$ state for the circuit with $\Sigma_\alpha=e^{\pm i \frac{\pi}{4} \Sigma_\alpha}$ evolved from time $0$ to $t$. The time dependence of $P^\pm_{\alpha}(b)$ is omitted for notational simplicity.\\

We now summarize the quantum algorithm: After preparing the thermal state, we first apply the gates for $e^{\pm i\frac{\pi}{4}\Sigma_\alpha}$, as shown in Fig.~\ref{fig:final_qc} [$+$ for the first line and $-$ for the second line in Eq.~\eqref{eqn:Txy_commutator_expand}]. Then we evolve the resulting state in real time $e^{-iHt}$. Finally, we perform projective measurements in the computational basis. The measurement results allow us to reconstruct the thermal expectation value of the commutator in Eq.~\eqref{eqn:Txy_commutator_expand} by post-measurement processing and taking the difference of the results obtained from the two different circuits that differ in the sign of the $\Sigma_\alpha$ term.
Because we have two different Pauli strings in order to evaluate Eq.~\eqref{eqn:Txy_commutator}, we have to run four different quantum circuits that differ in the Pauli 
operators (different $\alpha$) and the signs of the term $i\frac{\pi}{4}\Sigma_\alpha$.

To conclude this section, we prove that the action of the proposed quantum circuit gives the correct value of $G^{xy}_r(t)$.
The first set of gates prepares the thermal state density matrix, $\rho_T=\frac{e^{-\beta H}}{Z}$. 
Then, after applying the gates for $e^{\pm i \frac{\pi}{4} \Sigma_\alpha}$ and the real time evolution gates $U_t$, the density matrix of the quantum processor becomes
\begin{align}
\rho^{\pm}_\alpha(t) = \frac{1}{Z} U_t  e^{\pm i \frac{\pi}{4} \Sigma_\alpha}  e^{-\beta H} e^{\mp i \frac{\pi}{4} \Sigma_\alpha} U_t^\dagger \,.
\end{align}
Measuring the quantum processors in the computational basis, we obtain the diagonal part of the density matrix. Using the cyclic property of the trace, and the fact that $T^{xy}(0)$ is diagonal in the computational basis, we can write:
\begin{align}
&  \sum_b \langle b |T_{\rm sum}^{xy}(0) | b \rangle P^{\pm}_{\alpha}(b) = {\rm Tr}[T_{\rm sum}^{xy}(0) \rho_\alpha^{\pm}(t)] \nn\\
&= \frac{1}{Z} {\rm Tr}[ e^{\mp i \frac{\pi}{4} \Sigma_\alpha} U_t^\dagger T_{\rm sum}^{xy}(0) U_t  e^{\pm i \frac{\pi}{4} \Sigma_\alpha}  e^{-\beta H} ] \nn\\
&= \frac{1}{Z} {\rm Tr}[ e^{\mp i \frac{\pi}{4} \Sigma_\alpha} T_{\rm sum}^{xy}(t) e^{\pm i \frac{\pi}{4} \Sigma_\alpha}  e^{-\beta H} ] \,.
\end{align}
Using Eq.~\eqref{eq:comm_eq}, the thermal expectation value of the commutator $[T_{\rm sum}^{xy}(t), \Sigma_\alpha]$ is obtained from
\begin{align}
&{\rm Tr}[T_{\rm sum}^{xy}(0)\rho^{+}_\alpha(t)] - {\rm Tr}[T_{\rm sum}^{xy}(0)\rho^{-}_\alpha(t)] \nn\\
& \!\! = {\rm Tr} ( [e^{- i \frac{\pi}{4} \Sigma_\alpha}   T_{\rm sum}^{xy}(t)  e^{i \frac{\pi}{4} \Sigma_\alpha}  - e^{i \frac{\pi}{4} \Sigma_\alpha}   T_{\rm sum}^{xy}(t)  e^{- i \frac{\pi}{4} \Sigma_\alpha} ]  \rho_T ) \nn\\ 
& \!\! = \frac{i}{Z} {\rm Tr}( \left[T_{\rm sum}^{xy}(t), \Sigma_\alpha \right] e^{-\beta H} ) \,.
\end{align}
A final usage of Eq.~\eqref{eqn:Txy_commutator} leads to the retarded Green's function of $T^{xy}$.

\section{Calculation Systematics}
\label{sec:sys}
Before presenting results, we discuss various systematics of the calculation. The analysis we will present in this section is important to understand whether one can obtain the physical quantity (i.e., shear viscosity) at a given accuracy with a given amount of quantum computing resource. This analysis will be useful for large scale quantum computation of transport coefficients in the future.

\subsection{Continuum Limit and Renormalization}
For physical limits, one needs to take the continuum $a\to0$ and the infinite volume limits and remove the local Hilbert space truncation by setting $j_{\rm max}\to\infty$. In the continuum limit $a\to0$, the coupling of the theory needs proper renormalization. In 2+1$D$ SU(2) pure gauge theory, the mass dimension of $g$ is $0.5$, so a unitless quantity for the coupling is $ag^2$. What needs to be done is to tune $ag^2$ as $a\to0$ such that a physical observable is invariant. An example of physical observables is the correlation length of the ground state, which can be extracted from the subsystem size dependence of the entanglement entropy~\cite{Ebner:2024mee}. In a renormalization scheme where both the pressure and the trace of the stress-energy tensor $T^\mu_\mu$ do not require any additional renormalization other than the running coupling, the renormalization of the coupling is given by~\cite{Romatschke:2019nmo}
\begin{align}
\label{eqn:RG}
\frac{{\rm d}\ln(ag^2)}{{\rm d}\ln a} = 1 \,, 
\end{align}
which means the rescaled coupling $ag^2 \propto a$. In the Hamiltonian approach, the lattice gauge theory also has a truncation in the local Hilbert space, labeled by $j_{\rm max}$. Whether the running coupling has a non-trivial dependence on $j_{\rm max}$ should be studied analytically and tested against numerical calculations, which are left for future work.

Besides the running coupling, the stress-energy tensor component $T^{xy}$ may need additional renormalization since the lattice breaks the Lorentz invariance of the continuum theory. Generally, one can write
\be
\eta^R = \lim_{a\to0,\,j_{\rm max}\to\infty}Z(\mu, a, j_{\rm max}) \eta^B(a, j_{\rm max}) \,,
\ee
where $R$ and $B$ represent renormalized and bare quantities. The bare quantity is the direct numerical result obtained by using the truncated lattice gauge theory with a lattice spacing $a$ and local Hilbert space truncation $j_{\rm max}$. $Z(\mu, a, j_{\rm max})$ is the additional operator renormalization factor needed when taking the continuum limit, which should be distinguished from the partition function introduced in Eq.~\eqref{eqn:rho_T} and depends on the final renormalization scheme in which we want to obtain the quantity $\eta^R$, e.g., the $\overline{\rm MS}$ scheme with the renormalization scale $\mu$. One way to obtain $Z(\mu, a, j_{\rm max})$ is to perform a lattice perturbation theory calculation. Alternatively, one can develop gradient flow methods for the Hamiltonian approach to regularize the stress-energy tensor operator, as done in Euclidean lattice calculations~\cite{Luscher:2010iy,Luscher:2011bx}. Compared with Euclidean lattice calculations, the potential dependence on $j_{\rm max}$ is new and needs systematic understanding.

As can be seen, taking the continuum limit and renormalizing the quantity properly can be complicated. Since our current study is on a small lattice with a low-$j_{\rm max}$ truncation, we will only take into account the coupling renormalization here and leave studies of additional operator renormalization to future work.

\subsection{Local Truncation Effect}
\label{sec:jmax_effect}
We then consider local Hilbert space (i.e., $j_{\rm max}$) truncation effect with a given lattice spacing $a$, which means the coupling $ag^2$ is fixed, and a fixed lattice size. The truncation effect can lead to artifacts in thermodynamic description of the system, since they constrain the energy spectrum of the system. This has been considered on a SU(2) plaquette chain~\cite{Ebner:2023ixq}, and here, we generalize it to the honeycomb lattice case.

Specifically, we first consider the internal energy density $\varepsilon$ and the entropy density $s$ as a function of $j_{\rm max}$ on a $2\times2$ honeycomb lattice with $ag^2=1$, which are defined as
\begin{align}
\varepsilon \equiv -\frac{1}{\mathcal{A}} \frac{\partial\ln Z}{\partial\beta} \,,\qquad
s \equiv \frac{1}{\mathcal{A}} \frac{\partial (T\ln Z)}{\partial T} \,, \label{eq:entropy}
\end{align}
where the partition function $Z$ can be evaluated from ${\rm Tr}\,e^{-\beta H}$ by exactly diagonalizing the Hamiltonian. 

\begin{figure}
\subfloat[Internal energy density.\label{fig:u_jmax}]{%
  \includegraphics[height=2.3in]{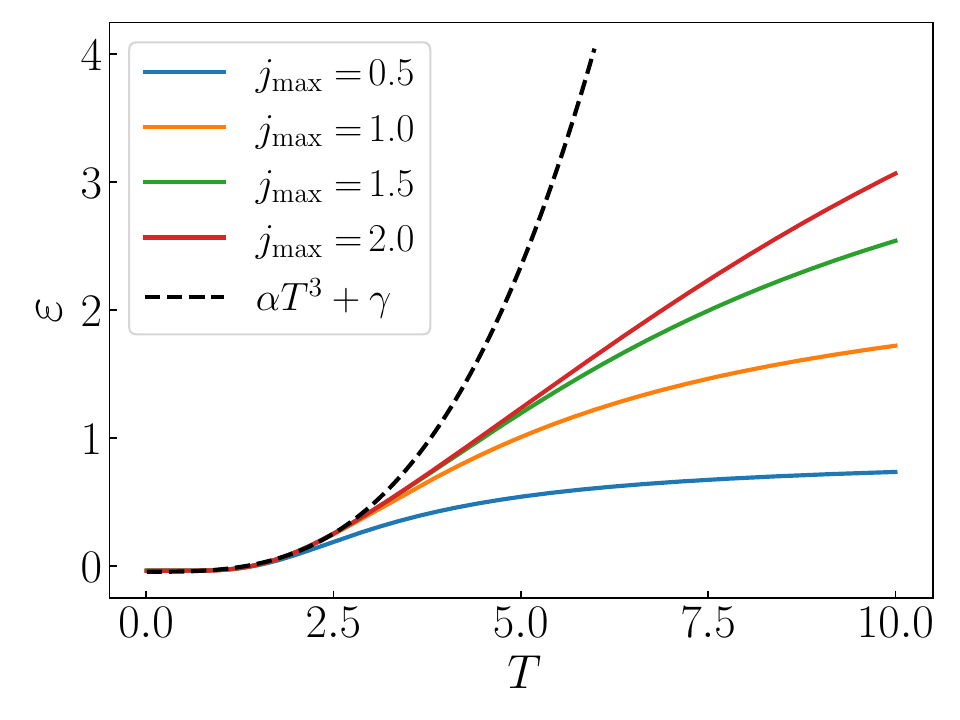}%
}\hfill
\subfloat[Entropy density.\label{fig:s_jmax}]{%
  \includegraphics[height=2.3in]{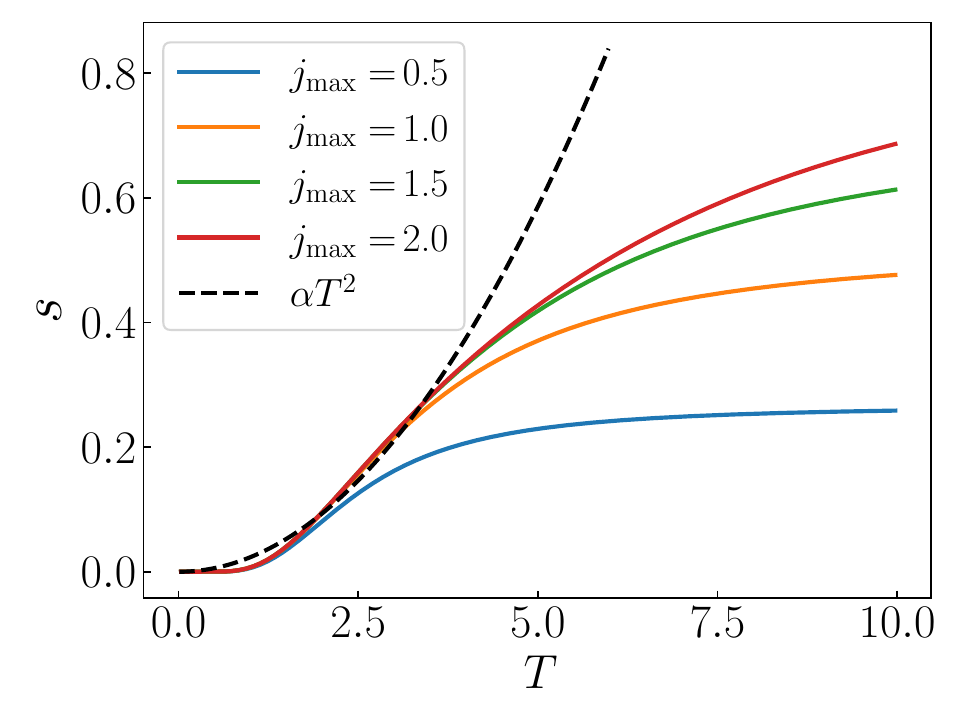}%
}
\caption{Internal energy and entropy densities in lattice units as functions of temperature for several values of $j_{\rm max}$ on a $2\times2$ lattice with $ag^2=1$. The dashed lines are expectations from the continuum theory.}
\label{fig:U_T_s_jmax}
\end{figure}

The results of the internal energy and entropy densities are shown in Fig.~\ref{fig:U_T_s_jmax}, together with fits of the forms $\alpha_\varepsilon T^3+\gamma$ and $\alpha_s T^2$, respectively, which are expectations from the continuum theory. The fits use the results with $j_{\rm max}=2$ and in the temperature range $T<2.5$ (in lattice units). The fitted parameters are $\alpha_\varepsilon=0.0191$, $\gamma=-0.0466$ and $\alpha_s=0.0234$. We see that the $2\times2$ lattice with $j_{\rm max}=1$ already shows continuum behavior for $ag^2=1$ in the temperature range $T\lesssim 3$. In the region $3<T<5$, the curves of $\varepsilon$ and $s$ already converge with $j_{\rm max}=1.5$. Further increasing $j_{\rm max}$ is of no help for continuum physics, and one has to use a bigger lattice.

\begin{figure}
\subfloat[Internal energy density.\label{fig:u_N}]{%
  \includegraphics[height=2.3in]{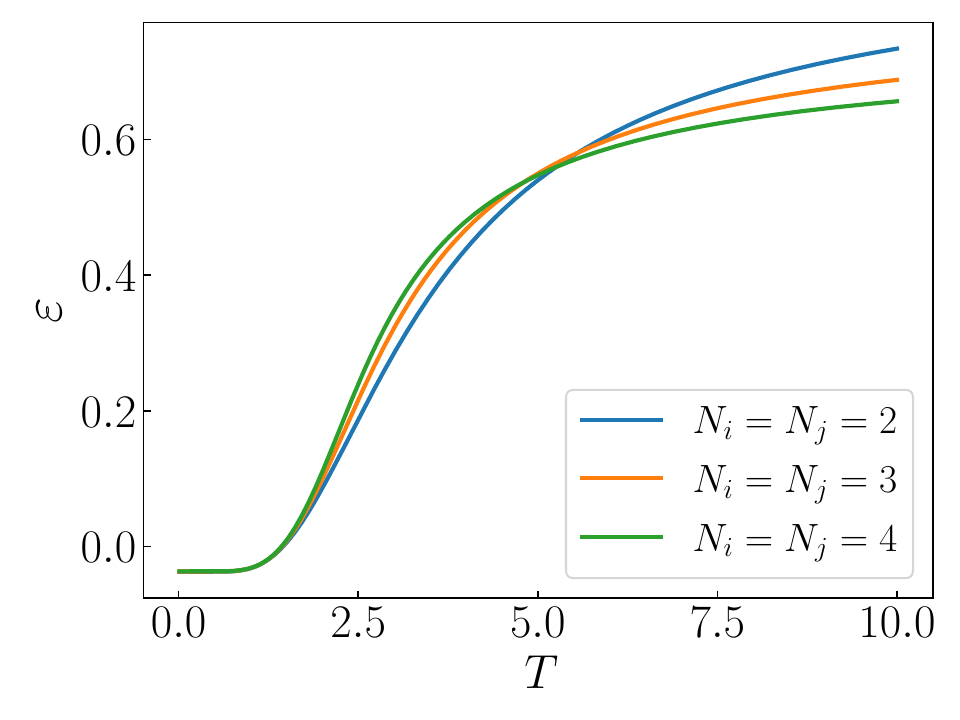}%
}\hfill
\subfloat[Entropy density.\label{fig:s_N}]{%
  \includegraphics[height=2.3in]{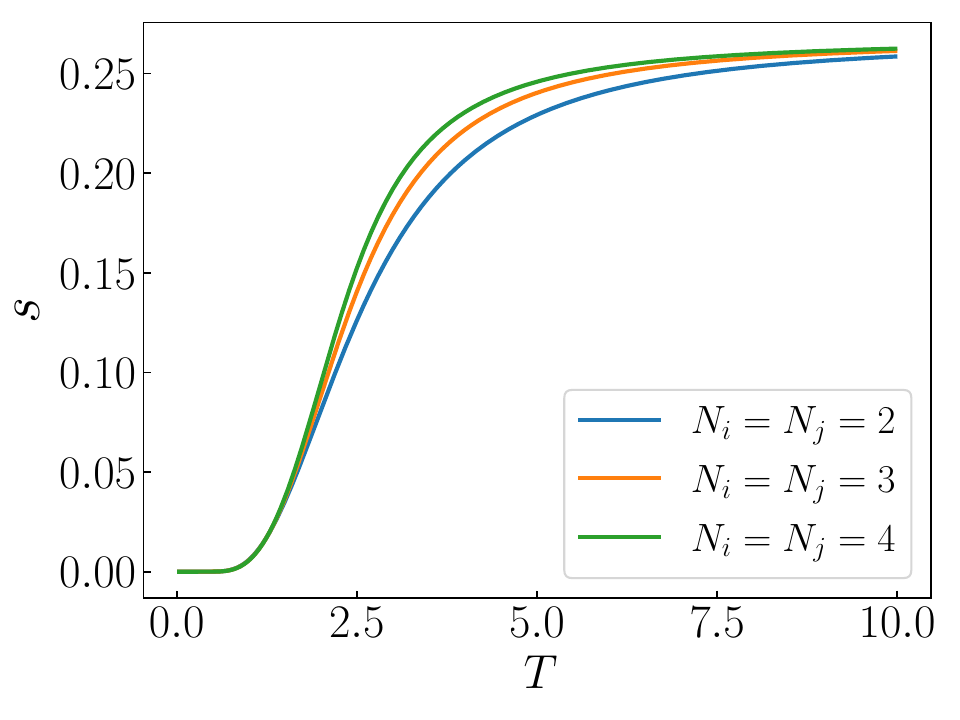}%
}
\caption{Internal energy and entropy densities in lattice units as functions of temperature for several lattice sizes with $j_{\rm max}=\frac{1}{2}$ and $ag^2=1$. $N_i$ and $N_j$ specify the size of the honeycomb lattice along the $i$ and $j$ axes respectively, as shown in Fig.~\ref{fig:honeycomb_lattice}.}
\label{fig:u_s_N}
\end{figure}

At last, we provide an analytic estimate of the $j_{\rm max}$ needed. At a given lattice size, if we want to describe all the states below a fixed energy $E$ [corresponding to the Hamiltonian in Eq.~\eqref{eq:HKS}] with an accuracy $1-\epsilon$, the minimum $j_{\rm max}$ needed is at most
\begin{align}
\label{eqn:jmax_needed}
j_{\rm max} = \frac{4N_l\widetilde{E}}{3\sqrt{3}g^2\epsilon} \,,
\end{align}
where $N_l$ is the number of links on the lattice, and $\widetilde{E} = E+\frac{8\sqrt{3}}{9g^2a^2}N_{p}$, where $N_p$ denotes the number of plaquettes on the lattice.
We provide a proof of this formula in Appendix~\ref{app:jmax}.

\subsection{Finite Volume Effect}
\label{sec:V_effect}
\begin{figure}
    \centering
    \includegraphics[width=3in]{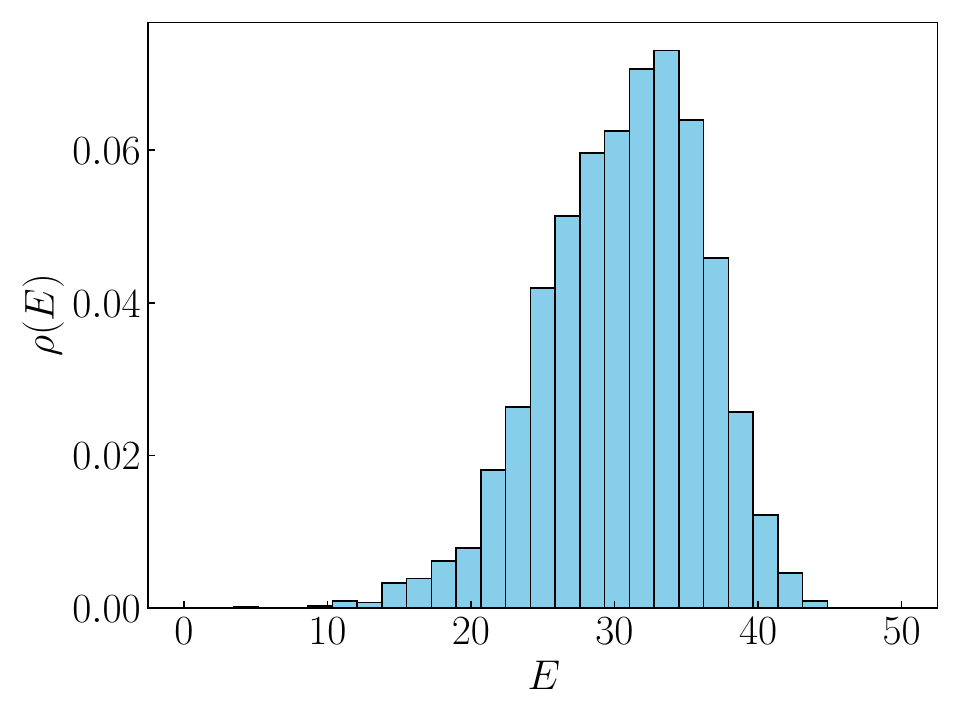}
    \caption{Energy eigenstate density on a $4\times4$ lattice with $j_{\rm max}=\frac{1}{2}$ and $ag^2=1$.}
    \label{fig:state_density}
\end{figure}

Next we study the lattice size dependence of the internal energy and entropy densities with $ag^2=1$ and $j_{\rm max}=\frac{1}{2}$. The results are shown in Fig.~\ref{fig:u_s_N}, where it can be seen that these two densities change little with the lattice size. Increasing the lattice size leads to a decrease of $\varepsilon$ at high temperature, which indicates that one needs to increase $j_{\rm max}$ to maintain the same energy density at a given temperature. Both energy and entropy densities saturate at high temperature, which is a finite size effect. Figure~\ref{fig:state_density} shows the energy eigenstate density on a $4\times4$ lattice with $j_{\rm max}=\frac{1}{2}$ and $ag^2=1$. The total number of states is $2^{16}=65536$. When the energy is below $35$ (in lattice units), the state density $\rho(E)$ keeps increasing with $E$. This is qualitatively similar to the continuum theory. However, once the energy exceeds $35$, $\rho(E)$ starts to drop with $E$, which is an artifact originated from the finite size of the Hilbert space.

\begin{figure}[p]
\subfloat[Different commutators at $\beta=0.3$.\label{fig:eta_sum_x0}]{%
  \includegraphics[height=2.3in]{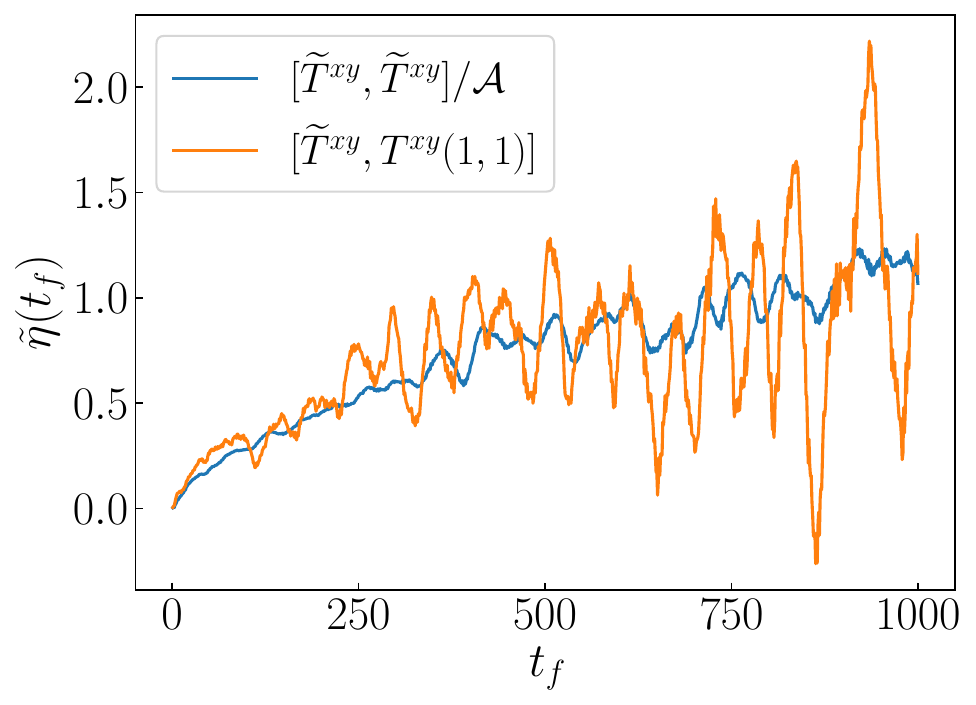}%
}\hfill
\subfloat[Different lattice sizes at $\beta=0.3$.\label{fig:eta_N}]{%
  \includegraphics[height=2.3in]{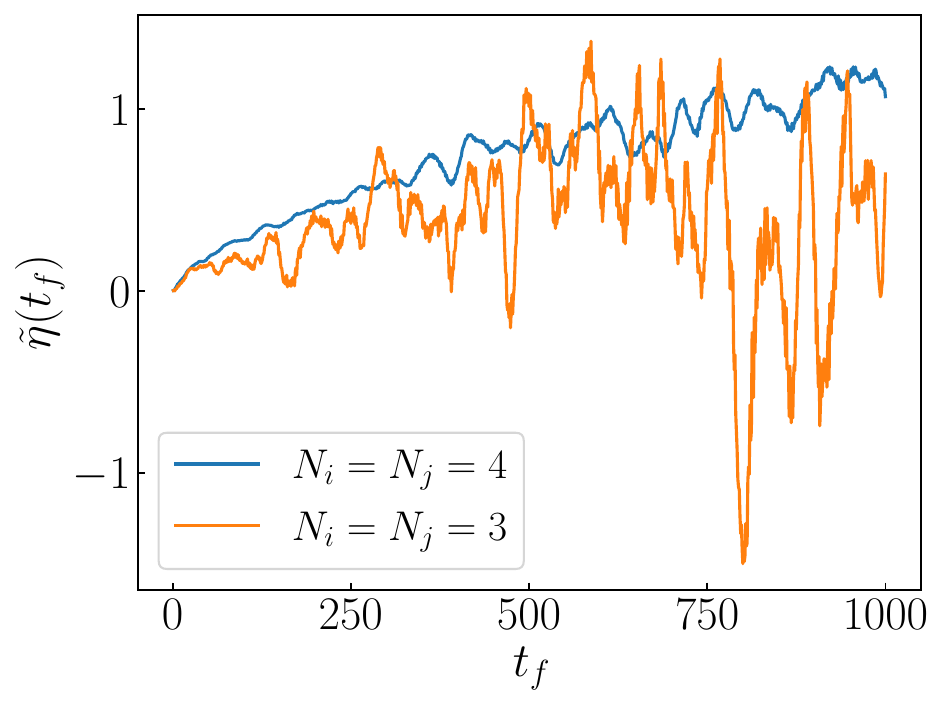}%
}\hfill
\subfloat[Different temperatures.\label{fig:eta_beta}]{%
  \includegraphics[height=2.3in]{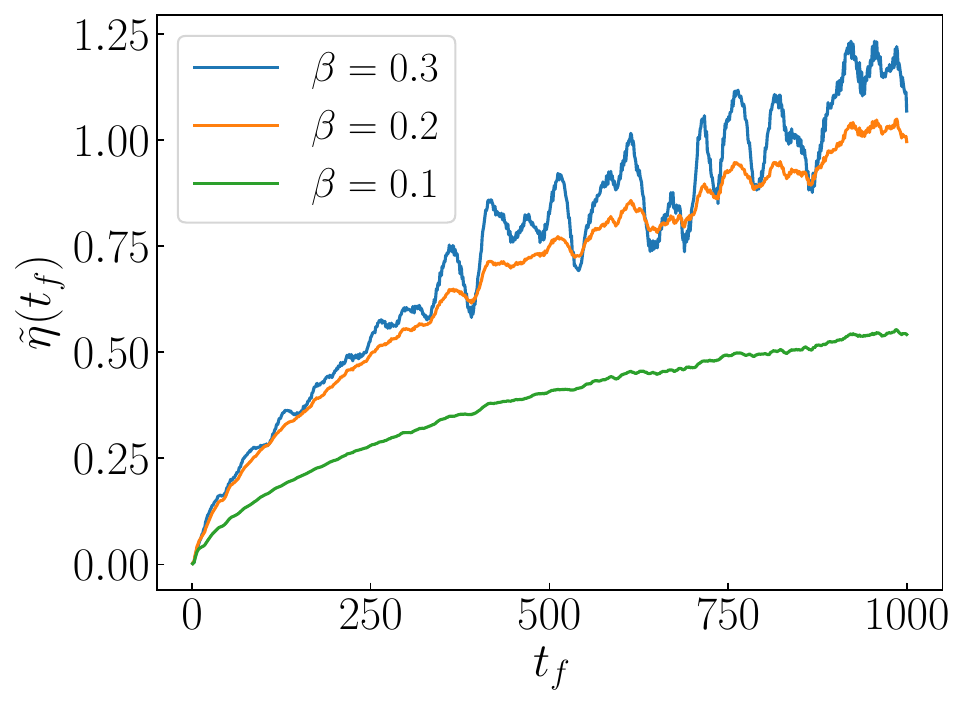}%
}
\caption{$\tilde{\eta}$ defined in Eq.~\eqref{eqn:eta_to_use} as a function of time. (a) Results on a $4\times4$ lattice with $j_{\rm max}=\frac{1}{2}$ and $ag^2=1$ at $\beta=0.3$ in lattice units, obtained from two ways that are equivalent if the system obeys translational invariance. (b) Results on two lattices with different sizes but the same $j_{\rm max}=\frac{1}{2}$ and $ag^2=1$ at $\beta=0.3$. (c) Results on a $4\times4$ lattice with $j_{\rm max}=\frac{1}{2}$ and $ag^2=1$ at different temperatures.}
\label{fig:u_s_jmax}
\end{figure}

Another way of inspecting the finite size effect is to compare $\tilde{\eta}(t_f)$ calculated via different methods on different lattices. In Fig.~\ref{fig:eta_sum_x0}, we compare the result of $\tilde{\eta}(t_f)$ obtained from the commutator $[\widetilde{T}^{xy}(t), {T}^{xy}(0)]$ as in the third line of Eq.~\eqref{eqn:eta_to_use}, where ${T}^{xy}(0)$ is located at $(i,j)=(1,1)$, with that from the commutator $[\widetilde{T}^{xy}(t), \widetilde{T}^{xy}(0)]/\ml{A}$ as in the second-to-last line of Eq.~\eqref{eqn:eta_to_use} on a $4\times4$ lattice with the cutoff $j_{\rm max}=\frac{1}{2}$ and $ag^2=1$ at $\beta=0.3$ (in lattice units). We see that $\tilde{\eta}(t_f)$ oscillates in both results with the $[\widetilde{T}^{xy}(t), {T}^{xy}(0)]$ case oscillating more severely. The oscillation is caused by the finite state density. Inspecting Eq.~\eqref{eqn:eta_to_use}, we can see the oscillating factor $f(t_f)$ only smooths out if the energy levels are dense enough. Thus we expect at lower temperatures that the oscillating is more severe when the Hilbert space is of fixed size. The reason why the calculation with the $[\widetilde{T}^{xy}(t), {T}^{xy}(0)]$ commutator oscillates more severely is the additional non-positive definite term in the double sums over $n,m$, i.e., $\langle n|\widetilde{T}^{xy}|m \rangle \langle m|{T}^{xy}|n \rangle$, which can be positive or negative. On the other hand, the $|\langle n|\widetilde{T}^{xy}|m \rangle|^2$ term is positive semi-definite so it leads to a smoother result. If we had used periodic boundary conditions, spatial translation invariance would have been preserved and as a result, the two methods would have given the same result. We use a closed boundary condition here since it leads to a more straightforward construction of the relevant quantum circuits due to the absence of an overall spin-flipping degeneracy explained earlier and long-range qubit interaction in the quantum hardware. In the following, we will mainly use the method with the $[\widetilde{T}^{xy}(t), \widetilde{T}^{xy}(0)]/\ml{A}$ commutator, unless explicitly mentioned otherwise.

\begin{figure*}[t]
    \centering
    \includegraphics[width=0.96\textwidth]{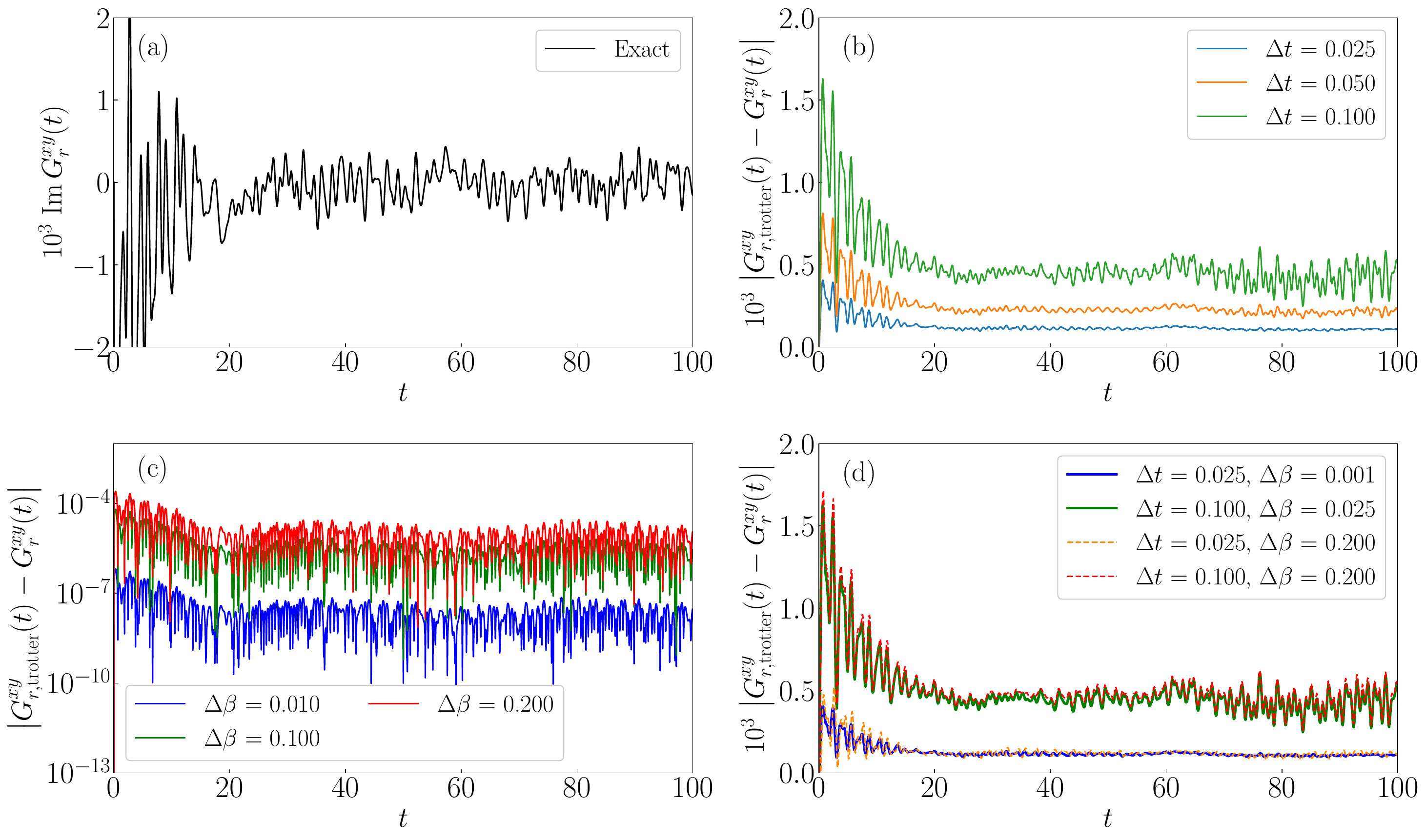}
    \caption{Trotter error in computing $G_r^{xy}(t)$ as a function of time on a $3\times 3 $ lattice with $j_{\rm max}=\frac{1}{2}$ and $ag^2=1$ for $\beta=0.2$ in lattice units. (a) Exact $G_r^{xy}(t)$ results obtained from Eq.~\eqref{eq:Gtot_defin}. (b) Uncertainty when Trotter decomposition is only applied in real time evolution. (c) Uncertainty when Trotter decomposition is solely applied in preparing the thermal state. (d) Uncertainty when  Trotter decomposition is applied in both real time evolution and thermal state preparation.}
    \label{fig:trotter}
\end{figure*}

In Fig.~\ref{fig:eta_N} we compare the results of $\tilde{\eta}(t_f)$ calculated on two lattices of different sizes at the same conditions: $j_{\rm max}=\frac{1}{2}$, $\beta=0.3$. We see the smaller lattice leads to a bigger finite size effect due to the much smaller state density. It is expected that the physical state density will increase exponentially with the lattice size (e.g.~$N_p$ plaquettes) and $j_{\rm max}$ as roughly $\frac{(2j_{\rm max}+1)^{3N_p}}{2^{2N_p}}$~\cite{berndt}, where the denominator is a rough estimate of the Gauss's law constraint effect.

We also find that the oscillation becomes less severe at higher temperatures due to the increase of the density of active states, as shown in Fig.~\ref{fig:eta_beta}.

\subsection{Trotter Error\label{sec:trotter_error}}

In this subsection, we show the effect of implementing the Trotter decomposition in calculating the transport coefficients. In our case, we have two possible contributions: the error deriving from applying the Trotter decomposition in the real time evolution and that when preparing the thermal density matrix.

In order to quantify the different Trotter errors, we compute the $G_r^{xy}(t)$, defined in Eq.~\eqref{eq:Gtot_defin}, as a function of time $t$ on a $3\times 3 $ lattice with $j_{\rm max}=\frac{1}{2}$ for different values of the real time and imaginary time Trotter steps. We set $ag^2=1$ and $\beta=0.2$ in lattice units.

Panel (a) of Fig.~\ref{fig:trotter} depicts the behavior of $G_r^{xy}(t)$ as a function of time using the exact time evolution operators (without applying the Trotter decomposition). We constrain the $y$ axis range in order to have a better visual comparison with the uncertainty plots. Panel (b) of Fig.~\ref{fig:trotter} shows the absolute difference between the exact $G_r^{xy}$ results\footnote{We do not use the relative error because the exact curve oscillates around zero, causing a divergent relative error even though the difference is small.} and those when we apply the Trotter decomposition for real time evolution. The thermal density matrix is computed exactly. We observe that for $\Delta t >0.1$, we get a significant error: The magnitude of the green line at late time in (b) is of the order of $5\times10^{-4}$, which is comparable with the black line in (a). Using smaller time steps is needed to improve accuracy, which may result in a deeper quantum circuit. Panel (c) of Fig.~\ref{fig:trotter} shows the absolute error in $G_r^{xy}$ as a function of time, when the Trotter decomposition is implemented for preparing the thermal state. The real time propagator is computed exactly. We observe that we have a negligible uncertainty. Panel (d) of Fig.~\ref{fig:trotter} illustrates the results of combining the Trotter decomposition for the real time evolution and that for preparing the thermal state. As these plots show, the major Trotter error comes from the real time evolution.

\subsection{Thermal State Preparation Success Probability}
\label{sect:ps}
Earlier in Sec.~\ref{sect:Tstate}, we mentioned that in practical application for QCD, only the deconfined temperature regime roughly above $150$ MeV is of interest. So the exponential decay of success probability in the thermal state preparation at low temperature may not be a problem. However, one may worry that even at high temperature, the success probability still decreases exponentially with the system size, just with a small prefactor in the exponent. Here, we provide a detailed analysis of this dependence.

\begin{figure}
    \centering
    \includegraphics[width=3in]{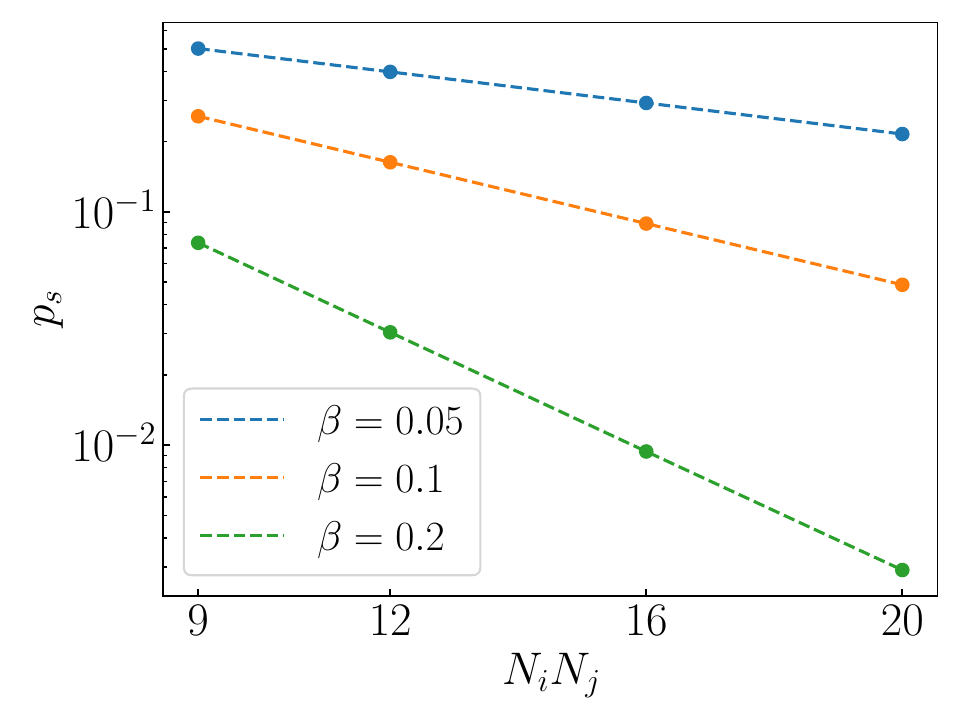}
    \caption{Success probability of thermal state preparation as a function of the honeycomb lattice size for several temperatures. The lattice spacing $a$ is fixed corresponding to $ag^2=1$ and $j_{\rm max}=\frac{1}{2}$ is used.}
    \label{fig:ps}
\end{figure}

The success probability of the thermal state preparation is given by
\begin{align}
p_s = \frac{1}{d_\mathcal{H}} \sum_n e^{-\beta(E_n-E_0)} \,, 
\end{align}
where we have chosen $E_T=E_0$, and $d_\mathcal{H}$ denotes the total Hilbert space dimension. We plot the success probability as a function of the lattice size on a periodic\footnote{This is the only place where we use the periodic boundary condition throughout the paper. It allows us to obtain all the eigenenergies and eigenstates on a $5\times4$ honeycomb lattice.} honeycomb lattice with $j_{\rm max}=\frac{1}{2}$ in Fig.~\ref{fig:ps}, where the lattice spacing $a$ is fixed such that the coupling is $ag^2=1$. At this fixed coupling, the first excited state has an energy of about $6.2$ above the ground state, i.e., $E_1-E_0\approx6.2$, which does not change much as the lattice size increases. High temperature in this case should at least be $T \gtrsim 1/6.2\approx 0.16$. Three temperatures are shown for comparison in the plot. We see that the success probability decays exponentially with the system size, but the high temperature case has a very small prefactor in the exponent.

For physical application, one must make sure the energy density $\varepsilon$ does not change with the lattice size at fixed $a$ and $\beta$, since it captures the local equilibrium physics. Figure~\ref{fig:u_s_N} and Eq.~\eqref{eqn:jmax_needed} imply that one has to increase $j_{\rm max}$ as the lattice becomes bigger, in order to keep the energy density fixed. Our current computing resources forbid us to analyze this case so we leave it to future work.

\subsection{Numerical Integration Error}
The uncertainty control in numerical integration is well known. We take it as an individual issue to discuss because it plays an important role in practical calculations. In quantum computing, one will calculate $G_r^{xy}(t)$ at many time points and then integrate to obtain $\eta$ as in Eq.~\eqref{eqn:eta_to_use}. Under a given accuracy, if one can reduce the number of time points at which to compute $G_r^{xy}(t)$ on quantum computers, one will reduce the amount of quantum resources needed to achieve the accuracy. 

\begin{figure}
    \centering
    \includegraphics[width=3in]{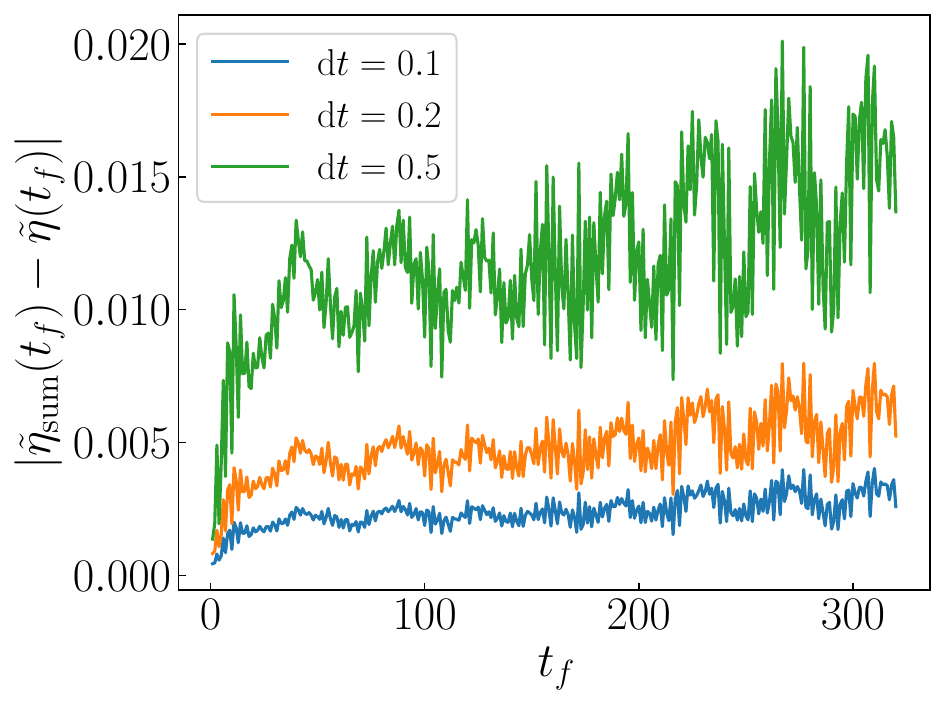}
    \caption{Numerical integration uncertainty as a function of final time for three different finite elements of time.}
    \label{fig:int_error}
\end{figure}

To demonstrate the integration error, we define the Riemann sum version of $\tilde{\eta}(t_f)$ as
\begin{align}
\tilde{\eta}_{\rm sum}(t_f) \equiv -(\Delta t)^2 \sum_{k=1}^{N_t} k\, {\rm Im}G_r^{xy}(k\Delta t) \,,
\end{align}
where $N_t = t_f/\Delta t$. Its difference from the exact result $|\tilde{\eta}_{\rm sum}(t_f)-\tilde{\eta}(t_f)|$ is shown in Fig.~\ref{fig:int_error} for a $4\times4$ lattice with $ag^2=1$ , $j_{\rm max}=\frac{1}{2}$ and $\beta=0.3$ (in lattice units). Comparing with Fig.~\ref{fig:eta_sum_x0}, we find the relative error up to $t_f=300$ is less than $5\%$ for ${\rm d}t=0.5$ and is roughly $1\%$ for ${\rm d}t=0.2$.

It is clear that the error grows with $\Delta t$, and its magnitude scales as $t_f^2/N_t = t_f \Delta t$. One can easily improve this by using the midpoint value in the Riemann sum, i.e., $\frac{2k-1}{2}G_r^{xy}(\frac{2k-1}{2}\Delta t)$, whose error magnitude scales as $t_f^3/N_t^2 = t_f (\Delta t)^2$.

\subsection{Fitting Uncertainty\label{sec:fitting_uncertainty}}
Next we discuss the uncertainty when one extracts the bare value of the shear viscosity from the infinite time limit of $\tilde{\eta}(t_f)$. As we have seen in Secs.~\ref{sec:jmax_effect} and~\ref{sec:V_effect}, the result of $\tilde{\eta}(t_f)$ oscillates at late time due to the finite volume and local truncation effects, and thus, the infinite time limit is not well defined. To overcome this issue, we fit the time dependence of $\tilde{\eta}(t_f)$ via some function that becomes constant at large $t_f$. We consider two fitting functions
\begin{align}
f_1(t_f) &= a_1+c_1 e^{-b_1 t_f} \nn\\
f_2(t_f) &= a_2+\Big(c_2+\frac{d_2}{t_f}\Big) e^{-b_2 t_f} \,,
\end{align}
where $a_{1,2},b_{1,2},c_{1,2}$ and $d_2$ are parameters.

\begin{figure}
    \centering
    \includegraphics[width=\columnwidth  ]{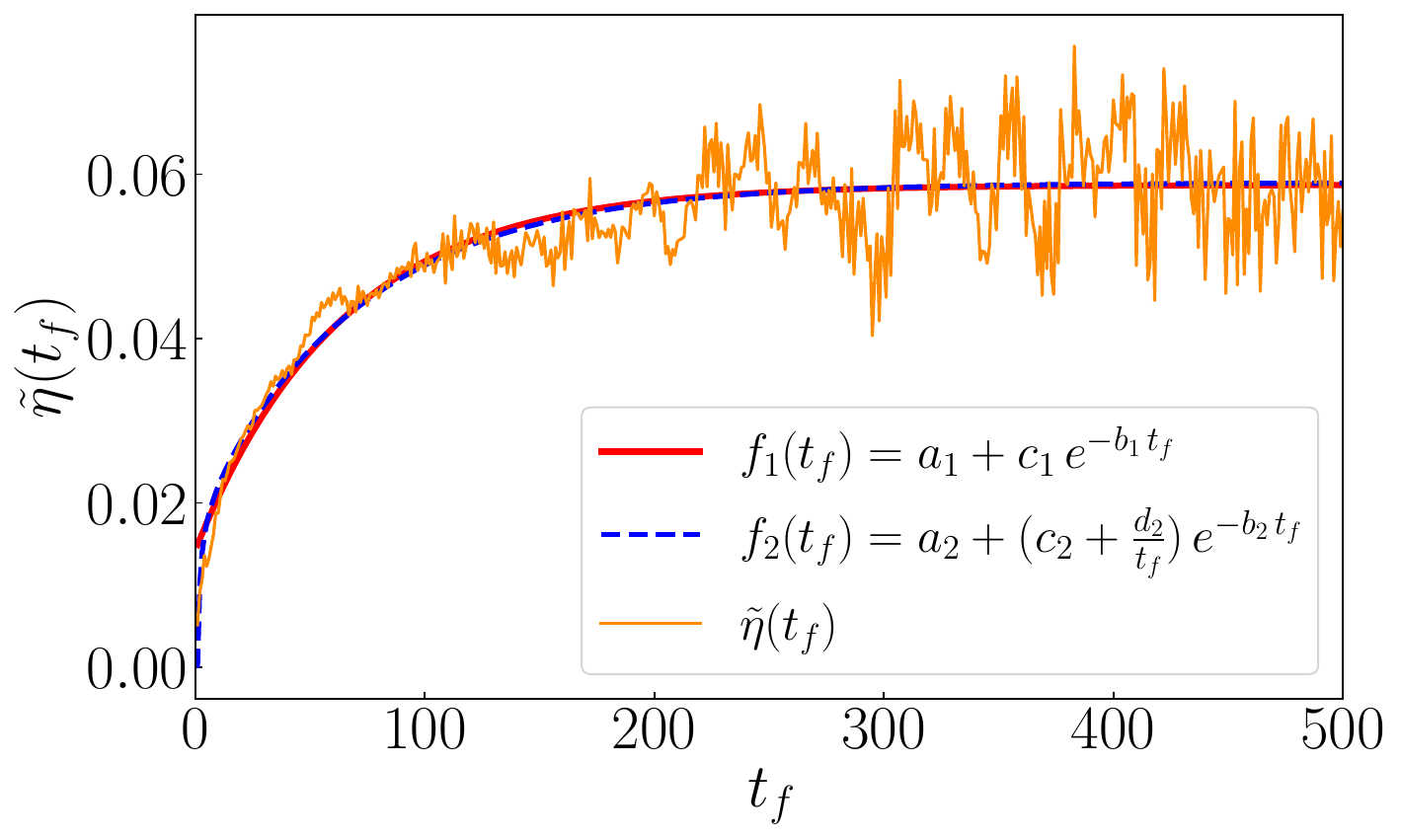}
    \caption{$\tilde{\eta}(t_f)$ as a function of time, fitted by two different functional forms on a $4\times4$ lattice with $j_{\rm max}=\frac{1}{2}$, $ag^2=0.6$ and $\beta=0.2$ in lattice units.}
    \label{fig:fittingdifference}
\end{figure}

Figure~\ref{fig:fittingdifference} shows the fitting results of $\tilde{\eta}(t_f)$ as a function of $t_{f}$ on a $4\times4$ lattice with $j_{\rm max}=\frac{1}{2}$, $ag^2=0.6$ and $\beta=0.2$ (in lattice units), where we can observe the exponential behavior. We use all the 500 time points up to $t_f=500$ (two neighboring time points are separated by $\Delta t_f=1$) in the fit. The fitting is implemented through the \texttt{python scipy} \textit{curve\textunderscore fit} function.
From Eq.~\eqref{eqn:eta_to_use}, the bare value of $\eta$ is given by the plateaus coefficient ($a_{1,2}$) value. We see that the two functions are very similar. Indeed, the $a_i$ values fitted in Fig.~\ref{fig:fittingdifference} are identical up to fitting uncertainties: For $f_1$, we find $a_1=0.0587(3)$, while for $f_2$, we have $a_2=0.0590(4)$.

At late time, due to the finite volume and local truncation effects, the signal becomes noisy. In order to reflect this real time fluctuation in our fitting uncertainty for the plateau value, we choose different ranges of $t_f$ used in the fit and estimate the average and uncertainty associated with them. For example, if we choose $t_f\in[0,400], [0,500], [0,600], [0,700]$ or $[0,1000]$ in the fit, we find the values shown in Table~\ref{tab:plateau_values_tf} when using the function $f_1$. In the result section, we will use the function $f_1$ and apply this procedure to estimate the uncertainties\footnote{For $\beta_0=0.225$ to be introduced in Sec.~\ref{sec:results}, we use $t_f\in[0,t_f']$ with $t_f'=500,600,700,1000$ because the fitting function does not work properly when $t_f'=400$.}.

\begin{table}[ht]
    \centering
\begin{tabular}{|c|c|}
\hline
Range of $t_f$ in fitting & Result of $a_1$\\
\hline
$[0,400]$  & 0.0586(4)\\
$[0,500]$  & 0.0587(3)\\
$[0,600]$  & 0.0590(4)\\
$[0,700]$  & 0.0585(4)\\
$[0,1000]$ & 0.0624(4)\\
\hline
\end{tabular}
\caption{Different fitting results of $a_1$ for different fitting time ranges.}
\label{tab:plateau_values_tf}
\end{table}

\subsection{Number of Shots and CNOT Gates}
Finally, we estimate the number of times (shots) that one needs to repeat simulating the same quantum circuit and performing measurements, as well as the number of CNOT gates needed.
It turns out that a huge number of shots is necessary to evaluate correctly the retarded Green's function from the commutator $[T_{\rm sum}^{xy}(t),T^{xy}_{ij}(0)]$ because its thermal expectation value is very small, around $10^{-4}-10^{-2}$. We assume that we prepare the thermal density matrix with an efficient quantum algorithm and will neglect the thermal state preparation part in the following estimate of the number of shots. As we mentioned previously, the implemented ${\rm QITP}_{\rm th}$ becomes inefficient when studying big systems at low temperature due to the large drop of the success probability. If we include the proposed ${\rm QITP}_{\rm th}$ algorithm for the thermal state preparation, the number of shots estimated below must be divided by the ${\rm QITP}_{\rm th}$ success probability.

Given a number of shots $n_{\rm shot}$ and a probability of measuring the wavefunction of the quantum circuit containing the Pauli string $\Sigma_\alpha$ of the sign $s$ at time $t$ to be in the computational basis state $\ket{b}$, i.e., $P_\alpha^s(b)$, the measurement uncertainty can be computed by the binomial distribution $\delta P_\alpha^s(b) = \sqrt{\frac{P_\alpha^s(b) [1-P_\alpha^s(b)]}{n_{\rm shot}}}$. The uncertainty of the measured retarded Green's function can be obtained from Eq.~\eqref{eq:comm_prob}
\begin{equation}
    \delta G^{xy}_r(t) = \sqrt{ \sum_{\alpha} \sum_{s=\pm}  \sum_b \left[\langle b| T^{xy}_{\rm sum}(0) |b\rangle \delta P^{s}_\alpha(b) \right]^2 } \,.
\end{equation}

The upper bound of this equation can be obtained by setting $P^{s}_\alpha(b)=\frac{1}{d_\ml{H}}\,, \forall b$  where $d_\ml{H}$ is the dimension of the system Hilbert space. Hence, we obtain the upper bound at a fixed $t$ as
\begin{align}
\max \delta G^{xy}_{r}(t) & =  \sqrt{ \sum_{\alpha} \sum_{s=\pm} \sum_b \langle b| T^{xy}_{\rm sum}(0) |b\rangle ^2 \frac{ \frac{1}{d_\ml{H}} (1-\frac{1}{d_\ml{H}})} {n_{\rm shot}} } \nn \\
& \le \sqrt{ \sum_{\alpha} \sum_{s=\pm} 
\frac{d_\ml{H}\,d_T^2}{d_\ml{H}} \frac{ (1-\frac{1}{d_\ml{H}})}{n_{\rm shot}} } \nn \\ 
& \leq\sqrt{ \sum_{\alpha} \sum_{s=\pm} \frac{d_T^2}{n_{\rm shot}} } \leq\sqrt{ 4 \frac{d_T^2}{n_{\rm shot}} } \,, 
\end{align}
where, in the second line, we use that $T^{xy}_{\rm sum}(0)$ is given by a sum of Pauli-$z$ strings, so $\left|\langle b | T^{xy}_{\rm sum}(0) | b\rangle\right|\leq d_T$ [note $\langle b | T^{xy}_{\rm sum}(0) | b\rangle$ is real], and in the third line, $(1-\frac{1}{d_\ml{H}})\leq 1$. We estimate the absolute value upper bound of the $T^{xy}_{\rm sum}(0)$ operator to be $d_T=3$ for a $2\times2$ lattice, $d_T=6$ for a $3\times3$ lattice, $d_T=12$ for a $4\times4$ lattice, $d_T=19$ for a $5\times5$ lattice, where $j_{\rm max}=\frac{1}{2}$ is used for all cases. A rough estimation gives that $d_T$ is smaller than $2 N_xN_y$.

Given the retarded Green's function $G^{xy}_{r}(t)$, the required number of shots to achieve a relative error $\epsilon = \frac{{\rm max}\delta G^{xy}_{r}(t)}{G^{xy}_{r}(t)}$ is given by
\begin{equation}
    n_{\rm shot} \simeq \frac{4\,d_T^2}{\epsilon^2 [G^{xy}_{r}(t)]^2} \sim \frac{4\times10^6\,d_T^2}{\epsilon^2} \,, \label{eq:shots_estimation}
\end{equation}
where we use that $G^{xy}_{r}(t) \sim 10^{-3}$. The number of shots needed increases polynomially with the lattice size because of the $d_T^2$ factor in Eq.~\eqref{eq:shots_estimation}. 

The maximum number of CNOT gates that need to be implemented for studying the real time evolution is given by $70$ per time step and lattice point: $64$ for implementing the evolution driven by the magnetic part of the Hamiltonian and six for the electric part. Implementing the $\frac{\pi}{4}\Sigma_\alpha$ gate requires two CNOT gates for one $\alpha$ and one sign. Hence, on a $N_x\times N_y$ honeycomb lattice with $j_{\rm max}=\frac{1}{2}$, to implement the $\frac{\pi}{4}\Sigma_\alpha$ gate and real time evolution with a number of time steps $N_t$, the number of CNOT gates we need is
\begin{equation}
    \# {\rm CNOT} \le 2 + 70N_t N_x N_y \,. \label{eq:cnot_estimation}
\end{equation}

\subsection{Classical Computational Resource Estimate}
In Appendix~\ref{app:mps}, we study the computational time to classically simulate the quantum circuit for the real time evolution via the matrix product state method. The computational time grows exponentially with the Trotter steps when the bond dimension is large, which is expected to be necessary to describe states with volume-law entanglement, as is usually the case when the system thermalizes at late time. Performing such real time calculations by exact diagonalization on a large lattice with high $j_{\rm max}$ truncation also requires exponentially growing resources. For example, the Hilbert space on a $3\times 3$ lattice with $j_{\rm max}=1$ is $519233$ and that on a $5\times 5$ lattice with $j_{\rm max}=\frac{1}{2}$ is $2^{25}$, when all external links are in $j=0$ states. The needed space to store a single $T^{xy}_{ij}$ matrix in the energy eigenbasis is about $1.1$ TB and $4.5$ PB, respectively (single-precision float). Quantum computing may be able to overcome this difficulty.

\section{Results}
\label{sec:results}
\subsection{Classical Computing Results}
\subsubsection{\texorpdfstring{$\eta/s$}{Lg} in the continuum}

We will show results of the ratio between the shear viscosity $\eta$ and the entropy density $s$ as a function of temperature in the continuum limit. The continuum limit is taken via extrapolating toward $a=0$. In this work, we will only consider the renormalization group equation of the coupling as $a$ varies as written in Eq.~\eqref{eqn:RG}. Additional operator renormalization is left for future work which may lead to a $20\%$ change roughly~\cite{steve}.

\begin{table*}[t]
    \centering
    \begin{tabular}{|c|ccc|c|}
    \hline
        $ag^2$ set for the fitting & $c_0$ & $c_1$ & $c_2$ &  $\frac{\eta}{s} (ag^2=0)$   \\
        \hline
        $\{0.5,0.55,0.6,0.65 \}$  & 0.07(2)   &  14(12)$\cdot 10^{-4}$& 81(12)$\cdot 10^{-1}$ & 0.07(2) \\
        $\{0.4,0.5,0.55,0.6,0.65 \}$ & 0.068(16) & 14(9)$\cdot 10^{-4}$ & 80(8)$\cdot 10^{-1}$  &0.070(16) \\
        $\{0.4,0.5,0.55,0.6,0.65,0.7 \}$ & 0.118(14) & 9(6)$\cdot 10^{-5}$ &12(1) & 0.118(14)\\
    \hline
    \end{tabular}
    \caption{Obtained parameter values from fitting the $ag^2$ dependence of $\frac{\eta}{s}$ on a $4\times4$ lattice with $j_{\rm max}=\frac{1}{2}$ for $\beta_0=0.2$ using the exponential function in Eq.~\eqref{eq:exp_ga2}. The last column lists the obtained values in the continuum limit.}
    \label{tab:fit_param_eta_S}
\end{table*}

We start this procedure by fixing a ``physical'' temperature, which is kept invariant as lattice spacing $a$ changes. We set this ``physical'' temperature $T_0$ ($\beta_0=\frac{1}{T_0}$) to be the temperature when $ag^2=1$, which is a number in the lattice unit corresponding to $ag^2=1$. Then the temperatures at other couplings (i.e., other lattice spacings) are $T=ag^2T_0$ ($\beta=\frac{\beta_0}{ag^2}$). It would also be useful to fix the temperature scale by the confinement-deconfinement crossover temperature, which is left for future work. 

Using the procedure described in Sec.~\ref{sec:fitting_uncertainty}, we obtain the $\eta(\beta_{0},ag^2)$ values as a function of $ag^2$ for a $4\times4$ lattice with $j_{\rm max}=\frac{1}{2}$. Then, using Eq.~\eqref{eq:entropy}, we compute the entropy density $s(\beta_{0},ag^2)$.

\begin{figure}
    \centering
    \includegraphics[width=\columnwidth]{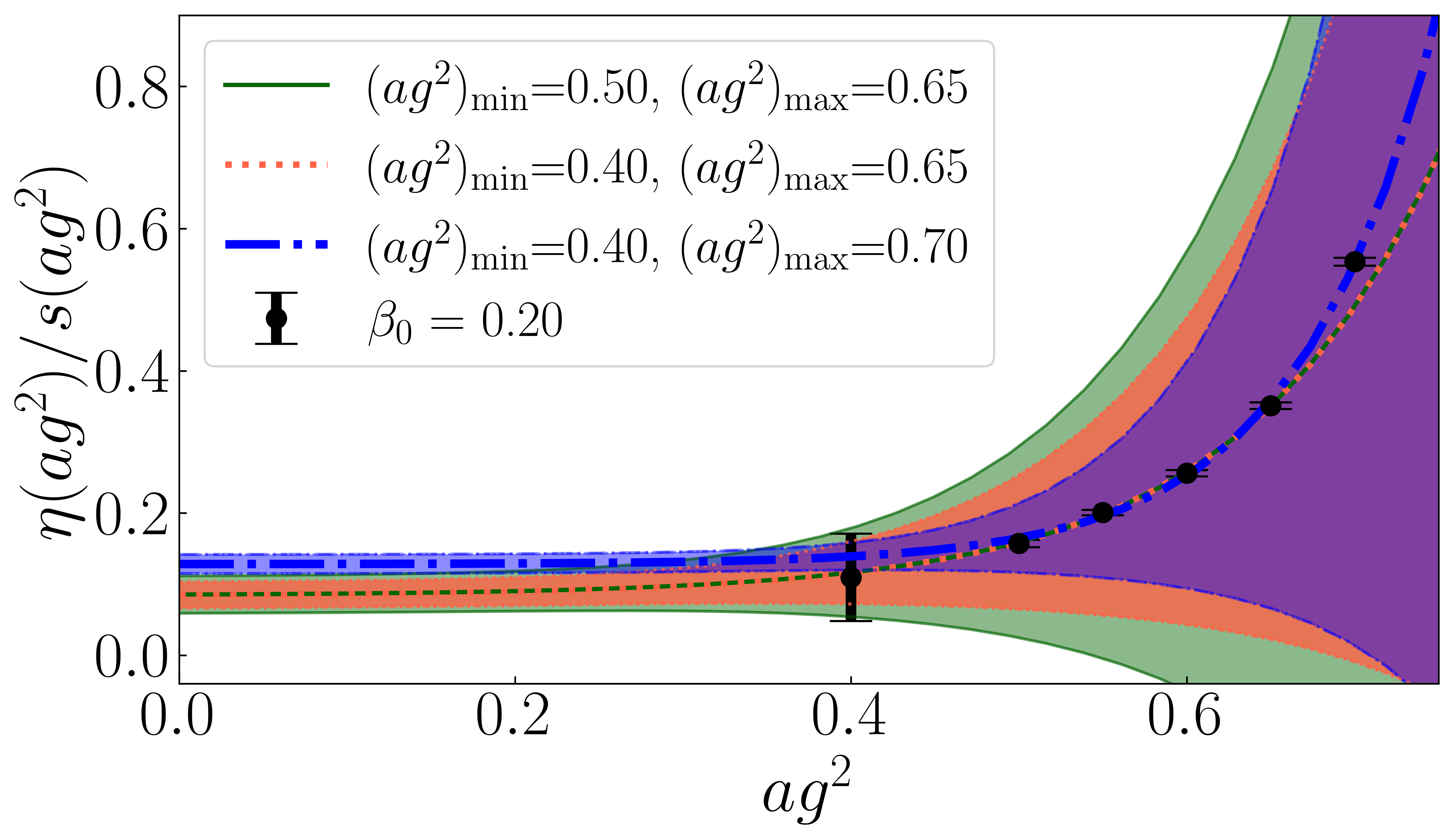}
    \caption{Results of the coupling dependence of $\frac{\eta}{s}$ for $\beta_0=0.2$ on a $4\times4$ lattice with $j_{\rm max}=\frac{1}{2}$. Black points represent the calculated $\frac{\eta}{s}$ at different couplings, lines indicate the fitting results and bands describe one sigma uncertainty of each fitting.}
    \label{fig:eta_over_S_vs_a}
\end{figure}

The black circles of Fig.~\ref{fig:eta_over_S_vs_a} represent the obtained results for the ratios $\frac{\eta(\beta_{0},ag^2)}{s(\beta_{0},ag^2)}$ at six different couplings $ag^2=\{0.4,0.5,0.55,0.6,0.65,0.7\}$ for $\beta_{0}=0.2$. We stop at $ag^2=0.4$ rather than going to smaller couplings since at such a small coupling, the low-$j_{\rm max}$ truncation effect is very large, signaling large oscillation of $\tilde{\eta}(t_f)$ at late time, which deteriorates the fitting of the plateau value. The vertical uncertainty bars associated with the black points describe the fitting uncertainties explained in Sec.~\ref{sec:fitting_uncertainty}. The fitting uncertainties are very small except for $ag^2=0.4$. It can be seen that an exponential function can describe the trend of the black points,
\begin{equation}
    f(ag^2)=c_0 + c_1e^{c_2 ag^2}\,,\label{eq:exp_ga2}
\end{equation}
where $c_0,c_1,c_2$ are fitting parameters. To quantify the systematic uncertainty of using the function $f(ag^2)$, we choose three different data sets to perform the fit: $\{ag^2\}_1=\{0.5,0.55,0.6,0.65\}$, $\{ag^2\}_2=\{0.4,0.5,0.55,0.6,0.65\}$ and $\{ag^2\}_3=\{0.4,0.5,0.55,0.6,0.65,0.7\}$, which are shown in Fig.~\ref{fig:eta_over_S_vs_a} in green, orange and blue, respectively. The fitted parameter values are listed in Table~\ref{tab:fit_param_eta_S}. The band with the same color of the line represents the relative uncertainty at one sigma of the fits. The band grows at large $ag^2$ since the fitting function grows exponentially there and a small uncertainty in either $c_1$ or $c_2$ leads to a big uncertainty after proper error propagation.

The continuum limit for $\frac{\eta}{s}$ at $ag^2=0$ is obtained as $c_0+c_1$. 
The results for $\beta_0=0.2$ are shown in the last column of Table~\ref{tab:fit_param_eta_S} for the three different fitting ranges of $ag^2$. The three values are compatible within two-sigma error bars. However, we observe a change in the continuum limit value when the data point at $ag^2=0.7$ is included in the fitting. We attribute this to the potentially larger lattice discretization effect at bigger lattice spacing.

\begin{figure}
    \centering
    \includegraphics[width=\columnwidth  ]{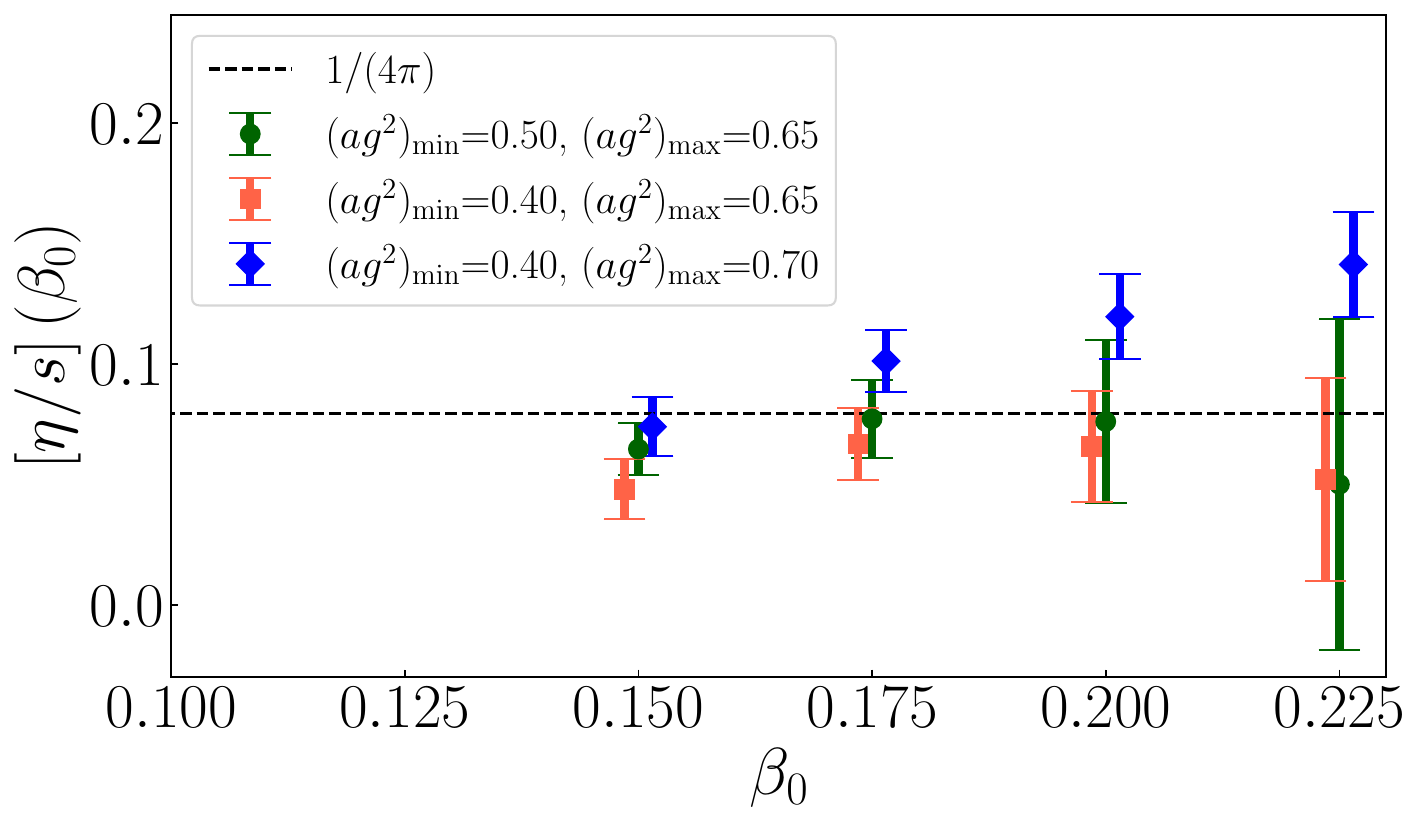}
    \caption{Obtained $\frac{\eta}{s}$ results as a function of $\beta_0$ on a $4\times4$ lattice with $j_{\rm max}=\frac{1}{2}$. We slightly shift the data horizontally for better visualization of the three fittings using different data sets.}
    \label{fig:etaS_vs_beta}
\end{figure}

Iterating this procedure for different ``physical'' temperatures $T_0$, we obtain the temperature dependence of $\frac{\eta}{s}$ in the continuum. We show in Fig.~\ref{fig:etaS_vs_beta} results from all the three fitting data sets. The blue and orange dots are slightly shifted horizontally for better visualization. We do not study temperatures higher than $\beta_0=0.15$ since higher $j_{\rm max}$ truncation is needed to accurately describe highly excited states, as discussed in Sec.~\ref{sec:jmax_effect}. The uncertainty grows rapidly at lower temperatures (e.g., $\beta_0=0.225$), since not many states of the theory are effectively contributing to the retarded Green's function (suppressed by $e^{-\beta_0E}$) and then the density of contributing states is not large enough to suppress the real time fluctuation in $\tilde{\eta}(t_f)$ due to our small lattice and local Hilbert space truncation, as seen in Sec.~\ref{sec:V_effect}. 

Our results of $\frac{\eta}{s}$ are consistent with the holographic result $\frac{1}{4\pi}$ within uncertainties, which is shown as the dashed line in Fig.~\ref{fig:etaS_vs_beta}. We also observe a trend of decrease in $\frac{\eta}{s}$ as temperature increases from the blue dots. However, this trend is not obvious from the green and orange dots. All these should be further studied on bigger lattices with higher $j_{\rm max}$ truncation in the future, to better understand the finite volume and local Hilbert space truncation effects.

\subsubsection{Structure of Spectral Function}
We also study the off-diagonal matrix elements of $\widetilde{T}^{xy}$ in small frequency $\omega$ ranges, which are related to the spectral function that is defined as
\begin{align}
\rho^{xy}(\omega) \equiv \,& \frac{1}{\ml{A}} \int {\rm d}t \, e^{i\omega t} {\rm Tr}\big( [\widetilde{T}^{xy}(t), \widetilde{T}^{xy}(0)] \rho_T \big) \nn\\
= \,& \frac{1}{\ml{A}Z}\sum_{n}\sum_m  2\pi \delta(\omega+E_n-E_m) |\langle n|\widetilde{T}^{xy}|m\rangle|^2 \nn\\
& \times (e^{-\beta E_n} - e^{-\beta E_m}) \,.
\end{align}
When $\omega$ is small, $|\langle n|\widetilde{T}^{xy}|m\rangle|^2$ is closely related to $\frac{\rho^{xy}(\omega)}{\omega}$:
\begin{align}
\rho^{xy}(\omega) = & \, \frac{1}{\ml{A}Z}\sum_{n}\sum_m  2\pi \delta(\omega+E_n-E_m) |\langle n|\widetilde{T}^{xy}|m\rangle|^2 \nn\\
& \times e^{-\beta E_n} [ \beta\omega+O(\omega^2)] \,.
\end{align}

\begin{figure}
\subfloat[$ag^2=0.6$.\label{fig:Toff0.6}]{%
  \includegraphics[height=2.3in]{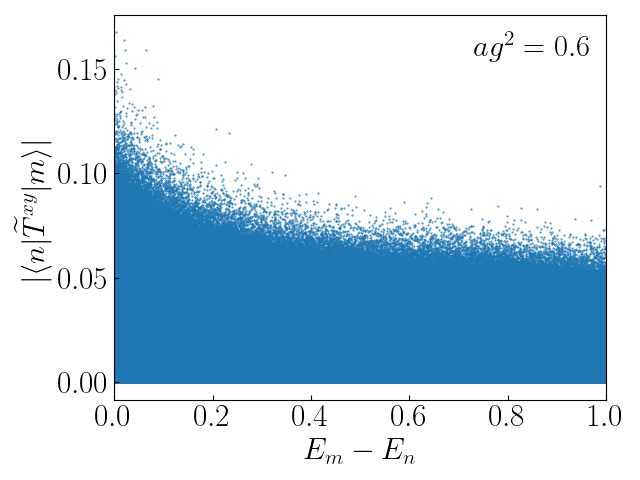}%
}\hfill
\subfloat[$ag^2=1$.\label{fig:Toff1.0}]{%
  \includegraphics[height=2.3in]{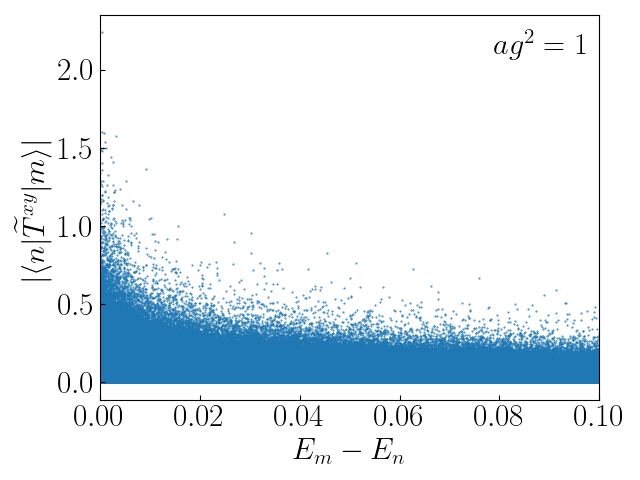}%
}
\caption{Magnitudes of off-diagonal matrix elements of $\widetilde{T}^{xy}$ on a $4\times4$ lattice with $j_{\rm max}=\frac{1}{2}$ and two different couplings: (a) $ag^2=0.6$, (b) $ag^2=1$.}
\label{fig:Toff}
\end{figure}

Our results of $|\langle n|\widetilde{T}^{xy}|m\rangle| $ on a $4\times4$ lattice with $j_{\rm}=\frac{1}{2}$ are shown in Fig.~\ref{fig:Toff} for two values of $ag^2$: 0.6 and 1.0, where we use eigenstates in the energy windows $15<E_n,E_m<17$ and $26<E_n,E_m<28$, respectively. Previous calculations showed no structure in $\frac{\rho^{xy}(\omega)}{\omega}$ when $\omega$ is small (which means it is flat in $\omega$) for strongly coupled $\ml{N}=4$ supersymmetric Yang-Mills theory while there is a peak structure in perturbative QCD results~\cite{Moore:2008ws,Zhu:2012be}. Our results exhibit peak structures at small $\omega$ at the two couplings studied. The peak is broader when the coupling is smaller (note the $x$-axis scale is different in the two plots in Fig.~\ref{fig:Toff}). Whether these peak structures persist with higher $j_{\rm max}$ truncation should be studied in the future.\\

\subsection{Quantum Computing Results}

\begin{figure*}
    \centering
    \includegraphics[width=\textwidth]{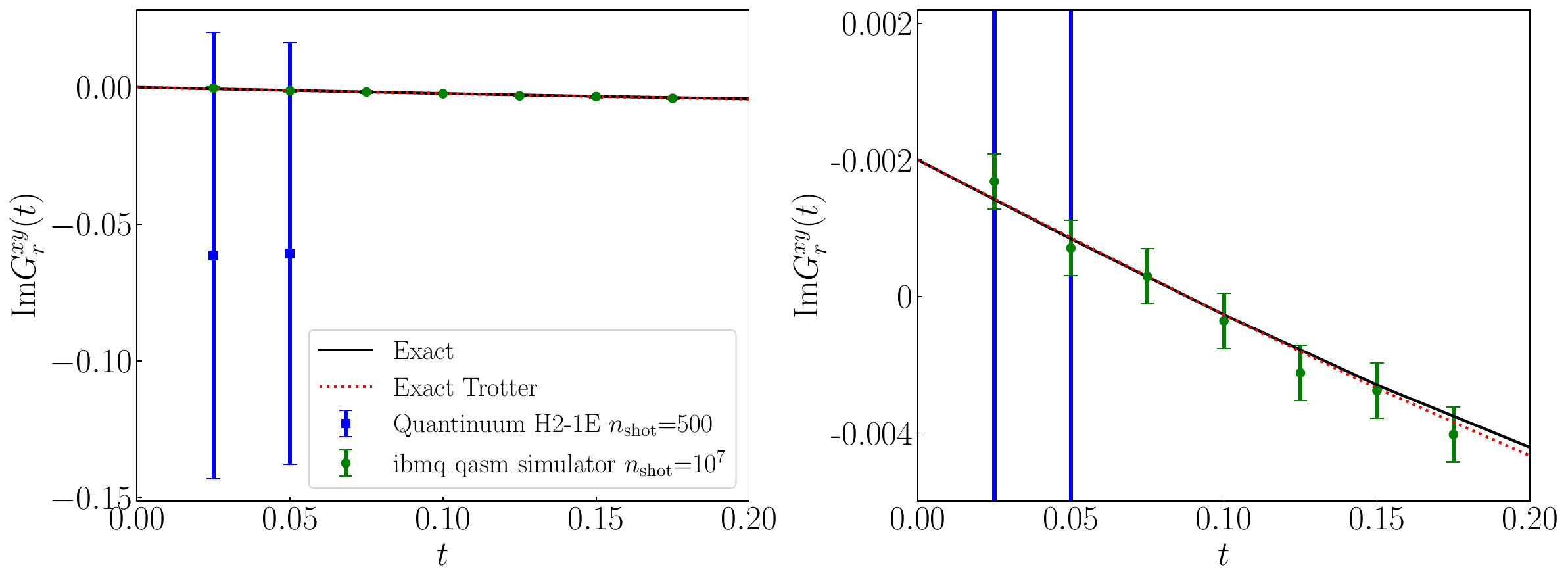}
    \caption{Imaginary part of the retarded Green's function as a function of time, obtained from exact diagonalization (black solid line), classically evolved quantum circuits (red dashed line), Quantinuum \texttt{H2-1E} emulator with $n_{\rm shots}=500$ shots (blue dots) and noiseless IBM simulator using $10^7$ shots (green dots). The red dashed line is obtained by connecting nearest points at intervals of $\Delta t=0.025$ with straight lines.
    In the right panel, we zoom in on the $y$-axis so the statistical uncertainty associated with the green dots can be seen.}
    \label{fig:quantinuum}
\end{figure*}

We test the quantum circuit proposed in Sec.~\ref{sec:qc} to evaluate $G_r^{xy}(t)$ on a $2\times 2$ lattice with $j_{\rm max}=\frac{1}{2}$ and $ag^2=1$ for $\beta=0.15$ (in lattice units). Parts of the thermal state preparation and real time evolution quantum circuits can be found in Appendix~\ref{app:circuit}.

We will show results of the imaginary part of the retarded Green's function (it is purely imaginary) at different times, which are calculated from the commutator $[T_{\rm sum}^{xy}(t),T^{xy}_{10}(0)]$. We do not evaluate the commutator by using $T_{\rm sum}^{xy}(0)$ at $t=0$ because it requires 16 quantum circuits per time step. Indeed, for a $N_i \times N_j$ lattice, the decomposition of $T_{\rm sum}^{xy}$ contains $2 N_i N_j$ Pauli strings [see Eqs.~\eqref{eqn:T12_spin} and~\eqref{eqn:Txy_commutator}]. Since the commutator formula in Eq.~\eqref{eq:comm_eq} is valid only for a single Pauli string, we need to prepare different circuits for each of the $2 N_i N_j$ terms in $T_{\rm sum}^{xy}$; for each commutator, we have two quantum circuits with different signs in the exponent.

We first run the quantum circuits on the Quantinuum \texttt{H2-1E} emulator~\cite{quantinuum} to compute the first two time points with a real time Trotter step of $\Delta t=0.025$ and a single imaginary time Trotter step of $\Delta \tau =\frac{\beta}{2}=0.075$. For each quantum circuit, we measure the evolved circuit, which collapses the wavefunction onto some basis state, and repeat for a total number of $n_{\rm shot}=500$ times\footnote{It is economically expensive to acquire more shots for the Quantinuum emulator.}. We expect that running the same quantum circuits on the real machine would give us very close results because the Quantinuum emulator is very close to the real hardware~\cite{quantinuum_iqus,neutrinos_marc}. For the given parameters, the success probability of obtaining the thermal density matrix is $0.189$, which means only $18.9\%$ of the total 500 shots give useful measurement outcomes. The required number of CNOT gates to prepare the thermal state and perform the $\frac{\pi}{4}\Sigma_\alpha$ gate is 76. For each real time step, the number of CNOT gates is $34$. So the total number of CNOT gates grow linearly as $76+34 N_t$ with the total real time steps $N_t$. The number of implemented CNOT gates differs from Eq.~\eqref{eq:cnot_estimation} that describes the general case when we have a seven-body term in the Hamiltonian. Here, for a $2\times 2$ lattice, we only have four-body interactions. 

The results obtained from the Quantinuum \texttt{H2-1E} are shown in blue squares in Fig.~\ref{fig:quantinuum}, where the exact diagonalization results are also shown for comparison, as well as the results obtained from the classically evolved quantum circuits with the same real and imaginary time Trotter steps, which we call ``exact Trotter''. The perfect agreement between the exact diagonalization results and the classically evolved quantum circuit results show the Trotter errors are negligible. The associated statistical uncertainties with $500$ shots are huge since the small expectation values of the commutator (which are $10^{-4}-10^{-3}$) require a significant number of shots (more than a million) to reconstruct the observable from the projective samples of the state.
To confirm this, we run the same quantum circuits on the noiseless \texttt{ibmq\textunderscore qasm\textunderscore simulator} IBM simulator~\cite{IBM,Qiskit} with $n_{\rm shot}=10^7$ and obtain results shown as green circles in Fig.~\ref{fig:quantinuum}. We observe that the IBM simulator results obtained from $10^7$ shots have much smaller statistical uncertainties and are compatible with the exact results, demonstrating the validity of the proposed quantum circuit in computing the retarded Green's function to obtain the shear viscosity.

\section{Conclusions}
\label{sec:conclusions}
In this work, we considered calculating the shear viscosity nonperturbatively by using the Hamiltonian lattice approach for the $2+1$-dimensional SU(2) Yang-Mills theory. The shear viscosity is obtained from the real time retarded Green's function of the stress-energy tensor via the Kubo formula. We included the renormalization of the coupling when taking the continuum limit but not additional operator renormalization. By exactly diagonalizing the theory on a $4\times4$ honeycomb lattice with $j_{\rm max}=\frac{1}{2}$, we found the ratio of the shear viscosity and the entropy density is consistent with the well-known holographic result $\frac{1}{4\pi}$ at several temperatures. On the other hand, our results showed a peak structure in the spectral function divided by the frequency (i.e., $\frac{\rho^{xy}(\omega)}{\omega}$) when the frequency is small, which is qualitatively different from the holographic result but similar to the perturbative one.

The finite volume and local Hilbert space truncation effects are probably large in our current studies, which motivate future calculations with higher $j_{\rm max}$ values in the Hilbert space truncation on bigger lattices. These calculations probably can neither be easily done by exact diagonalization due to the memory limitation nor by the matrix product state classical simulation method due to the exponential growth of the computational time.

So we propose a quantum algorithm to evaluate the shear viscosity on quantum devices. We analyzed various systematics of the calculation. 
We tested the reliability of the quantum algorithm on a $2\times2$ lattice with $j_{\rm max}=\frac{1}{2}$ and found the quantum results agree with the classical ones, despite the huge number of shots needed to accurately evaluate the retarded Green's function. Moreover, the success probability of the ${\rm QITP}_{\rm th}$ algorithm used in preparing the thermal state decreases with the lattice size exponentially, with a very small prefactor in the exponent at high temperature. Therefore, it would be desirable to upgrade or develop more efficient methods to prepare the  thermal state for the evaluation of the retarded Green's function. 
Future physical calculations of the shear viscosity beyond our current study would probably require robust large-scale quantum computers that are capable of performing a large number of shots.

Future studies should also investigate the shear viscosity calculations for $3+1D$ SU(2) pure gauge theory, theories with dynamical fermions and SU(3) theories by using the Hamiltonians studied in Refs.~\cite{Raychowdhury:2019iki,Ciavarella:2021nmj,Kadam:2022ipf,Cataldi:2023xki,Kavaki:2024ijd}. It is also interesting to calculate other transport coefficients such as the bulk viscosity, heavy quark diffusion coefficient~\cite{Casalderrey-Solana:2006fio,Caron-Huot:2009ncn,Brambilla:2022xbd,Altenkort:2023oms} and quarkonium transport coefficients~\cite{Brambilla:2016wgg,Brambilla:2017zei,Scheihing-Hitschfeld:2023tuz,Scheihing-Hitschfeld:2022xqx,Nijs:2023dbc,Leino:2024pen}. All of these studies will deepen our understanding of the nonperturbative real time dynamics in QCD.

\begin{acknowledgements}
We thank David Kaplan, Berndt M\"uller, Saurabh Vasant Kadam, Martin Savage, Paul Romatschke, Steve Sharpe and Larry Yaffe for useful discussions. This work 
is supported by the U.S. Department of Energy, Office of Science, Office of Nuclear Physics, InQubator for Quantum Simulation (IQuS) (https://iqus.uw.edu) under Award Number DOE (NP) Award DE-SC0020970 via the program on Quantum Horizons: QIS Research and Innovation for Nuclear Science. A.C. acknowledges support from the U.S. Department of Energy, Office of Science under contract DE-AC02-05CH11231, partially through Quantum Information Science Enabled Discovery (QuantISED) for High Energy Physics (KA2401032).
This research used resources of the Oak Ridge Leadership Computing Facility at the Oak Ridge National Laboratory, which is supported by the Office of Science of the U.S. Department of Energy under Contract No. DE-AC05-00OR22725.
We acknowledge the use of Quantinnum and IBM Quantum services for this work. The views expressed are those of the authors, and do not reflect the official policy or position of IBM or the IBM Quantum team.

\end{acknowledgements}

\appendix

\section{An Estimate of $j_{\rm max}$ Needed}
\label{app:jmax}
For the purpose of this estimate, we undo the energy shift in Eq.~\eqref{eq:HKS} and write the Hamiltonian as
\begin{align}
\label{eq:HKS2}
H =   \frac{3\sqrt{3}g^2}{4}  \sum_{\rm links} E_i^a E_i^a
+ \frac{4\sqrt{3}}{9 g^2a^2}  \sum_{\bs x} ( 2-\varhexagon({\bs x}) ) \,,
\end{align}
where $\varhexagon({\bs x})$ is the trace of six lattice Wilson lines over the fundamental SU(2) indices. So $2-\varhexagon({\bs x})$ is positive semi-definite for any plaquette at ${\bs x}$.

Some of the techniques we will use have been used in early work studying quantum computing for scalar field theory~\cite{Jordan:2012xnu}.
We assume the lattice has $N_{l}$ links labeled as $1,2,\dots, N_{l}$, where the electric basis is represented by $j_i$ with $i\in[1,2,\dots, N_{l}]$. We consider an arbitrary wavefunction given in the untruncated basis as
\begin{align}
|\psi\rangle = \sum_{j_1=0}^{\infty} \sum_{j_2=0}^{\infty} \cdots \sum_{j_{N_l}=0}^{\infty} \psi(j_1,j_2,\dots, j_{N_l}) | j_1, j_2, \dots, j_{N_l}\rangle \,.
\end{align}
If the electric basis is truncated at $j_{\rm max}$, the truncated wavefunction is then
\begin{align}
|\psi_{\rm cut}\rangle = \sum_{j_1=0}^{j_{\rm max}} \sum_{j_2=0}^{j_{\rm max}} \cdots \sum_{j_{N_l}=0}^{j_{\rm max}} \psi(j_1,j_2,\dots, j_{N_l}) | j_1, j_2, \dots, j_{N_l}\rangle \,.
\end{align}
Its overlap with the untruncated wavefunction is
\begin{align}
\label{eqn:overlap}
\langle \psi | \psi_{\rm cut}\rangle = \sum_{j_1=0}^{j_{\rm max}} \sum_{j_2=0}^{j_{\rm max}} \cdots \sum_{j_{N_l}=0}^{j_{\rm max}} |\psi(j_1,j_2,\dots, j_{N_l})|^2 \,.
\end{align}
The lower bound of the overlap can be estimated as
\begin{align}
\label{eqn:overlap_lower}
\langle \psi | \psi_{\rm cut}\rangle \geq 1 - N_l \max_{i\in{\rm links}} P(j_i>j_{\rm max}) \,,
\end{align}
where $P(j_i>j_{\rm max})$ is the probability of $j_i>j_{\rm max}$. Using the Markov's inequality, we can show
\begin{align}
\label{eqn:P_ji}
P(j_i>j_{\rm max}) \leq P(j_i \geq j_{\rm max}) \leq \frac{\langle j_i\rangle}{j_{\rm max}} \leq \frac{\langle j_i(j_i+1) \rangle}{j_{\rm max}} \,,
\end{align}
where $\langle \ml{O} \rangle$ denotes the expectation value of the observable $\ml{O}$.

If we want to describe all states below the energy $\widetilde{E}$ [corresponding to the Hamiltonian in Eq.~\eqref{eq:HKS2}], we require
\begin{align}
\label{eqn:Ebound}
\widetilde{E} & \geq \langle \psi | H | \psi \rangle  \geq \langle \psi | \frac{3\sqrt{3}g^2}{4}\sum_{\rm links} E_i^a E_i^a | \psi \rangle \nn\\
& \geq \frac{3\sqrt{3}g^2}{4}  \langle \psi | E_i^a E_i^a | \psi \rangle = \frac{3\sqrt{3}g^2}{4} \langle j_i(j_i+1) \rangle \,,
\end{align}
where we have used the fact that in Eq.~\eqref{eq:HKS2}, the magnetic term is positive semi-definite and so is each electric term in the sum over all links. Combining Eqs.~\eqref{eqn:overlap_lower},~\eqref{eqn:P_ji} and~\eqref{eqn:Ebound} leads to
\begin{align}
\langle \psi | \psi_{\rm cut}\rangle \geq 1 - N_l \frac{4\widetilde{E}}{3\sqrt{3}g^2j_{\rm max}} \,.
\end{align}
If we require the error of describing the wavefunction at most to be $\epsilon$, i.e., $\langle \psi | \psi_{\rm cut}\rangle \geq 1-\epsilon$, we find the minimum of $j_{\rm max}$ needed is at most
\begin{align}
j_{\rm max} = \frac{4N_l\widetilde{E}}{3\sqrt{3}g^2\epsilon} \,.
\end{align}

\section{Quantum circuit for $2\times2$ lattice}
\label{app:circuit}

\begin{figure*}
    \centering
    $$\scalebox{0.6}{\Qcircuit @C=1.0em @R=0.2em @!R { \\
	 	\nghost{{q}_{0} :  } & \lstick{{q}_{0} :  } & \qw& \qw & \qw & \qw & \qw & \qw & \qw & \qw & \qw & \qw & \qw & \qw\\
	 	\nghost{{q}_{1} :  } & \lstick{{q}_{1} :  } & \qw& \qw & \ctrl{2} & \qw & \qw & \qw & \ctrl{2} & \qw & \qw & \qw & \qw & \qw\\
	 	\nghost{{q}_{2} :  } & \lstick{{q}_{2} :  } & \qw& \qw & \qw & \qw & \ctrl{1} & \qw & \qw & \qw & \ctrl{1} & \qw & \qw & \qw\\
	 	\nghost{{q}_{3} :  } & \lstick{{q}_{3} :  } & \gate{H} & \gate{\mathrm{R_Z}(-0.009623)} & \targ & \gate{\mathrm{R_Z}\,(\mathrm{0.02887})} & \targ & \gate{\mathrm{R_Z}\,(\mathrm{0.02887})} & \targ & \gate{\mathrm{R_Z}\,(\mathrm{0.02887})} & \targ & \gate{\mathrm{H}} & \qw & \qw\\
\\ } }$$
\caption{Quantum circuit for the real time evolution driven by the $(i,j)=(1,1)$ magnetic Hamiltonian term for a $2 \times 2$ lattice with $j_{\rm max}=\frac{1}{2}$, $ag^2=1$, and $\Delta t=0.05$. $q_i=0,1,2,3$ represent the qubits where we map our physical system.}
    \label{fig:real_time_2x2_qc}
\end{figure*}
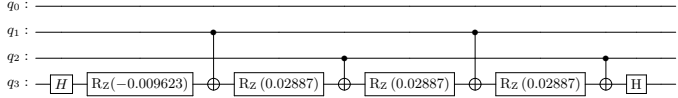

This appendix presents two of the implemented quantum circuits for the magnetic Hamiltonian term at position $(1,1)$ when $j_{\rm max}=\frac{1}{2}$ and $ag^2=1$ are chosen. Figure~\ref{fig:QITP_2x2_qc} shows the quantum circuit for the ${\rm QITP}_{\rm th}$ algorithm when $\beta=0.15$ and Fig.~\ref{fig:real_time_2x2_qc} for the real time evolution with $\Delta t=0.05$. By implementing single qubit rotations given by the Hadamard gate $H$ and $S=R_z(\frac{\pi}{2})$, we can transform an arbitrary tensor product of Pauli operators into a tensor product of Pauli-$z$ operators.

\begin{figure*}
\centering
$$ \scalebox{0.6}{
\Qcircuit @C=1.0em @R=0.2em @!R { \\
	 	\nghost{{q}_{0} :  } & \lstick{{q}_{0} :  } & \qw & \qw & \qw & \qw & \qw & \qw & \qw & \qw & \qw & \qw & \qw & \qw & \qw & \qw & \qw & \qw & \qw & \qw & \qw & \qw & \qw & \qw\\
	 	\nghost{{q}_{1} :  } & \lstick{{q}_{1} :  } & \qw & \qw & \qw & \qw & \qw & \qw & \qw & \qw & \qw & \ctrl{3} & \qw & \qw & \qw & \qw & \qw & \qw & \qw & \ctrl{3} & \qw & \qw & \qw & \qw\\
	 	\nghost{{q}_{2} :  } & \lstick{{q}_{2} :  } & \qw & \qw & \qw & \qw & \qw & \ctrl{2} & \qw & \qw & \qw & \qw & \qw & \qw & \qw & \ctrl{2} & \qw & \qw & \qw & \qw & \qw & \qw & \qw & \qw\\
	 	\nghost{{q}_{3} :  } & \lstick{{q}_{3} :  } & \gate{\mathrm{H}} & \qw & \qw & \ctrl{1} & \qw & \qw & \qw & \ctrl{1} & \qw & \qw & \qw & \ctrl{1} & \qw & \qw & \qw & \ctrl{1} & \gate{\mathrm{H}} & \qw & \qw & \qw & \qw & \qw\\
	 	\nghost{{q}_{4} :  } & \lstick{{q}_{4} :  } & \gate{\mathrm{S^\dagger}} & \gate{\mathrm{H}} & \gate{\mathrm{R_Z}\,(\mathrm{0.8447})} & \targ & \gate{\mathrm{R_Z}\,(\mathrm{-0.01294})} & \targ & \gate{\mathrm{R_Z}\,(\mathrm{0.2222})} & \targ & \gate{\mathrm{R_Z}\,(\mathrm{-0.06362})} & \targ & \gate{\mathrm{R_Z}\,(\mathrm{-0.06362})} & \targ & \gate{\mathrm{R_Z}\,(\mathrm{0.2222})} & \targ & \gate{\mathrm{R_Z}\,(\mathrm{0.2222})} & \targ & \gate{\mathrm{R_Z}\,(\mathrm{-0.06362})} & \targ & \gate{\mathrm{H}} & \gate{\mathrm{S}} & \meter & \qw\\
\\ }}$$
\caption{${\rm QITP}_{\rm th}$ quantum circuit for the magnetic Hamiltonian term at $(1,1)$ position for a $2 \times 2$ lattice with $j_{\rm max}=\frac{1}{2}$, $ag^2=1$, and $\beta=0.15$. $q_i=0,1,2,3$ represent the qubits where we map our physical system, and the $q_4$ qubit indicates the ancilla qubit to be measured.}
    \label{fig:QITP_2x2_qc}
\end{figure*}
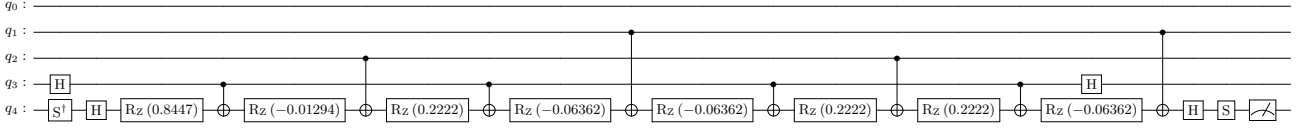

\section{Matrix Product State Classical Simulation}
\label{app:mps}

\begin{figure}[ht]
    \centering
    \includegraphics[width=\columnwidth]{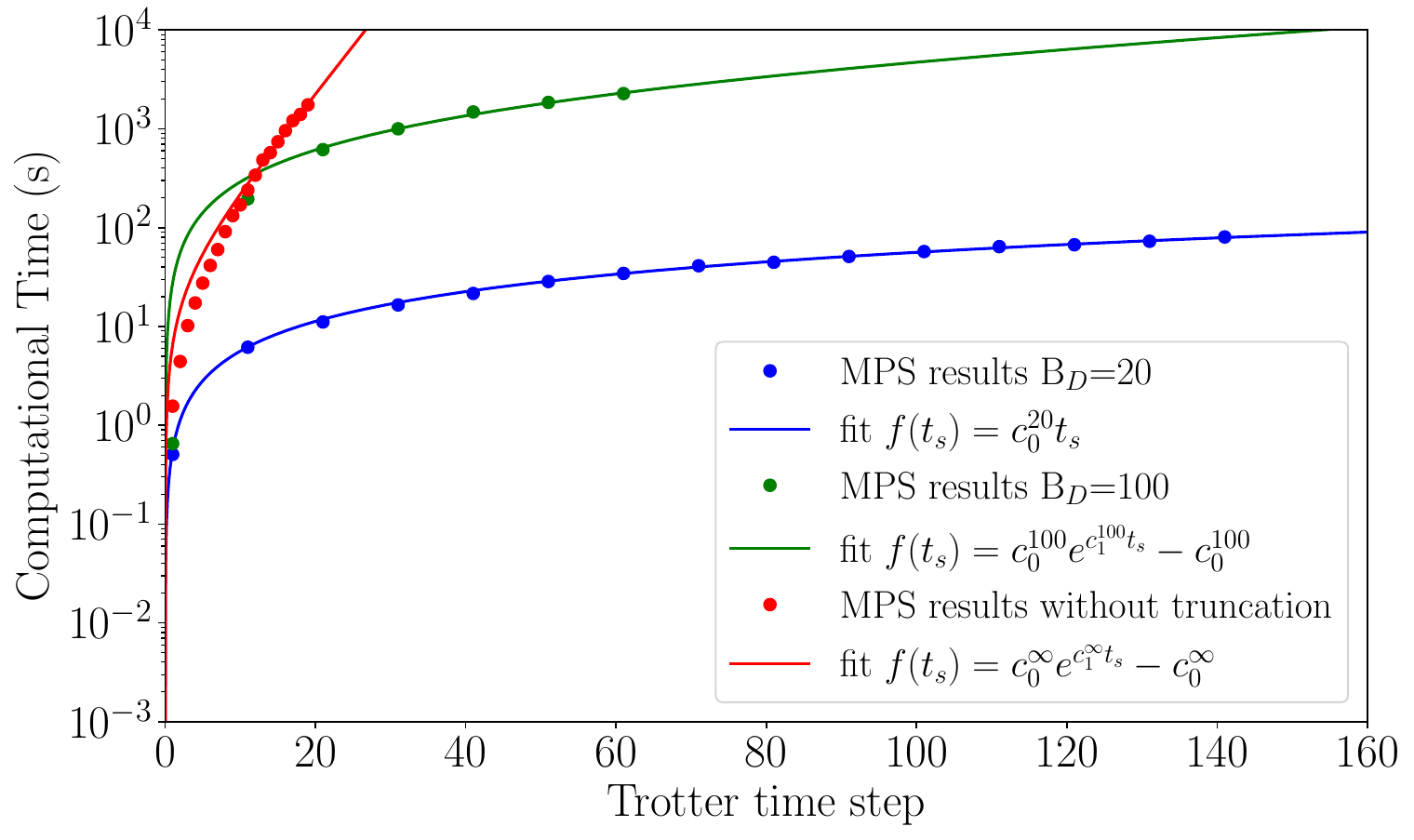}
    \caption{Computational time for $1024$ shots using \texttt{qiskit MPS Aer emulator} as a function of the Trotter time step ($t_s$) on a $5 \times 5$ lattice with $j_{\rm max}=\frac{1}{2}$, $ag^2=1$. Here, we only implement the real time evolution. Blue, green and red circles indicate the obtained results for $B_D=20$, $B_D=100$ and $B_D=\infty$, respectively. Lines with the same color represent the fitting results.}
    \label{fig:MPS_computational_time}
\end{figure}

\begin{table}[ht]
\centering
\begin{tabular}{|c|c|c|c|}
\hline
 &\multicolumn{3}{|c|}{Fit results}\\
 \hline
&  $B_D=20$ & $B_D=100$ &$B_D=\infty$\\
\hline
$c^{B_D}_0$ &$0.563(3)$ & $3(1)\cdot10^3$& $26(3)$ \\
$c^{B_D}_1$ & & $0.010(3)$ & $0.227(7)$\\

\hline
Trotter time steps &\multicolumn{3}{|c|}{ Computational time estimation}\\
\hline
& $B_D=20$ & $B_D=100$ & $B_D=\infty$\\
\hline
20 &  & 11 min&$\sim$ 40 min\\
50 & & 30 min& $\sim$ 500 hour\\
75  & &1 hour& $\sim$ 10 year\\
100  & &80 min & $\sim$ 4000 year\\
125  & 1 min & 2 hours& $\sim$ $10^5$ year \\
150  & 2 min &3 hours&$\sim$ 3$\times 10^8$ year\\ 
300 & 3 min& 15 hour&\\
500 & 6 min& 12 days&\\ 
1000 & 10 min  &2 years&\\ 
\hline
\end{tabular}
\caption{Using the fitting result of Fig.~\ref{fig:MPS_computational_time}, we estimate the order of magnitude of the required computational time to implement the real time evolution for specific time steps and $1024$ shots\footnote{For more accurate results, we have to enhance the number of shots with a further increase of the computational time.} on a laptop for different bond dimension values on a $5\times 5$ lattice with $ag^2=1$ and $j_{\rm max}=\frac{1}{2}$.}
 \label{tab:MPS_estimation}
\end{table}

\begin{figure}[ht]
    \centering
    \includegraphics[width=\columnwidth]{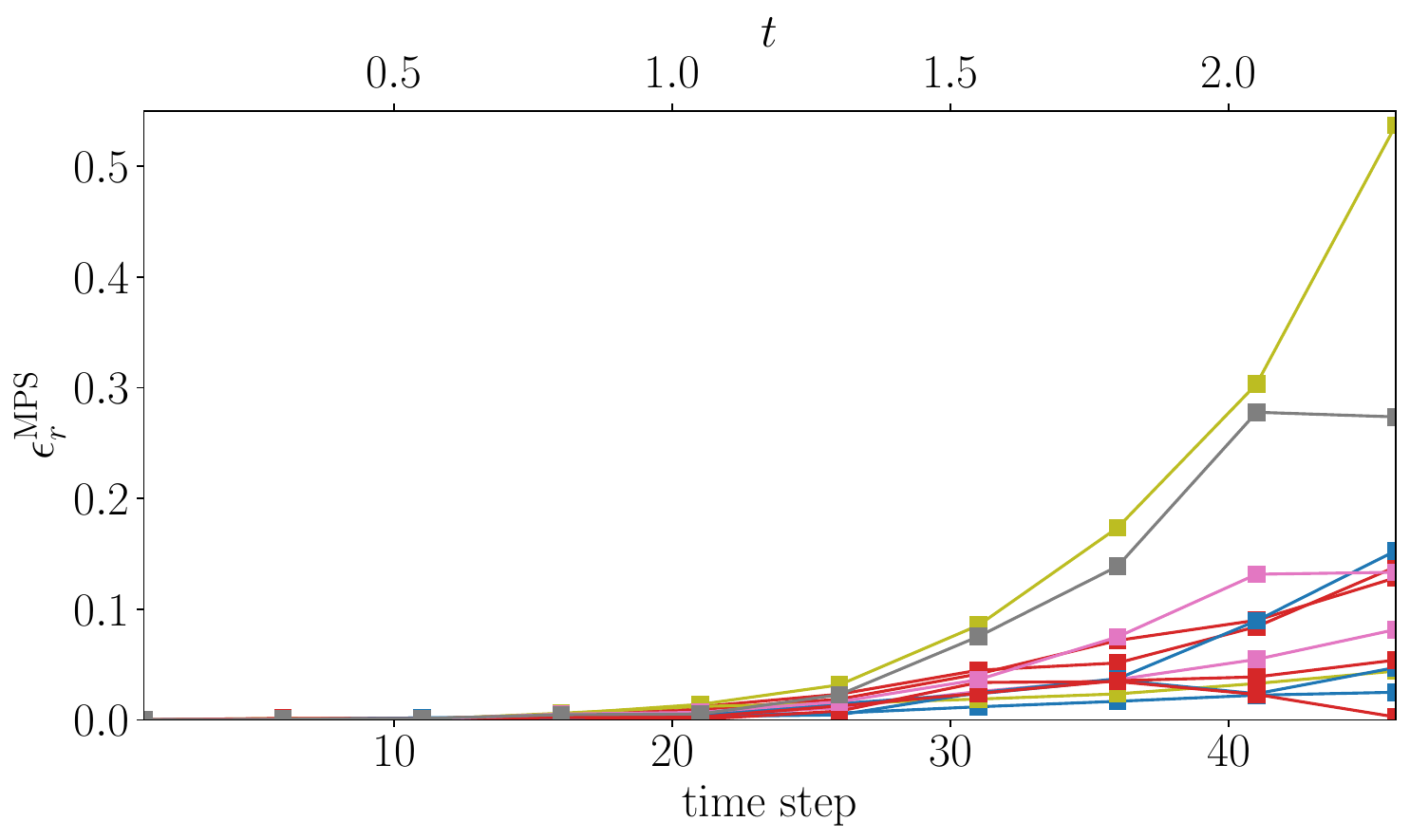}
    \caption{Relative error [defined in Eq.~\eqref{eq:MPS_relative_error}] using \texttt{qiskit MPS} method for the real time evolution starting from a random bit state, which are generated by the applications of random $X$ gate on the default initial  state, on a $5\times5$ lattice with $ag^2=1$, $j_{\rm max}=\frac{1}{2}$ and $\Delta t=0.05$.
    Bottom $x$-axis illustrates the time step while the top one the time of the real time evolution.}
    \label{fig:MPS_relative_error}
\end{figure}

Classically, one can simulate the quantum circuit by implementing the matrix product state (MPS) algorithms~\cite{MPS_review_vidal,MPS_review_Schollwock}. We simulate just the real time evolution part of the quantum circuit on the \texttt{Aer qiskit emulator} with the MPS flag activated, starting from the default initial state of the quantum processor $\ket{000\dots 0}$ on a $5\times5$ lattice with $j_{\rm max}=\frac{1}{2}$, $ag^2=1$ and $\Delta t =0.05$. We investigate the dependence of the computational time on the bond dimension. Its value corresponds to a trade-off between faster calculations (lower value) and better accuracy (higher value). 

Figure~\ref{fig:MPS_computational_time} shows the computational time as a function of the number of Trotter steps ($t_s$) for different \texttt{MPS qiskit emulators}~\cite{IBM,Qiskit} with bond dimensions $B_D=20,100$ and without truncation, namely $B_D=\infty$, which indicates the exact result. All the simulations are implemented on a laptop\footnote{We implement the simulations in jupyter notebook with the multi-threads options on $13$-th Intel Gen i7-13620H chip.} with 1024 number of shots. Increasing the number of shots to obtain more accurate results leads to a further increase of the computational time.

The blue, green and red circles represent the results with $B_D=20$, $B_D=100$, and $B_D=\infty$, respectively. We interpolate these data with two different functions: a line for $B_D=20$, $f(t_s)=c_0^{20} t_s$, and an exponential for $B_D=100$ and $B_D=\infty$, $f(t_s)=c_0^{B_D} e^{c_1^{B_D} t_s}-c_0^{B_D}$. In both cases, we impose that at zero time step, the computational time is $0$. The obtained fitting parameters are reported in the first two rows of Table~\ref{tab:MPS_estimation}. 

In the same table, using the fitted parameters, we estimate the order of magnitude of the needed computational time to perform $t_s$ time steps. We have to evolve our system for long time in order to evaluate the shear viscosity by integrating over time. The total computational time would be given by the sum of computational times for all the time steps. Moreover, we have to add the computational time to prepare the thermal state in realistic analyses.  
The exponential trend and the numbers shown in Table~\ref{tab:MPS_estimation} suggest that it is almost impossible to use the classical MPS simulation of the quantum circuit to accurately calculate the shear viscosity (imposing higher $B_D$ values). 

Indeed, in the quick calculations with a lower bond dimension in the MPS algorithm, the obtained results become less accurate after a long evolution time. We implement the real time evolution using the \texttt{MPS qiskit} function fixing the bond dimension $B_D=20$ and $B_D=100$ on a $5\times 5$ honeycomb lattice with $j_{\rm max}=\frac{1}{2}$, $ag^2=1$ and $\Delta t=0.05$. Figure~\ref{fig:MPS_relative_error} shows the obtained relative error defined as
\begin{equation}
    \epsilon_r^{\rm MPS}=\frac{\left|p_{100}(t)-p_{20}(t)\right|}{p_{100}(t)}\,,\label{eq:MPS_relative_error}
\end{equation}
where $p_{B_D}(t)$ indicates the probability of the initial state $|i\rangle$ at time $t$ using $B_D$ bond dimension, i.e., $p_{B_D}(t) = \langle i| U_t|i\rangle$. 
Different lines correspond to the obtained results starting from a random bit state, generated by random applications of $X$ gate on the default initial state. We can observe a rapid increase of the relative error with the time step. Therefore, the shear viscosity calculations that require long time evolution would be extremely imprecise using MPS algorithms with low bond dimensions.

\bibliography{main.bib}

\begin{thebibliography}{76}%
\makeatletter
\providecommand \@ifxundefined [1]{%
 \@ifx{#1\undefined}
}%
\providecommand \@ifnum [1]{%
 \ifnum #1\expandafter \@firstoftwo
 \else \expandafter \@secondoftwo
 \fi
}%
\providecommand \@ifx [1]{%
 \ifx #1\expandafter \@firstoftwo
 \else \expandafter \@secondoftwo
 \fi
}%
\providecommand \natexlab [1]{#1}%
\providecommand \enquote  [1]{``#1''}%
\providecommand \bibnamefont  [1]{#1}%
\providecommand \bibfnamefont [1]{#1}%
\providecommand \citenamefont [1]{#1}%
\providecommand \href@noop [0]{\@secondoftwo}%
\providecommand \href [0]{\begingroup \@sanitize@url \@href}%
\providecommand \@href[1]{\@@startlink{#1}\@@href}%
\providecommand \@@href[1]{\endgroup#1\@@endlink}%
\providecommand \@sanitize@url [0]{\catcode `\\12\catcode `\$12\catcode
  `\&12\catcode `\#12\catcode `\^12\catcode `\_12\catcode `\%12\relax}%
\providecommand \@@startlink[1]{}%
\providecommand \@@endlink[0]{}%
\providecommand \url  [0]{\begingroup\@sanitize@url \@url }%
\providecommand \@url [1]{\endgroup\@href {#1}{\urlprefix }}%
\providecommand \urlprefix  [0]{URL }%
\providecommand \Eprint [0]{\href }%
\providecommand \doibase [0]{http://dx.doi.org/}%
\providecommand \selectlanguage [0]{\@gobble}%
\providecommand \bibinfo  [0]{\@secondoftwo}%
\providecommand \bibfield  [0]{\@secondoftwo}%
\providecommand \translation [1]{[#1]}%
\providecommand \BibitemOpen [0]{}%
\providecommand \bibitemStop [0]{}%
\providecommand \bibitemNoStop [0]{.\EOS\space}%
\providecommand \EOS [0]{\spacefactor3000\relax}%
\providecommand \BibitemShut  [1]{\csname bibitem#1\endcsname}%
\let\auto@bib@innerbib\@empty
\bibitem [{\citenamefont {Song}\ \emph {et~al.}(2011)\citenamefont {Song},
  \citenamefont {Bass}, \citenamefont {Heinz}, \citenamefont {Hirano},\ and\
  \citenamefont {Shen}}]{Song:2010mg}%
  \BibitemOpen
  \bibfield  {author} {\bibinfo {author} {\bibfnamefont {H.}~\bibnamefont
  {Song}}, \bibinfo {author} {\bibfnamefont {S.~A.}\ \bibnamefont {Bass}},
  \bibinfo {author} {\bibfnamefont {U.}~\bibnamefont {Heinz}}, \bibinfo
  {author} {\bibfnamefont {T.}~\bibnamefont {Hirano}}, \ and\ \bibinfo {author}
  {\bibfnamefont {C.}~\bibnamefont {Shen}},\ }\href {\doibase
  10.1103/PhysRevLett.106.192301} {\bibfield  {journal} {\bibinfo  {journal}
  {Phys. Rev. Lett.}\ }\textbf {\bibinfo {volume} {106}},\ \bibinfo {pages}
  {192301} (\bibinfo {year} {2011})},\ \bibinfo {note} {[Erratum:
  Phys.Rev.Lett. 109, 139904 (2012)]},\ \Eprint
  {http://arxiv.org/abs/1011.2783} {arXiv:1011.2783 [nucl-th]} \BibitemShut
  {NoStop}%
\bibitem [{\citenamefont {Schenke}\ \emph {et~al.}(2011)\citenamefont
  {Schenke}, \citenamefont {Jeon},\ and\ \citenamefont
  {Gale}}]{Schenke:2010rr}%
  \BibitemOpen
  \bibfield  {author} {\bibinfo {author} {\bibfnamefont {B.}~\bibnamefont
  {Schenke}}, \bibinfo {author} {\bibfnamefont {S.}~\bibnamefont {Jeon}}, \
  and\ \bibinfo {author} {\bibfnamefont {C.}~\bibnamefont {Gale}},\ }\href
  {\doibase 10.1103/PhysRevLett.106.042301} {\bibfield  {journal} {\bibinfo
  {journal} {Phys. Rev. Lett.}\ }\textbf {\bibinfo {volume} {106}},\ \bibinfo
  {pages} {042301} (\bibinfo {year} {2011})},\ \Eprint
  {http://arxiv.org/abs/1009.3244} {arXiv:1009.3244 [hep-ph]} \BibitemShut
  {NoStop}%
\bibitem [{\citenamefont {Bernhard}\ \emph {et~al.}(2019)\citenamefont
  {Bernhard}, \citenamefont {Moreland},\ and\ \citenamefont
  {Bass}}]{Bernhard:2019bmu}%
  \BibitemOpen
  \bibfield  {author} {\bibinfo {author} {\bibfnamefont {J.~E.}\ \bibnamefont
  {Bernhard}}, \bibinfo {author} {\bibfnamefont {J.~S.}\ \bibnamefont
  {Moreland}}, \ and\ \bibinfo {author} {\bibfnamefont {S.~A.}\ \bibnamefont
  {Bass}},\ }\href {\doibase 10.1038/s41567-019-0611-8} {\bibfield  {journal}
  {\bibinfo  {journal} {Nature Phys.}\ }\textbf {\bibinfo {volume} {15}},\
  \bibinfo {pages} {1113} (\bibinfo {year} {2019})}\BibitemShut {NoStop}%
\bibitem [{\citenamefont {Nijs}\ \emph {et~al.}(2021)\citenamefont {Nijs},
  \citenamefont {van~der Schee}, \citenamefont {G\"ursoy},\ and\ \citenamefont
  {Snellings}}]{Nijs:2020ors}%
  \BibitemOpen
  \bibfield  {author} {\bibinfo {author} {\bibfnamefont {G.}~\bibnamefont
  {Nijs}}, \bibinfo {author} {\bibfnamefont {W.}~\bibnamefont {van~der Schee}},
  \bibinfo {author} {\bibfnamefont {U.}~\bibnamefont {G\"ursoy}}, \ and\
  \bibinfo {author} {\bibfnamefont {R.}~\bibnamefont {Snellings}},\ }\href
  {\doibase 10.1103/PhysRevLett.126.202301} {\bibfield  {journal} {\bibinfo
  {journal} {Phys. Rev. Lett.}\ }\textbf {\bibinfo {volume} {126}},\ \bibinfo
  {pages} {202301} (\bibinfo {year} {2021})},\ \Eprint
  {http://arxiv.org/abs/2010.15130} {arXiv:2010.15130 [nucl-th]} \BibitemShut
  {NoStop}%
\bibitem [{\citenamefont {Policastro}\ \emph {et~al.}(2001)\citenamefont
  {Policastro}, \citenamefont {Son},\ and\ \citenamefont
  {Starinets}}]{Policastro:2001yc}%
  \BibitemOpen
  \bibfield  {author} {\bibinfo {author} {\bibfnamefont {G.}~\bibnamefont
  {Policastro}}, \bibinfo {author} {\bibfnamefont {D.~T.}\ \bibnamefont {Son}},
  \ and\ \bibinfo {author} {\bibfnamefont {A.~O.}\ \bibnamefont {Starinets}},\
  }\href {\doibase 10.1103/PhysRevLett.87.081601} {\bibfield  {journal}
  {\bibinfo  {journal} {Phys. Rev. Lett.}\ }\textbf {\bibinfo {volume} {87}},\
  \bibinfo {pages} {081601} (\bibinfo {year} {2001})},\ \Eprint
  {http://arxiv.org/abs/hep-th/0104066} {arXiv:hep-th/0104066} \BibitemShut
  {NoStop}%
\bibitem [{\citenamefont {Moore}(2020)}]{Moore:2020pfu}%
  \BibitemOpen
  \bibfield  {author} {\bibinfo {author} {\bibfnamefont {G.~D.}\ \bibnamefont
  {Moore}},\ }in\ \href@noop {} {\emph {\bibinfo {booktitle} {{Criticality in
  QCD and the Hadron Resonance Gas}}}}\ (\bibinfo {year} {2020})\ \Eprint
  {http://arxiv.org/abs/2010.15704} {arXiv:2010.15704 [hep-ph]} \BibitemShut
  {NoStop}%
\bibitem [{\citenamefont {Jeon}(1995)}]{Jeon:1994if}%
  \BibitemOpen
  \bibfield  {author} {\bibinfo {author} {\bibfnamefont {S.}~\bibnamefont
  {Jeon}},\ }\href {\doibase 10.1103/PhysRevD.52.3591} {\bibfield  {journal}
  {\bibinfo  {journal} {Phys. Rev. D}\ }\textbf {\bibinfo {volume} {52}},\
  \bibinfo {pages} {3591} (\bibinfo {year} {1995})},\ \Eprint
  {http://arxiv.org/abs/hep-ph/9409250} {arXiv:hep-ph/9409250} \BibitemShut
  {NoStop}%
\bibitem [{\citenamefont {Arnold}\ \emph {et~al.}(2000)\citenamefont {Arnold},
  \citenamefont {Moore},\ and\ \citenamefont {Yaffe}}]{Arnold:2000dr}%
  \BibitemOpen
  \bibfield  {author} {\bibinfo {author} {\bibfnamefont {P.}~\bibnamefont
  {Arnold}}, \bibinfo {author} {\bibfnamefont {G.~D.}\ \bibnamefont {Moore}}, \
  and\ \bibinfo {author} {\bibfnamefont {L.~G.}\ \bibnamefont {Yaffe}},\ }\href
  {\doibase 10.1088/1126-6708/2000/11/001} {\bibfield  {journal} {\bibinfo
  {journal} {JHEP}\ }\textbf {\bibinfo {volume} {2000}},\ \bibinfo {pages}
  {001} (\bibinfo {year} {2000})},\ \Eprint
  {http://arxiv.org/abs/hep-ph/0010177} {arXiv:hep-ph/0010177} \BibitemShut
  {NoStop}%
\bibitem [{\citenamefont {Arnold}\ \emph {et~al.}(2003)\citenamefont {Arnold},
  \citenamefont {Moore},\ and\ \citenamefont {Yaffe}}]{Arnold:2003zc}%
  \BibitemOpen
  \bibfield  {author} {\bibinfo {author} {\bibfnamefont {P.}~\bibnamefont
  {Arnold}}, \bibinfo {author} {\bibfnamefont {G.~D.}\ \bibnamefont {Moore}}, \
  and\ \bibinfo {author} {\bibfnamefont {L.~G.}\ \bibnamefont {Yaffe}},\ }\href
  {\doibase 10.1088/1126-6708/2003/05/051} {\bibfield  {journal} {\bibinfo
  {journal} {JHEP}\ }\textbf {\bibinfo {volume} {2003}},\ \bibinfo {pages}
  {051} (\bibinfo {year} {2003})},\ \Eprint
  {http://arxiv.org/abs/hep-ph/0302165} {arXiv:hep-ph/0302165} \BibitemShut
  {NoStop}%
\bibitem [{\citenamefont {Ghiglieri}\ \emph {et~al.}(2018)\citenamefont
  {Ghiglieri}, \citenamefont {Moore},\ and\ \citenamefont
  {Teaney}}]{Ghiglieri:2018dib}%
  \BibitemOpen
  \bibfield  {author} {\bibinfo {author} {\bibfnamefont {J.}~\bibnamefont
  {Ghiglieri}}, \bibinfo {author} {\bibfnamefont {G.~D.}\ \bibnamefont
  {Moore}}, \ and\ \bibinfo {author} {\bibfnamefont {D.}~\bibnamefont
  {Teaney}},\ }\href {\doibase 10.1007/JHEP03(2018)179} {\bibfield  {journal}
  {\bibinfo  {journal} {JHEP}\ ,\ \bibinfo {pages} {1}} (\bibinfo {year}
  {2018})},\ \Eprint {http://arxiv.org/abs/1802.09535} {arXiv:1802.09535
  [hep-ph]} \BibitemShut {NoStop}%
\bibitem [{\citenamefont {Meyer}(2007)}]{Meyer:2007ic}%
  \BibitemOpen
  \bibfield  {author} {\bibinfo {author} {\bibfnamefont {H.~B.}\ \bibnamefont
  {Meyer}},\ }\href {\doibase 10.1103/PhysRevD.76.101701} {\bibfield  {journal}
  {\bibinfo  {journal} {Phys. Rev. D}\ }\textbf {\bibinfo {volume} {76}},\
  \bibinfo {pages} {101701} (\bibinfo {year} {2007})},\ \Eprint
  {http://arxiv.org/abs/0704.1801} {arXiv:0704.1801 [hep-lat]} \BibitemShut
  {NoStop}%
\bibitem [{\citenamefont {Mages}\ \emph {et~al.}(2015)\citenamefont {Mages},
  \citenamefont {Bors\'anyi}, \citenamefont {Fodor}, \citenamefont
  {Sch\"afer},\ and\ \citenamefont {Szab\'o}}]{Mages:2015rea}%
  \BibitemOpen
  \bibfield  {author} {\bibinfo {author} {\bibfnamefont {S.~W.}\ \bibnamefont
  {Mages}}, \bibinfo {author} {\bibfnamefont {S.}~\bibnamefont {Bors\'anyi}},
  \bibinfo {author} {\bibfnamefont {Z.}~\bibnamefont {Fodor}}, \bibinfo
  {author} {\bibfnamefont {A.}~\bibnamefont {Sch\"afer}}, \ and\ \bibinfo
  {author} {\bibfnamefont {K.}~\bibnamefont {Szab\'o}},\ }\href {\doibase
  10.22323/1.214.0232} {\bibfield  {journal} {\bibinfo  {journal} {PoS}\
  }\textbf {\bibinfo {volume} {LATTICE2014}},\ \bibinfo {pages} {232} (\bibinfo
  {year} {2015})}\BibitemShut {NoStop}%
\bibitem [{\citenamefont {Altenkort}\ \emph
  {et~al.}(2023{\natexlab{a}})\citenamefont {Altenkort}, \citenamefont {Eller},
  \citenamefont {Francis}, \citenamefont {Kaczmarek}, \citenamefont {Mazur},
  \citenamefont {Moore},\ and\ \citenamefont {Shu}}]{Altenkort:2022yhb}%
  \BibitemOpen
  \bibfield  {author} {\bibinfo {author} {\bibfnamefont {L.}~\bibnamefont
  {Altenkort}}, \bibinfo {author} {\bibfnamefont {A.~M.}\ \bibnamefont
  {Eller}}, \bibinfo {author} {\bibfnamefont {A.}~\bibnamefont {Francis}},
  \bibinfo {author} {\bibfnamefont {O.}~\bibnamefont {Kaczmarek}}, \bibinfo
  {author} {\bibfnamefont {L.}~\bibnamefont {Mazur}}, \bibinfo {author}
  {\bibfnamefont {G.~D.}\ \bibnamefont {Moore}}, \ and\ \bibinfo {author}
  {\bibfnamefont {H.-T.}\ \bibnamefont {Shu}},\ }\href {\doibase
  10.1103/PhysRevD.108.014503} {\bibfield  {journal} {\bibinfo  {journal}
  {Phys. Rev. D}\ }\textbf {\bibinfo {volume} {108}},\ \bibinfo {pages}
  {014503} (\bibinfo {year} {2023}{\natexlab{a}})}\BibitemShut {NoStop}%
\bibitem [{\citenamefont {Banuls}\ \emph {et~al.}(2020)\citenamefont {Banuls},
  \citenamefont {Blatt}, \citenamefont {Catani}, \citenamefont {Celi},
  \citenamefont {Cirac}, \citenamefont {Dalmonte}, \citenamefont {Fallani},
  \citenamefont {Jansen}, \citenamefont {Lewenstein}, \citenamefont
  {Montangero} \emph {et~al.}}]{banuls2020simulating}%
  \BibitemOpen
  \bibfield  {author} {\bibinfo {author} {\bibfnamefont {M.~C.}\ \bibnamefont
  {Banuls}}, \bibinfo {author} {\bibfnamefont {R.}~\bibnamefont {Blatt}},
  \bibinfo {author} {\bibfnamefont {J.}~\bibnamefont {Catani}}, \bibinfo
  {author} {\bibfnamefont {A.}~\bibnamefont {Celi}}, \bibinfo {author}
  {\bibfnamefont {J.~I.}\ \bibnamefont {Cirac}}, \bibinfo {author}
  {\bibfnamefont {M.}~\bibnamefont {Dalmonte}}, \bibinfo {author}
  {\bibfnamefont {L.}~\bibnamefont {Fallani}}, \bibinfo {author} {\bibfnamefont
  {K.}~\bibnamefont {Jansen}}, \bibinfo {author} {\bibfnamefont
  {M.}~\bibnamefont {Lewenstein}}, \bibinfo {author} {\bibfnamefont
  {S.}~\bibnamefont {Montangero}},  \emph {et~al.},\ }\href {\doibase
  10.1140/epjd/e2020-100571-8} {\bibfield  {journal} {\bibinfo  {journal} {Eur.
  Phys. J.}\ }\textbf {\bibinfo {volume} {74}},\ \bibinfo {pages} {1} (\bibinfo
  {year} {2020})}\BibitemShut {NoStop}%
\bibitem [{\citenamefont {Klco}\ \emph {et~al.}(2022)\citenamefont {Klco},
  \citenamefont {Roggero},\ and\ \citenamefont {Savage}}]{klco2022standard}%
  \BibitemOpen
  \bibfield  {author} {\bibinfo {author} {\bibfnamefont {N.}~\bibnamefont
  {Klco}}, \bibinfo {author} {\bibfnamefont {A.}~\bibnamefont {Roggero}}, \
  and\ \bibinfo {author} {\bibfnamefont {M.~J.}\ \bibnamefont {Savage}},\
  }\href {\doibase 10.1088/1361-6633/ac58a4} {\bibfield  {journal} {\bibinfo
  {journal} {Rep. Prog. Phys.}\ }\textbf {\bibinfo {volume} {85}},\ \bibinfo
  {pages} {064301} (\bibinfo {year} {2022})}\BibitemShut {NoStop}%
\bibitem [{\citenamefont {Bauer}\ \emph
  {et~al.}(2023{\natexlab{a}})\citenamefont {Bauer}, \citenamefont {Davoudi},
  \citenamefont {Balantekin}, \citenamefont {Bhattacharya}, \citenamefont
  {Carena}, \citenamefont {de~Jong}, \citenamefont {Draper}, \citenamefont
  {El-Khadra}, \citenamefont {Gemelke}, \citenamefont {Hanada}, \citenamefont
  {Kharzeev}, \citenamefont {Lamm}, \citenamefont {Li}, \citenamefont {Liu},
  \citenamefont {Lukin}, \citenamefont {Meurice}, \citenamefont {Monroe},
  \citenamefont {Nachman}, \citenamefont {Pagano}, \citenamefont {Preskill},
  \citenamefont {Rinaldi}, \citenamefont {Roggero}, \citenamefont {Santiago},
  \citenamefont {Savage}, \citenamefont {Siddiqi}, \citenamefont {Siopsis},
  \citenamefont {Van~Zanten}, \citenamefont {Wiebe}, \citenamefont {Yamauchi},
  \citenamefont {Yeter-Aydeniz},\ and\ \citenamefont {Zorzetti}}]{Bauer_2023}%
  \BibitemOpen
  \bibfield  {author} {\bibinfo {author} {\bibfnamefont {C.~W.}\ \bibnamefont
  {Bauer}}, \bibinfo {author} {\bibfnamefont {Z.}~\bibnamefont {Davoudi}},
  \bibinfo {author} {\bibfnamefont {A.~B.}\ \bibnamefont {Balantekin}},
  \bibinfo {author} {\bibfnamefont {T.}~\bibnamefont {Bhattacharya}}, \bibinfo
  {author} {\bibfnamefont {M.}~\bibnamefont {Carena}}, \bibinfo {author}
  {\bibfnamefont {W.~A.}\ \bibnamefont {de~Jong}}, \bibinfo {author}
  {\bibfnamefont {P.}~\bibnamefont {Draper}}, \bibinfo {author} {\bibfnamefont
  {A.}~\bibnamefont {El-Khadra}}, \bibinfo {author} {\bibfnamefont
  {N.}~\bibnamefont {Gemelke}}, \bibinfo {author} {\bibfnamefont
  {M.}~\bibnamefont {Hanada}}, \bibinfo {author} {\bibfnamefont
  {D.}~\bibnamefont {Kharzeev}}, \bibinfo {author} {\bibfnamefont
  {H.}~\bibnamefont {Lamm}}, \bibinfo {author} {\bibfnamefont {Y.-Y.}\
  \bibnamefont {Li}}, \bibinfo {author} {\bibfnamefont {J.}~\bibnamefont
  {Liu}}, \bibinfo {author} {\bibfnamefont {M.}~\bibnamefont {Lukin}}, \bibinfo
  {author} {\bibfnamefont {Y.}~\bibnamefont {Meurice}}, \bibinfo {author}
  {\bibfnamefont {C.}~\bibnamefont {Monroe}}, \bibinfo {author} {\bibfnamefont
  {B.}~\bibnamefont {Nachman}}, \bibinfo {author} {\bibfnamefont
  {G.}~\bibnamefont {Pagano}}, \bibinfo {author} {\bibfnamefont
  {J.}~\bibnamefont {Preskill}}, \bibinfo {author} {\bibfnamefont
  {E.}~\bibnamefont {Rinaldi}}, \bibinfo {author} {\bibfnamefont
  {A.}~\bibnamefont {Roggero}}, \bibinfo {author} {\bibfnamefont {D.~I.}\
  \bibnamefont {Santiago}}, \bibinfo {author} {\bibfnamefont {M.~J.}\
  \bibnamefont {Savage}}, \bibinfo {author} {\bibfnamefont {I.}~\bibnamefont
  {Siddiqi}}, \bibinfo {author} {\bibfnamefont {G.}~\bibnamefont {Siopsis}},
  \bibinfo {author} {\bibfnamefont {D.}~\bibnamefont {Van~Zanten}}, \bibinfo
  {author} {\bibfnamefont {N.}~\bibnamefont {Wiebe}}, \bibinfo {author}
  {\bibfnamefont {Y.}~\bibnamefont {Yamauchi}}, \bibinfo {author}
  {\bibfnamefont {K.}~\bibnamefont {Yeter-Aydeniz}}, \ and\ \bibinfo {author}
  {\bibfnamefont {S.}~\bibnamefont {Zorzetti}},\ }\href {\doibase
  10.1103/prxquantum.4.027001} {\bibfield  {journal} {\bibinfo  {journal} {PRX
  Quantum}\ }\textbf {\bibinfo {volume} {4}} (\bibinfo {year}
  {2023}{\natexlab{a}}),\ 10.1103/prxquantum.4.027001}\BibitemShut {NoStop}%
\bibitem [{\citenamefont {Beck}\ and\ \citenamefont
  {et~al}(2023)}]{beck2023quantum}%
  \BibitemOpen
  \bibfield  {author} {\bibinfo {author} {\bibfnamefont {D.}~\bibnamefont
  {Beck}}\ and\ \bibinfo {author} {\bibnamefont {et~al}},\ }\href {\doibase
  https://arxiv.org/abs/2303.00113} {\enquote {\bibinfo {title} {Quantum
  information science and technology for nuclear physics. input into u.s.
  long-range planning, 2023},}\ } (\bibinfo {year} {2023}),\ \Eprint
  {http://arxiv.org/abs/2303.00113} {arXiv:2303.00113 [nucl-ex]} \BibitemShut
  {NoStop}%
\bibitem [{\citenamefont {Bauer}\ \emph
  {et~al.}(2023{\natexlab{b}})\citenamefont {Bauer}, \citenamefont {Davoudi},
  \citenamefont {Klco},\ and\ \citenamefont {Savage}}]{bauer2023quantum}%
  \BibitemOpen
  \bibfield  {author} {\bibinfo {author} {\bibfnamefont {C.~W.}\ \bibnamefont
  {Bauer}}, \bibinfo {author} {\bibfnamefont {Z.}~\bibnamefont {Davoudi}},
  \bibinfo {author} {\bibfnamefont {N.}~\bibnamefont {Klco}}, \ and\ \bibinfo
  {author} {\bibfnamefont {M.~J.}\ \bibnamefont {Savage}},\ }\href {\doibase
  https://doi.org/10.1038/s42254-023-00599-8} {\bibfield  {journal} {\bibinfo
  {journal} {Nat. Rev. Phys.}\ ,\ \bibinfo {pages} {1}} (\bibinfo {year}
  {2023}{\natexlab{b}})}\BibitemShut {NoStop}%
\bibitem [{\citenamefont {Lamm}\ \emph {et~al.}(2019)\citenamefont {Lamm},
  \citenamefont {Lawrence},\ and\ \citenamefont {Yamauchi}}]{Lamm:2019bik}%
  \BibitemOpen
  \bibfield  {author} {\bibinfo {author} {\bibfnamefont {H.}~\bibnamefont
  {Lamm}}, \bibinfo {author} {\bibfnamefont {S.}~\bibnamefont {Lawrence}}, \
  and\ \bibinfo {author} {\bibfnamefont {Y.}~\bibnamefont {Yamauchi}} (\bibinfo
  {collaboration} {NuQS}),\ }\href {\doibase 10.1103/PhysRevD.100.034518}
  {\bibfield  {journal} {\bibinfo  {journal} {Phys. Rev. D}\ }\textbf {\bibinfo
  {volume} {100}},\ \bibinfo {pages} {034518} (\bibinfo {year} {2019})},\
  \Eprint {http://arxiv.org/abs/1903.08807} {arXiv:1903.08807 [hep-lat]}
  \BibitemShut {NoStop}%
\bibitem [{\citenamefont {Feynman}(1982)}]{feynman}%
  \BibitemOpen
  \bibfield  {author} {\bibinfo {author} {\bibfnamefont {R.~P.}\ \bibnamefont
  {Feynman}},\ }\href {\doibase 10.1007/BF02650179} {\bibfield  {journal}
  {\bibinfo  {journal} {Int J Theor Phys}\ }\textbf {\bibinfo {volume} {21}},\
  \bibinfo {pages} {467} (\bibinfo {year} {1982})}\BibitemShut {NoStop}%
\bibitem [{\citenamefont {Cohen}\ \emph {et~al.}(2021)\citenamefont {Cohen},
  \citenamefont {Lamm}, \citenamefont {Lawrence},\ and\ \citenamefont
  {Yamauchi}}]{Cohen:2021imf}%
  \BibitemOpen
  \bibfield  {author} {\bibinfo {author} {\bibfnamefont {T.~D.}\ \bibnamefont
  {Cohen}}, \bibinfo {author} {\bibfnamefont {H.}~\bibnamefont {Lamm}},
  \bibinfo {author} {\bibfnamefont {S.}~\bibnamefont {Lawrence}}, \ and\
  \bibinfo {author} {\bibfnamefont {Y.}~\bibnamefont {Yamauchi}} (\bibinfo
  {collaboration} {NuQS}),\ }\href {\doibase 10.1103/PhysRevD.104.094514}
  {\bibfield  {journal} {\bibinfo  {journal} {Phys. Rev. D}\ }\textbf {\bibinfo
  {volume} {104}},\ \bibinfo {pages} {094514} (\bibinfo {year} {2021})},\
  \Eprint {http://arxiv.org/abs/2104.02024} {arXiv:2104.02024 [hep-lat]}
  \BibitemShut {NoStop}%
\bibitem [{\citenamefont {Baier}\ \emph {et~al.}(2008)\citenamefont {Baier},
  \citenamefont {Romatschke}, \citenamefont {Son}, \citenamefont {Starinets},\
  and\ \citenamefont {Stephanov}}]{Baier:2007ix}%
  \BibitemOpen
  \bibfield  {author} {\bibinfo {author} {\bibfnamefont {R.}~\bibnamefont
  {Baier}}, \bibinfo {author} {\bibfnamefont {P.}~\bibnamefont {Romatschke}},
  \bibinfo {author} {\bibfnamefont {D.~T.}\ \bibnamefont {Son}}, \bibinfo
  {author} {\bibfnamefont {A.~O.}\ \bibnamefont {Starinets}}, \ and\ \bibinfo
  {author} {\bibfnamefont {M.~A.}\ \bibnamefont {Stephanov}},\ }\href {\doibase
  10.1088/1126-6708/2008/04/100} {\bibfield  {journal} {\bibinfo  {journal}
  {JHEP}\ ,\ \bibinfo {pages} {100}} (\bibinfo {year} {2008})},\ \Eprint
  {http://arxiv.org/abs/0712.2451} {arXiv:0712.2451 [hep-th]} \BibitemShut
  {NoStop}%
\bibitem [{\citenamefont {Moore}\ and\ \citenamefont
  {Sohrabi}(2011)}]{Moore:2010bu}%
  \BibitemOpen
  \bibfield  {author} {\bibinfo {author} {\bibfnamefont {G.~D.}\ \bibnamefont
  {Moore}}\ and\ \bibinfo {author} {\bibfnamefont {K.~A.}\ \bibnamefont
  {Sohrabi}},\ }\href {\doibase 10.1103/PhysRevLett.106.122302} {\bibfield
  {journal} {\bibinfo  {journal} {Phys. Rev. Lett.}\ }\textbf {\bibinfo
  {volume} {106}},\ \bibinfo {pages} {122302} (\bibinfo {year}
  {2011})}\BibitemShut {NoStop}%
\bibitem [{\citenamefont {Binder}\ \emph {et~al.}(2022)\citenamefont {Binder},
  \citenamefont {Mukaida}, \citenamefont {Scheihing-Hitschfeld},\ and\
  \citenamefont {Yao}}]{Binder:2021otw}%
  \BibitemOpen
  \bibfield  {author} {\bibinfo {author} {\bibfnamefont {T.}~\bibnamefont
  {Binder}}, \bibinfo {author} {\bibfnamefont {K.}~\bibnamefont {Mukaida}},
  \bibinfo {author} {\bibfnamefont {B.}~\bibnamefont {Scheihing-Hitschfeld}}, \
  and\ \bibinfo {author} {\bibfnamefont {X.}~\bibnamefont {Yao}},\ }\href
  {\doibase 10.1007/JHEP01(2022)137} {\bibfield  {journal} {\bibinfo  {journal}
  {JHEP}\ }\textbf {\bibinfo {volume} {01}},\ \bibinfo {pages} {137} (\bibinfo
  {year} {2022})},\ \Eprint {http://arxiv.org/abs/2107.03945} {arXiv:2107.03945
  [hep-ph]} \BibitemShut {NoStop}%
\bibitem [{\citenamefont {Kovtun}\ and\ \citenamefont
  {Yaffe}(2003)}]{Kovtun:2003vj}%
  \BibitemOpen
  \bibfield  {author} {\bibinfo {author} {\bibfnamefont {P.}~\bibnamefont
  {Kovtun}}\ and\ \bibinfo {author} {\bibfnamefont {L.~G.}\ \bibnamefont
  {Yaffe}},\ }\href {\doibase 10.1103/PhysRevD.68.025007} {\bibfield  {journal}
  {\bibinfo  {journal} {Phys. Rev. D}\ }\textbf {\bibinfo {volume} {68}},\
  \bibinfo {pages} {025007} (\bibinfo {year} {2003})},\ \Eprint
  {http://arxiv.org/abs/hep-th/0303010} {arXiv:hep-th/0303010} \BibitemShut
  {NoStop}%
\bibitem [{\citenamefont {Kovtun}(2012)}]{Kovtun:2012rj}%
  \BibitemOpen
  \bibfield  {author} {\bibinfo {author} {\bibfnamefont {P.}~\bibnamefont
  {Kovtun}},\ }\href {\doibase 10.1088/1751-8113/45/47/473001} {\bibfield
  {journal} {\bibinfo  {journal} {J. Phys. A}\ }\textbf {\bibinfo {volume}
  {45}},\ \bibinfo {pages} {473001} (\bibinfo {year} {2012})},\ \Eprint
  {http://arxiv.org/abs/1205.5040} {arXiv:1205.5040 [hep-th]} \BibitemShut
  {NoStop}%
\bibitem [{\citenamefont {Romatschke}(2021)}]{Romatschke:2021imm}%
  \BibitemOpen
  \bibfield  {author} {\bibinfo {author} {\bibfnamefont {P.}~\bibnamefont
  {Romatschke}},\ }\href {\doibase 10.1103/PhysRevLett.127.111603} {\bibfield
  {journal} {\bibinfo  {journal} {Phys. Rev. Lett.}\ }\textbf {\bibinfo
  {volume} {127}},\ \bibinfo {pages} {111603} (\bibinfo {year} {2021})},\
  \Eprint {http://arxiv.org/abs/2104.06435} {arXiv:2104.06435 [hep-th]}
  \BibitemShut {NoStop}%
\bibitem [{\citenamefont {Kogut}\ and\ \citenamefont
  {Susskind}(1975)}]{PhysRevD.11.395}%
  \BibitemOpen
  \bibfield  {author} {\bibinfo {author} {\bibfnamefont {J.}~\bibnamefont
  {Kogut}}\ and\ \bibinfo {author} {\bibfnamefont {L.}~\bibnamefont
  {Susskind}},\ }\href {\doibase 10.1103/PhysRevD.11.395} {\bibfield  {journal}
  {\bibinfo  {journal} {Phys. Rev. D}\ }\textbf {\bibinfo {volume} {11}},\
  \bibinfo {pages} {395} (\bibinfo {year} {1975})}\BibitemShut {NoStop}%
\bibitem [{\citenamefont {M\"uller}\ and\ \citenamefont
  {Yao}(2023)}]{Muller:2023nnk}%
  \BibitemOpen
  \bibfield  {author} {\bibinfo {author} {\bibfnamefont {B.}~\bibnamefont
  {M\"uller}}\ and\ \bibinfo {author} {\bibfnamefont {X.}~\bibnamefont {Yao}},\
  }\href {\doibase 10.1103/PhysRevD.108.094505} {\bibfield  {journal} {\bibinfo
   {journal} {Phys. Rev. D}\ }\textbf {\bibinfo {volume} {108}},\ \bibinfo
  {pages} {094505} (\bibinfo {year} {2023})}\BibitemShut {NoStop}%
\bibitem [{\citenamefont {Byrnes}\ and\ \citenamefont
  {Yamamoto}(2006)}]{Byrnes:2005qx}%
  \BibitemOpen
  \bibfield  {author} {\bibinfo {author} {\bibfnamefont {T.}~\bibnamefont
  {Byrnes}}\ and\ \bibinfo {author} {\bibfnamefont {Y.}~\bibnamefont
  {Yamamoto}},\ }\href {\doibase 10.1103/PhysRevA.73.022328} {\bibfield
  {journal} {\bibinfo  {journal} {Phys. Rev. A}\ }\textbf {\bibinfo {volume}
  {73}},\ \bibinfo {pages} {022328} (\bibinfo {year} {2006})}\BibitemShut
  {NoStop}%
\bibitem [{\citenamefont {Zohar}\ and\ \citenamefont
  {Burrello}(2015)}]{Zohar:2014qma}%
  \BibitemOpen
  \bibfield  {author} {\bibinfo {author} {\bibfnamefont {E.}~\bibnamefont
  {Zohar}}\ and\ \bibinfo {author} {\bibfnamefont {M.}~\bibnamefont
  {Burrello}},\ }\href {\doibase 10.1103/PhysRevD.91.054506} {\bibfield
  {journal} {\bibinfo  {journal} {Phys. Rev. D}\ }\textbf {\bibinfo {volume}
  {91}},\ \bibinfo {pages} {054506} (\bibinfo {year} {2015})},\ \Eprint
  {http://arxiv.org/abs/1409.3085} {arXiv:1409.3085 [quant-ph]} \BibitemShut
  {NoStop}%
\bibitem [{\citenamefont {Liu}\ and\ \citenamefont
  {Chandrasekharan}(2022)}]{Liu:2021tef}%
  \BibitemOpen
  \bibfield  {author} {\bibinfo {author} {\bibfnamefont {H.}~\bibnamefont
  {Liu}}\ and\ \bibinfo {author} {\bibfnamefont {S.}~\bibnamefont
  {Chandrasekharan}},\ }\href {\doibase 10.3390/sym14020305} {\bibfield
  {journal} {\bibinfo  {journal} {Symmetry}\ }\textbf {\bibinfo {volume}
  {14}},\ \bibinfo {pages} {305} (\bibinfo {year} {2022})},\ \Eprint
  {http://arxiv.org/abs/2112.02090} {arXiv:2112.02090 [hep-lat]} \BibitemShut
  {NoStop}%
\bibitem [{\citenamefont {Zache}\ \emph {et~al.}(2023)\citenamefont {Zache},
  \citenamefont {Gonz\'alez-Cuadra},\ and\ \citenamefont
  {Zoller}}]{Zache:2023dko}%
  \BibitemOpen
  \bibfield  {author} {\bibinfo {author} {\bibfnamefont {T.~V.}\ \bibnamefont
  {Zache}}, \bibinfo {author} {\bibfnamefont {D.}~\bibnamefont
  {Gonz\'alez-Cuadra}}, \ and\ \bibinfo {author} {\bibfnamefont
  {P.}~\bibnamefont {Zoller}},\ }\href {\doibase
  10.1103/PhysRevLett.131.171902} {\bibfield  {journal} {\bibinfo  {journal}
  {Phys. Rev. Lett.}\ }\textbf {\bibinfo {volume} {131}},\ \bibinfo {pages}
  {171902} (\bibinfo {year} {2023})},\ \Eprint
  {http://arxiv.org/abs/2304.02527} {arXiv:2304.02527 [quant-ph]} \BibitemShut
  {NoStop}%
\bibitem [{\citenamefont {Hayata}\ and\ \citenamefont
  {Hidaka}(2023)}]{Hayata:2023puo}%
  \BibitemOpen
  \bibfield  {author} {\bibinfo {author} {\bibfnamefont {T.}~\bibnamefont
  {Hayata}}\ and\ \bibinfo {author} {\bibfnamefont {Y.}~\bibnamefont
  {Hidaka}},\ }\href {\doibase 10.1007/JHEP09(2023)126} {\bibfield  {journal}
  {\bibinfo  {journal} {JHEP}\ ,\ \bibinfo {pages} {126}} (\bibinfo {year}
  {2023})},\ \Eprint {http://arxiv.org/abs/2305.05950} {arXiv:2305.05950
  [hep-lat]} \BibitemShut {NoStop}%
\bibitem [{\citenamefont {Klco}\ \emph {et~al.}(2020)\citenamefont {Klco},
  \citenamefont {Stryker},\ and\ \citenamefont {Savage}}]{Klco:2019evd}%
  \BibitemOpen
  \bibfield  {author} {\bibinfo {author} {\bibfnamefont {N.}~\bibnamefont
  {Klco}}, \bibinfo {author} {\bibfnamefont {J.~R.}\ \bibnamefont {Stryker}}, \
  and\ \bibinfo {author} {\bibfnamefont {M.~J.}\ \bibnamefont {Savage}},\
  }\href {\doibase 10.1103/PhysRevD.101.074512} {\bibfield  {journal} {\bibinfo
   {journal} {Phys. Rev. D}\ }\textbf {\bibinfo {volume} {101}},\ \bibinfo
  {pages} {074512} (\bibinfo {year} {2020})},\ \Eprint
  {http://arxiv.org/abs/1908.06935} {arXiv:1908.06935 [quant-ph]} \BibitemShut
  {NoStop}%
\bibitem [{\citenamefont {A~Rahman}\ \emph {et~al.}(2021)\citenamefont
  {A~Rahman}, \citenamefont {Lewis}, \citenamefont {Mendicelli},\ and\
  \citenamefont {Powell}}]{ARahman:2021ktn}%
  \BibitemOpen
  \bibfield  {author} {\bibinfo {author} {\bibfnamefont {S.}~\bibnamefont
  {A~Rahman}}, \bibinfo {author} {\bibfnamefont {R.}~\bibnamefont {Lewis}},
  \bibinfo {author} {\bibfnamefont {E.}~\bibnamefont {Mendicelli}}, \ and\
  \bibinfo {author} {\bibfnamefont {S.}~\bibnamefont {Powell}},\ }\href
  {\doibase 10.1103/PhysRevD.104.034501} {\bibfield  {journal} {\bibinfo
  {journal} {Phys. Rev. D}\ }\textbf {\bibinfo {volume} {104}},\ \bibinfo
  {pages} {034501} (\bibinfo {year} {2021})},\ \Eprint
  {http://arxiv.org/abs/2103.08661} {arXiv:2103.08661 [hep-lat]} \BibitemShut
  {NoStop}%
\bibitem [{\citenamefont {Hayata}\ \emph {et~al.}(2021)\citenamefont {Hayata},
  \citenamefont {Hidaka},\ and\ \citenamefont {Kikuchi}}]{Hayata:2021kcp}%
  \BibitemOpen
  \bibfield  {author} {\bibinfo {author} {\bibfnamefont {T.}~\bibnamefont
  {Hayata}}, \bibinfo {author} {\bibfnamefont {Y.}~\bibnamefont {Hidaka}}, \
  and\ \bibinfo {author} {\bibfnamefont {Y.}~\bibnamefont {Kikuchi}},\ }\href
  {\doibase 10.1103/PhysRevD.104.074518} {\bibfield  {journal} {\bibinfo
  {journal} {Phys. Rev. D}\ }\textbf {\bibinfo {volume} {104}},\ \bibinfo
  {pages} {074518} (\bibinfo {year} {2021})},\ \Eprint
  {http://arxiv.org/abs/2103.05179} {arXiv:2103.05179 [quant-ph]} \BibitemShut
  {NoStop}%
\bibitem [{\citenamefont {A~Rahman}\ \emph {et~al.}(2022)\citenamefont
  {A~Rahman}, \citenamefont {Lewis}, \citenamefont {Mendicelli},\ and\
  \citenamefont {Powell}}]{ARahman:2022tkr}%
  \BibitemOpen
  \bibfield  {author} {\bibinfo {author} {\bibfnamefont {S.}~\bibnamefont
  {A~Rahman}}, \bibinfo {author} {\bibfnamefont {R.}~\bibnamefont {Lewis}},
  \bibinfo {author} {\bibfnamefont {E.}~\bibnamefont {Mendicelli}}, \ and\
  \bibinfo {author} {\bibfnamefont {S.}~\bibnamefont {Powell}},\ }\href
  {\doibase 10.1103/PhysRevD.106.074502} {\bibfield  {journal} {\bibinfo
  {journal} {Phys. Rev. D}\ }\textbf {\bibinfo {volume} {106}},\ \bibinfo
  {pages} {074502} (\bibinfo {year} {2022})},\ \Eprint
  {http://arxiv.org/abs/2205.09247} {arXiv:2205.09247 [hep-lat]} \BibitemShut
  {NoStop}%
\bibitem [{\citenamefont {Yao}(2023)}]{Yao:2023pht}%
  \BibitemOpen
  \bibfield  {author} {\bibinfo {author} {\bibfnamefont {X.}~\bibnamefont
  {Yao}},\ }\href {\doibase 10.1103/PhysRevD.108.L031504} {\bibfield  {journal}
  {\bibinfo  {journal} {Phys. Rev. D}\ }\textbf {\bibinfo {volume} {108}},\
  \bibinfo {pages} {L031504} (\bibinfo {year} {2023})},\ \Eprint
  {http://arxiv.org/abs/2303.14264} {arXiv:2303.14264 [hep-lat]} \BibitemShut
  {NoStop}%
\bibitem [{\citenamefont {Turro}\ \emph {et~al.}(2022)\citenamefont {Turro},
  \citenamefont {Roggero}, \citenamefont {Amitrano}, \citenamefont {Luchi},
  \citenamefont {Wendt}, \citenamefont {Dubois}, \citenamefont {Quaglioni},\
  and\ \citenamefont {Pederiva}}]{turro_QITP_2022}%
  \BibitemOpen
  \bibfield  {author} {\bibinfo {author} {\bibfnamefont {F.}~\bibnamefont
  {Turro}}, \bibinfo {author} {\bibfnamefont {A.}~\bibnamefont {Roggero}},
  \bibinfo {author} {\bibfnamefont {V.}~\bibnamefont {Amitrano}}, \bibinfo
  {author} {\bibfnamefont {P.}~\bibnamefont {Luchi}}, \bibinfo {author}
  {\bibfnamefont {K.~A.}\ \bibnamefont {Wendt}}, \bibinfo {author}
  {\bibfnamefont {J.~L.}\ \bibnamefont {Dubois}}, \bibinfo {author}
  {\bibfnamefont {S.}~\bibnamefont {Quaglioni}}, \ and\ \bibinfo {author}
  {\bibfnamefont {F.}~\bibnamefont {Pederiva}},\ }\href {\doibase
  10.1103/PhysRevA.105.022440} {\bibfield  {journal} {\bibinfo  {journal}
  {Phys. Rev. A}\ }\textbf {\bibinfo {volume} {105}},\ \bibinfo {pages}
  {022440} (\bibinfo {year} {2022})}\BibitemShut {NoStop}%
\bibitem [{\citenamefont {Turro}(2023)}]{franceschino_thermal2023}%
  \BibitemOpen
  \bibfield  {author} {\bibinfo {author} {\bibfnamefont {F.}~\bibnamefont
  {Turro}},\ }\href {\doibase 10.48550/arXiv.2306.16580} {\enquote {\bibinfo
  {title} {Quantum imaginary time propagation algorithm for preparing thermal
  states},}\ } (\bibinfo {year} {2023}),\ \Eprint
  {http://arxiv.org/abs/2306.16580} {arXiv:2306.16580 [quant-ph]} \BibitemShut
  {NoStop}%
\bibitem [{\citenamefont {Nielsen}\ and\ \citenamefont
  {Chuang}(2012)}]{Nielsen:2012yss}%
  \BibitemOpen
  \bibfield  {author} {\bibinfo {author} {\bibfnamefont {M.~A.}\ \bibnamefont
  {Nielsen}}\ and\ \bibinfo {author} {\bibfnamefont {I.~L.}\ \bibnamefont
  {Chuang}},\ }\href {\doibase 10.1017/cbo9780511976667} {\emph {\bibinfo
  {title} {{Quantum Computation and Quantum Information}}}}\ (\bibinfo
  {publisher} {Cambridge University Press},\ \bibinfo {year}
  {2012})\BibitemShut {NoStop}%
\bibitem [{\citenamefont {Mitarai}\ \emph {et~al.}(2018)\citenamefont
  {Mitarai}, \citenamefont {Negoro}, \citenamefont {Kitagawa},\ and\
  \citenamefont {Fujii}}]{Gradient_pi_4}%
  \BibitemOpen
  \bibfield  {author} {\bibinfo {author} {\bibfnamefont {K.}~\bibnamefont
  {Mitarai}}, \bibinfo {author} {\bibfnamefont {M.}~\bibnamefont {Negoro}},
  \bibinfo {author} {\bibfnamefont {M.}~\bibnamefont {Kitagawa}}, \ and\
  \bibinfo {author} {\bibfnamefont {K.}~\bibnamefont {Fujii}},\ }\href
  {\doibase 10.1103/PhysRevA.98.032309} {\bibfield  {journal} {\bibinfo
  {journal} {Phys. Rev. A}\ }\textbf {\bibinfo {volume} {98}},\ \bibinfo
  {pages} {032309} (\bibinfo {year} {2018})}\BibitemShut {NoStop}%
\bibitem [{\citenamefont {Wierichs}\ \emph {et~al.}(2022)\citenamefont
  {Wierichs}, \citenamefont {Izaac}, \citenamefont {Wang},\ and\ \citenamefont
  {Lin}}]{Gradient_parameter_shift}%
  \BibitemOpen
  \bibfield  {author} {\bibinfo {author} {\bibfnamefont {D.}~\bibnamefont
  {Wierichs}}, \bibinfo {author} {\bibfnamefont {J.}~\bibnamefont {Izaac}},
  \bibinfo {author} {\bibfnamefont {C.}~\bibnamefont {Wang}}, \ and\ \bibinfo
  {author} {\bibfnamefont {C.~Y.-Y.}\ \bibnamefont {Lin}},\ }\href {\doibase
  10.22331/q-2022-03-30-677} {\bibfield  {journal} {\bibinfo  {journal}
  {{Quantum}}\ }\textbf {\bibinfo {volume} {6}},\ \bibinfo {pages} {677}
  (\bibinfo {year} {2022})}\BibitemShut {NoStop}%
\bibitem [{\citenamefont {Ebner}\ \emph
  {et~al.}(2024{\natexlab{a}})\citenamefont {Ebner}, \citenamefont {Sch\"afer},
  \citenamefont {Seidl}, \citenamefont {M\"uller},\ and\ \citenamefont
  {Yao}}]{Ebner:2024mee}%
  \BibitemOpen
  \bibfield  {author} {\bibinfo {author} {\bibfnamefont {L.}~\bibnamefont
  {Ebner}}, \bibinfo {author} {\bibfnamefont {A.}~\bibnamefont {Sch\"afer}},
  \bibinfo {author} {\bibfnamefont {C.}~\bibnamefont {Seidl}}, \bibinfo
  {author} {\bibfnamefont {B.}~\bibnamefont {M\"uller}}, \ and\ \bibinfo
  {author} {\bibfnamefont {X.}~\bibnamefont {Yao}},\ }\href@noop {} {\
  (\bibinfo {year} {2024}{\natexlab{a}})},\ \Eprint
  {http://arxiv.org/abs/2401.15184} {arXiv:2401.15184 [hep-lat]} \BibitemShut
  {NoStop}%
\bibitem [{\citenamefont {Romatschke}(2020)}]{Romatschke:2019nmo}%
  \BibitemOpen
  \bibfield  {author} {\bibinfo {author} {\bibfnamefont {P.}~\bibnamefont
  {Romatschke}},\ }\href {\doibase 10.1007/JHEP03(2020)174} {\bibfield
  {journal} {\bibinfo  {journal} {JHEP}\ ,\ \bibinfo {pages} {174}} (\bibinfo
  {year} {2020})},\ \Eprint {http://arxiv.org/abs/1910.09550} {arXiv:1910.09550
  [hep-lat]} \BibitemShut {NoStop}%
\bibitem [{\citenamefont {L\"uscher}(2010)}]{Luscher:2010iy}%
  \BibitemOpen
  \bibfield  {author} {\bibinfo {author} {\bibfnamefont {M.}~\bibnamefont
  {L\"uscher}},\ }\href {\doibase 10.1007/JHEP08(2010)071} {\bibfield
  {journal} {\bibinfo  {journal} {JHEP}\ }\textbf {\bibinfo {volume} {08}},\
  \bibinfo {pages} {071} (\bibinfo {year} {2010})},\ \bibinfo {note} {[Erratum:
  JHEP 03, 092 (2014)]},\ \Eprint {http://arxiv.org/abs/1006.4518}
  {arXiv:1006.4518 [hep-lat]} \BibitemShut {NoStop}%
\bibitem [{\citenamefont {Luscher}\ and\ \citenamefont
  {Weisz}(2011)}]{Luscher:2011bx}%
  \BibitemOpen
  \bibfield  {author} {\bibinfo {author} {\bibfnamefont {M.}~\bibnamefont
  {Luscher}}\ and\ \bibinfo {author} {\bibfnamefont {P.}~\bibnamefont
  {Weisz}},\ }\href {\doibase 10.1007/JHEP02(2011)051} {\bibfield  {journal}
  {\bibinfo  {journal} {JHEP}\ }\textbf {\bibinfo {volume} {02}},\ \bibinfo
  {pages} {051} (\bibinfo {year} {2011})},\ \Eprint
  {http://arxiv.org/abs/1101.0963} {arXiv:1101.0963 [hep-th]} \BibitemShut
  {NoStop}%
\bibitem [{\citenamefont {Ebner}\ \emph
  {et~al.}(2024{\natexlab{b}})\citenamefont {Ebner}, \citenamefont {M\"uller},
  \citenamefont {Sch\"afer}, \citenamefont {Seidl},\ and\ \citenamefont
  {Yao}}]{Ebner:2023ixq}%
  \BibitemOpen
  \bibfield  {author} {\bibinfo {author} {\bibfnamefont {L.}~\bibnamefont
  {Ebner}}, \bibinfo {author} {\bibfnamefont {B.}~\bibnamefont {M\"uller}},
  \bibinfo {author} {\bibfnamefont {A.}~\bibnamefont {Sch\"afer}}, \bibinfo
  {author} {\bibfnamefont {C.}~\bibnamefont {Seidl}}, \ and\ \bibinfo {author}
  {\bibfnamefont {X.}~\bibnamefont {Yao}},\ }\href {\doibase
  10.1103/PhysRevD.109.014504} {\bibfield  {journal} {\bibinfo  {journal}
  {Phys. Rev. D}\ }\textbf {\bibinfo {volume} {109}},\ \bibinfo {pages}
  {014504} (\bibinfo {year} {2024}{\natexlab{b}})},\ \Eprint
  {http://arxiv.org/abs/2308.16202} {arXiv:2308.16202 [hep-lat]} \BibitemShut
  {NoStop}%
\bibitem [{ber()}]{berndt}%
  \BibitemOpen
  \href@noop {} {}\bibinfo {note} {{Berndt M\"uller, private
  communication}}\BibitemShut {NoStop}%
\bibitem [{ste()}]{steve}%
  \BibitemOpen
  \href@noop {} {}\bibinfo {note} {{Steve Sharpe, private
  communication}}\BibitemShut {NoStop}%
\bibitem [{\citenamefont {Moore}\ and\ \citenamefont
  {Saremi}(2008)}]{Moore:2008ws}%
  \BibitemOpen
  \bibfield  {author} {\bibinfo {author} {\bibfnamefont {G.~D.}\ \bibnamefont
  {Moore}}\ and\ \bibinfo {author} {\bibfnamefont {O.}~\bibnamefont {Saremi}},\
  }\href {\doibase 10.1088/1126-6708/2008/09/015} {\bibfield  {journal}
  {\bibinfo  {journal} {JHEP}\ }\textbf {\bibinfo {volume} {2008}},\ \bibinfo
  {pages} {015} (\bibinfo {year} {2008})}\BibitemShut {NoStop}%
\bibitem [{\citenamefont {Zhu}\ and\ \citenamefont
  {Vuorinen}(2013)}]{Zhu:2012be}%
  \BibitemOpen
  \bibfield  {author} {\bibinfo {author} {\bibfnamefont {Y.}~\bibnamefont
  {Zhu}}\ and\ \bibinfo {author} {\bibfnamefont {A.}~\bibnamefont {Vuorinen}},\
  }\href {\doibase 10.1007/JHEP03(2013)002} {\bibfield  {journal} {\bibinfo
  {journal} {J. High Energ. Phys.}\ }\textbf {\bibinfo {volume} {2013}},\
  \bibinfo {pages} {1} (\bibinfo {year} {2013})}\BibitemShut {NoStop}%
\bibitem [{qua(2024)}]{quantinuum}%
  \BibitemOpen
  \href {https://www.quantinuum.com/} {\enquote {\bibinfo {title}
  {Quantinuum},}\ } (\bibinfo {year} {2024})\BibitemShut {NoStop}%
\bibitem [{\citenamefont {Farrell}\ \emph {et~al.}(2023)\citenamefont
  {Farrell}, \citenamefont {Chernyshev}, \citenamefont {Powell}, \citenamefont
  {Zemlevskiy}, \citenamefont {Illa},\ and\ \citenamefont
  {Savage}}]{quantinuum_iqus}%
  \BibitemOpen
  \bibfield  {author} {\bibinfo {author} {\bibfnamefont {R.~C.}\ \bibnamefont
  {Farrell}}, \bibinfo {author} {\bibfnamefont {I.~A.}\ \bibnamefont
  {Chernyshev}}, \bibinfo {author} {\bibfnamefont {S.~J.~M.}\ \bibnamefont
  {Powell}}, \bibinfo {author} {\bibfnamefont {N.~A.}\ \bibnamefont
  {Zemlevskiy}}, \bibinfo {author} {\bibfnamefont {M.}~\bibnamefont {Illa}}, \
  and\ \bibinfo {author} {\bibfnamefont {M.~J.}\ \bibnamefont {Savage}},\
  }\href {\doibase 10.1103/PhysRevD.107.054513} {\bibfield  {journal} {\bibinfo
   {journal} {Phys. Rev. D}\ }\textbf {\bibinfo {volume} {107}},\ \bibinfo
  {pages} {054513} (\bibinfo {year} {2023})}\BibitemShut {NoStop}%
\bibitem [{\citenamefont {Illa}\ and\ \citenamefont
  {Savage}(2023)}]{neutrinos_marc}%
  \BibitemOpen
  \bibfield  {author} {\bibinfo {author} {\bibfnamefont {M.}~\bibnamefont
  {Illa}}\ and\ \bibinfo {author} {\bibfnamefont {M.~J.}\ \bibnamefont
  {Savage}},\ }\href {\doibase 10.1103/PhysRevLett.130.221003} {\bibfield
  {journal} {\bibinfo  {journal} {Phys. Rev. Lett.}\ }\textbf {\bibinfo
  {volume} {130}},\ \bibinfo {pages} {221003} (\bibinfo {year}
  {2023})}\BibitemShut {NoStop}%
\bibitem [{IBM(2022)}]{IBM}%
  \BibitemOpen
  \href {https://quantum.ibm.com/} {\enquote {\bibinfo {title} {{IBM
  Quantum}},}\ } (\bibinfo {year} {2022})\BibitemShut {NoStop}%
\bibitem [{\citenamefont {{Qiskit contributors}}(2023)}]{Qiskit}%
  \BibitemOpen
  \bibfield  {author} {\bibinfo {author} {\bibnamefont {{Qiskit
  contributors}}},\ }\href {\doibase 10.5281/zenodo.2573505} {\enquote
  {\bibinfo {title} {Qiskit: An open-source framework for quantum computing},}\
  } (\bibinfo {year} {2023})\BibitemShut {NoStop}%
\bibitem [{\citenamefont {Raychowdhury}\ and\ \citenamefont
  {Stryker}(2020)}]{Raychowdhury:2019iki}%
  \BibitemOpen
  \bibfield  {author} {\bibinfo {author} {\bibfnamefont {I.}~\bibnamefont
  {Raychowdhury}}\ and\ \bibinfo {author} {\bibfnamefont {J.~R.}\ \bibnamefont
  {Stryker}},\ }\href {\doibase 10.1103/PhysRevD.101.114502} {\bibfield
  {journal} {\bibinfo  {journal} {Phys. Rev. D}\ }\textbf {\bibinfo {volume}
  {101}},\ \bibinfo {pages} {114502} (\bibinfo {year} {2020})},\ \Eprint
  {http://arxiv.org/abs/1912.06133} {arXiv:1912.06133 [hep-lat]} \BibitemShut
  {NoStop}%
\bibitem [{\citenamefont {Ciavarella}\ \emph {et~al.}(2021)\citenamefont
  {Ciavarella}, \citenamefont {Klco},\ and\ \citenamefont
  {Savage}}]{Ciavarella:2021nmj}%
  \BibitemOpen
  \bibfield  {author} {\bibinfo {author} {\bibfnamefont {A.}~\bibnamefont
  {Ciavarella}}, \bibinfo {author} {\bibfnamefont {N.}~\bibnamefont {Klco}}, \
  and\ \bibinfo {author} {\bibfnamefont {M.~J.}\ \bibnamefont {Savage}},\
  }\href {\doibase 10.1103/PhysRevD.103.094501} {\bibfield  {journal} {\bibinfo
   {journal} {Phys. Rev. D}\ }\textbf {\bibinfo {volume} {103}},\ \bibinfo
  {pages} {094501} (\bibinfo {year} {2021})},\ \Eprint
  {http://arxiv.org/abs/2101.10227} {arXiv:2101.10227 [quant-ph]} \BibitemShut
  {NoStop}%
\bibitem [{\citenamefont {Kadam}\ \emph {et~al.}(2023)\citenamefont {Kadam},
  \citenamefont {Raychowdhury},\ and\ \citenamefont {Stryker}}]{Kadam:2022ipf}%
  \BibitemOpen
  \bibfield  {author} {\bibinfo {author} {\bibfnamefont {S.~V.}\ \bibnamefont
  {Kadam}}, \bibinfo {author} {\bibfnamefont {I.}~\bibnamefont {Raychowdhury}},
  \ and\ \bibinfo {author} {\bibfnamefont {J.~R.}\ \bibnamefont {Stryker}},\
  }\href {\doibase 10.1103/PhysRevD.107.094513} {\bibfield  {journal} {\bibinfo
   {journal} {Phys. Rev. D}\ }\textbf {\bibinfo {volume} {107}},\ \bibinfo
  {pages} {094513} (\bibinfo {year} {2023})},\ \Eprint
  {http://arxiv.org/abs/2212.04490} {arXiv:2212.04490 [hep-lat]} \BibitemShut
  {NoStop}%
\bibitem [{\citenamefont {Cataldi}\ \emph {et~al.}(2023)\citenamefont
  {Cataldi}, \citenamefont {Magnifico}, \citenamefont {Silvi},\ and\
  \citenamefont {Montangero}}]{Cataldi:2023xki}%
  \BibitemOpen
  \bibfield  {author} {\bibinfo {author} {\bibfnamefont {G.}~\bibnamefont
  {Cataldi}}, \bibinfo {author} {\bibfnamefont {G.}~\bibnamefont {Magnifico}},
  \bibinfo {author} {\bibfnamefont {P.}~\bibnamefont {Silvi}}, \ and\ \bibinfo
  {author} {\bibfnamefont {S.}~\bibnamefont {Montangero}},\ }\href@noop {} {\
  (\bibinfo {year} {2023})},\ \Eprint {http://arxiv.org/abs/2307.09396}
  {arXiv:2307.09396 [hep-lat]} \BibitemShut {NoStop}%
\bibitem [{\citenamefont {Kavaki}\ and\ \citenamefont
  {Lewis}(2024)}]{Kavaki:2024ijd}%
  \BibitemOpen
  \bibfield  {author} {\bibinfo {author} {\bibfnamefont {A.~H.~Z.}\
  \bibnamefont {Kavaki}}\ and\ \bibinfo {author} {\bibfnamefont
  {R.}~\bibnamefont {Lewis}},\ }\href@noop {} {\  (\bibinfo {year} {2024})},\
  \Eprint {http://arxiv.org/abs/2401.14570} {arXiv:2401.14570 [hep-lat]}
  \BibitemShut {NoStop}%
\bibitem [{\citenamefont {Casalderrey-Solana}\ and\ \citenamefont
  {Teaney}(2006)}]{Casalderrey-Solana:2006fio}%
  \BibitemOpen
  \bibfield  {author} {\bibinfo {author} {\bibfnamefont {J.}~\bibnamefont
  {Casalderrey-Solana}}\ and\ \bibinfo {author} {\bibfnamefont
  {D.}~\bibnamefont {Teaney}},\ }\href {\doibase 10.1103/PhysRevD.74.085012}
  {\bibfield  {journal} {\bibinfo  {journal} {Phys. Rev. D}\ }\textbf {\bibinfo
  {volume} {74}},\ \bibinfo {pages} {085012} (\bibinfo {year} {2006})},\
  \Eprint {http://arxiv.org/abs/hep-ph/0605199} {arXiv:hep-ph/0605199}
  \BibitemShut {NoStop}%
\bibitem [{\citenamefont {Caron-Huot}\ \emph {et~al.}(2009)\citenamefont
  {Caron-Huot}, \citenamefont {Laine},\ and\ \citenamefont
  {Moore}}]{Caron-Huot:2009ncn}%
  \BibitemOpen
  \bibfield  {author} {\bibinfo {author} {\bibfnamefont {S.}~\bibnamefont
  {Caron-Huot}}, \bibinfo {author} {\bibfnamefont {M.}~\bibnamefont {Laine}}, \
  and\ \bibinfo {author} {\bibfnamefont {G.~D.}\ \bibnamefont {Moore}},\ }\href
  {\doibase 10.1088/1126-6708/2009/04/053} {\bibfield  {journal} {\bibinfo
  {journal} {JHEP}\ }\textbf {\bibinfo {volume} {04}},\ \bibinfo {pages} {053}
  (\bibinfo {year} {2009})},\ \Eprint {http://arxiv.org/abs/0901.1195}
  {arXiv:0901.1195 [hep-lat]} \BibitemShut {NoStop}%
\bibitem [{\citenamefont {Brambilla}\ \emph {et~al.}(2023)\citenamefont
  {Brambilla}, \citenamefont {Leino}, \citenamefont {Mayer-Steudte},\ and\
  \citenamefont {Petreczky}}]{Brambilla:2022xbd}%
  \BibitemOpen
  \bibfield  {author} {\bibinfo {author} {\bibfnamefont {N.}~\bibnamefont
  {Brambilla}}, \bibinfo {author} {\bibfnamefont {V.}~\bibnamefont {Leino}},
  \bibinfo {author} {\bibfnamefont {J.}~\bibnamefont {Mayer-Steudte}}, \ and\
  \bibinfo {author} {\bibfnamefont {P.}~\bibnamefont {Petreczky}} (\bibinfo
  {collaboration} {TUMQCD}),\ }\href {\doibase 10.1103/PhysRevD.107.054508}
  {\bibfield  {journal} {\bibinfo  {journal} {Phys. Rev. D}\ }\textbf {\bibinfo
  {volume} {107}},\ \bibinfo {pages} {054508} (\bibinfo {year} {2023})},\
  \Eprint {http://arxiv.org/abs/2206.02861} {arXiv:2206.02861 [hep-lat]}
  \BibitemShut {NoStop}%
\bibitem [{\citenamefont {Altenkort}\ \emph
  {et~al.}(2023{\natexlab{b}})\citenamefont {Altenkort}, \citenamefont
  {Kaczmarek}, \citenamefont {Larsen}, \citenamefont {Mukherjee}, \citenamefont
  {Petreczky}, \citenamefont {Shu},\ and\ \citenamefont
  {Stendebach}}]{Altenkort:2023oms}%
  \BibitemOpen
  \bibfield  {author} {\bibinfo {author} {\bibfnamefont {L.}~\bibnamefont
  {Altenkort}}, \bibinfo {author} {\bibfnamefont {O.}~\bibnamefont
  {Kaczmarek}}, \bibinfo {author} {\bibfnamefont {R.}~\bibnamefont {Larsen}},
  \bibinfo {author} {\bibfnamefont {S.}~\bibnamefont {Mukherjee}}, \bibinfo
  {author} {\bibfnamefont {P.}~\bibnamefont {Petreczky}}, \bibinfo {author}
  {\bibfnamefont {H.-T.}\ \bibnamefont {Shu}}, \ and\ \bibinfo {author}
  {\bibfnamefont {S.}~\bibnamefont {Stendebach}} (\bibinfo {collaboration}
  {HotQCD}),\ }\href {\doibase 10.1103/PhysRevLett.130.231902} {\bibfield
  {journal} {\bibinfo  {journal} {Phys. Rev. Lett.}\ }\textbf {\bibinfo
  {volume} {130}},\ \bibinfo {pages} {231902} (\bibinfo {year}
  {2023}{\natexlab{b}})},\ \Eprint {http://arxiv.org/abs/2302.08501}
  {arXiv:2302.08501 [hep-lat]} \BibitemShut {NoStop}%
\bibitem [{\citenamefont {Brambilla}\ \emph {et~al.}(2017)\citenamefont
  {Brambilla}, \citenamefont {Escobedo}, \citenamefont {Soto},\ and\
  \citenamefont {Vairo}}]{Brambilla:2016wgg}%
  \BibitemOpen
  \bibfield  {author} {\bibinfo {author} {\bibfnamefont {N.}~\bibnamefont
  {Brambilla}}, \bibinfo {author} {\bibfnamefont {M.~A.}\ \bibnamefont
  {Escobedo}}, \bibinfo {author} {\bibfnamefont {J.}~\bibnamefont {Soto}}, \
  and\ \bibinfo {author} {\bibfnamefont {A.}~\bibnamefont {Vairo}},\ }\href
  {\doibase 10.1103/PhysRevD.96.034021} {\bibfield  {journal} {\bibinfo
  {journal} {Phys. Rev. D}\ }\textbf {\bibinfo {volume} {96}},\ \bibinfo
  {pages} {034021} (\bibinfo {year} {2017})},\ \Eprint
  {http://arxiv.org/abs/1612.07248} {arXiv:1612.07248 [hep-ph]} \BibitemShut
  {NoStop}%
\bibitem [{\citenamefont {Brambilla}\ \emph {et~al.}(2018)\citenamefont
  {Brambilla}, \citenamefont {Escobedo}, \citenamefont {Soto},\ and\
  \citenamefont {Vairo}}]{Brambilla:2017zei}%
  \BibitemOpen
  \bibfield  {author} {\bibinfo {author} {\bibfnamefont {N.}~\bibnamefont
  {Brambilla}}, \bibinfo {author} {\bibfnamefont {M.~A.}\ \bibnamefont
  {Escobedo}}, \bibinfo {author} {\bibfnamefont {J.}~\bibnamefont {Soto}}, \
  and\ \bibinfo {author} {\bibfnamefont {A.}~\bibnamefont {Vairo}},\ }\href
  {\doibase 10.1103/PhysRevD.97.074009} {\bibfield  {journal} {\bibinfo
  {journal} {Phys. Rev. D}\ }\textbf {\bibinfo {volume} {97}},\ \bibinfo
  {pages} {074009} (\bibinfo {year} {2018})},\ \Eprint
  {http://arxiv.org/abs/1711.04515} {arXiv:1711.04515 [hep-ph]} \BibitemShut
  {NoStop}%
\bibitem [{\citenamefont {Scheihing-Hitschfeld}\ and\ \citenamefont
  {Yao}(2023{\natexlab{a}})}]{Scheihing-Hitschfeld:2023tuz}%
  \BibitemOpen
  \bibfield  {author} {\bibinfo {author} {\bibfnamefont {B.}~\bibnamefont
  {Scheihing-Hitschfeld}}\ and\ \bibinfo {author} {\bibfnamefont
  {X.}~\bibnamefont {Yao}},\ }\href {\doibase 10.1103/PhysRevD.108.054024}
  {\bibfield  {journal} {\bibinfo  {journal} {Phys. Rev. D}\ }\textbf {\bibinfo
  {volume} {108}},\ \bibinfo {pages} {054024} (\bibinfo {year}
  {2023}{\natexlab{a}})},\ \Eprint {http://arxiv.org/abs/2306.13127}
  {arXiv:2306.13127 [hep-ph]} \BibitemShut {NoStop}%
\bibitem [{\citenamefont {Scheihing-Hitschfeld}\ and\ \citenamefont
  {Yao}(2023{\natexlab{b}})}]{Scheihing-Hitschfeld:2022xqx}%
  \BibitemOpen
  \bibfield  {author} {\bibinfo {author} {\bibfnamefont {B.}~\bibnamefont
  {Scheihing-Hitschfeld}}\ and\ \bibinfo {author} {\bibfnamefont
  {X.}~\bibnamefont {Yao}},\ }\href {\doibase 10.1103/PhysRevLett.130.052302}
  {\bibfield  {journal} {\bibinfo  {journal} {Phys. Rev. Lett.}\ }\textbf
  {\bibinfo {volume} {130}},\ \bibinfo {pages} {052302} (\bibinfo {year}
  {2023}{\natexlab{b}})},\ \Eprint {http://arxiv.org/abs/2205.04477}
  {arXiv:2205.04477 [hep-ph]} \BibitemShut {NoStop}%
\bibitem [{\citenamefont {Nijs}\ \emph {et~al.}(2023)\citenamefont {Nijs},
  \citenamefont {Scheihing-Hitschfeld},\ and\ \citenamefont
  {Yao}}]{Nijs:2023dbc}%
  \BibitemOpen
  \bibfield  {author} {\bibinfo {author} {\bibfnamefont {G.}~\bibnamefont
  {Nijs}}, \bibinfo {author} {\bibfnamefont {B.}~\bibnamefont
  {Scheihing-Hitschfeld}}, \ and\ \bibinfo {author} {\bibfnamefont
  {X.}~\bibnamefont {Yao}},\ }\href@noop {} {\  (\bibinfo {year} {2023})},\
  \Eprint {http://arxiv.org/abs/2310.09325} {arXiv:2310.09325 [hep-ph]}
  \BibitemShut {NoStop}%
\bibitem [{\citenamefont {Leino}(2024)}]{Leino:2024pen}%
  \BibitemOpen
  \bibfield  {author} {\bibinfo {author} {\bibfnamefont {V.}~\bibnamefont
  {Leino}},\ }in\ \href@noop {} {\emph {\bibinfo {booktitle} {{40th
  International Symposium on Lattice Field Theory}}}}\ (\bibinfo {year}
  {2024})\ \Eprint {http://arxiv.org/abs/2401.06733} {arXiv:2401.06733
  [hep-lat]} \BibitemShut {NoStop}%
\bibitem [{\citenamefont {Jordan}\ \emph {et~al.}(2012)\citenamefont {Jordan},
  \citenamefont {Lee},\ and\ \citenamefont {Preskill}}]{Jordan:2012xnu}%
  \BibitemOpen
  \bibfield  {author} {\bibinfo {author} {\bibfnamefont {S.~P.}\ \bibnamefont
  {Jordan}}, \bibinfo {author} {\bibfnamefont {K.~S.~M.}\ \bibnamefont {Lee}},
  \ and\ \bibinfo {author} {\bibfnamefont {J.}~\bibnamefont {Preskill}},\
  }\href {\doibase 10.1126/science.1217069} {\bibfield  {journal} {\bibinfo
  {journal} {Science}\ }\textbf {\bibinfo {volume} {336}},\ \bibinfo {pages}
  {1130} (\bibinfo {year} {2012})},\ \Eprint {http://arxiv.org/abs/1111.3633}
  {arXiv:1111.3633 [quant-ph]} \BibitemShut {NoStop}%
\bibitem [{\citenamefont {Vidal}(2003)}]{MPS_review_vidal}%
  \BibitemOpen
  \bibfield  {author} {\bibinfo {author} {\bibfnamefont {G.}~\bibnamefont
  {Vidal}},\ }\href {\doibase 10.1103/PhysRevLett.91.147902} {\bibfield
  {journal} {\bibinfo  {journal} {Phys. Rev. Lett.}\ }\textbf {\bibinfo
  {volume} {91}},\ \bibinfo {pages} {147902} (\bibinfo {year}
  {2003})}\BibitemShut {NoStop}%
\bibitem [{\citenamefont {Schollw\"{o}ck}(2011)}]{MPS_review_Schollwock}%
  \BibitemOpen
  \bibfield  {author} {\bibinfo {author} {\bibfnamefont {U.}~\bibnamefont
  {Schollw\"{o}ck}},\ }\href {\doibase
  https://doi.org/10.1016/j.aop.2010.09.012} {\bibfield  {journal} {\bibinfo
  {journal} {Annals of Physics}\ }\textbf {\bibinfo {volume} {326}},\ \bibinfo
  {pages} {96} (\bibinfo {year} {2011})},\ \bibinfo {note} {january 2011
  Special Issue}\BibitemShut {NoStop}%
\end{thebibliography}%
\bibliographystyle{apsrev4-1}
\end{document}